\documentclass[12pt,a4paper]{article}
\usepackage{axodraw}
\usepackage{epsfig}

\let\log\ln

\advance\textheight by \topskip
\if@twoside \oddsidemargin -17pt \evensidemargin 00pt
\else \oddsidemargin 00pt \evensidemargin 00pt
\fi
\expandafter\ifx\csname mathrm\endcsname\relax\def\mathrm#1{{\rm #1}}\fi

\renewcommand{\theequation}{\thesection.\arabic{equation}}
\newcounter{saveeqn}

\makeatletter 
\@addtoreset{equation}{section}
\makeatother

\def\beq{\begin{equation}}
\def\eeq{\end{equation}}
\def\beqar{\begin{eqnarray}}
\def\eeqar{\end{eqnarray}}
\def\barr#1{\begin{array}{#1}}
\def\earr{\end{array}}
\def\bfi{\begin{figure}}
\def\efi{\end{figure}}
\def\btab{\begin{table}}
\def\etab{\end{table}}
\def\bce{\begin{center}}
\def\ece{\end{center}}
\def\nn{\nonumber}
\def\nl{\nonumber\\}
\def\nln{\nonumber\\*[-1ex]\phantom{\fbox{\rule{0em}{2ex}}}}
\def\co{,}
\def\nlc{\co\nonumber\\}

\def\al{\alpha}

\def\ga{\gamma}
\def\de{\delta}

\def\eps{\epsilon}
\def\la{\lambda}
\def\si{\sigma}
\def\Ga{\Gamma}
\def\De{\Delta}

\def\refeq#1{\mbox{(\ref{#1})}}
\def\refeqs#1{\mbox{(\ref{#1})}}
\def\refeqf#1{\mbox{(\ref{#1})}}
\def\reffi#1{\mbox{Fig.~\ref{#1}}}

\def\refta#1{\mbox{Table~\ref{#1}}}
\def\reftas#1{\mbox{Tables~\ref{#1}}}
\def\refse#1{\mbox{Sect.~\ref{#1}}}

\def\refapp#1{\mbox{Appendix~\ref{#1}}}
\def\citere#1{\mbox{Ref.~\cite{#1}}}
\def\citeres#1{\mbox{Refs.~\cite{#1}}}

\def\solid{\raise.9mm\hbox{\protect\rule{1.1cm}{.2mm}}}
\def\dash{\raise.9mm\hbox{\protect\rule{2mm}{.2mm}}\hspace*{1mm}}
\def\dot{\rlap{$\cdot$}\hspace*{2mm}}

\def\solid{\raise.9mm\hbox{\protect\rule{12mm}{.2mm}}}
\def\dash{\raise.9mm\hbox{\protect\rule{1.6mm}{.2mm}}\hspace*{1mm}}
\def\dot{\raise.9mm\hbox{\protect\rule{0.8mm}{.2mm}}\hspace*{0.8mm}}
\def\dashdot{\raise.9mm\hbox{\protect\rule{.3mm}{.2mm}}\hspace*{.8mm}\raise.9mm\hbox{\protect\rule{1.3mm}{.2mm}}\hspace*{.8mm}}

\newcommand{\GeV}{\unskip\,\mathrm{GeV}}

\newcommand{\TeV}{\unskip\,\mathrm{TeV}}
\newcommand{\fba}{\unskip\,\mathrm{fb}}

\def\mathswitchr#1{\relax\ifmmode{\mathrm{#1}}\else$\mathrm{#1}$\fi}

\newcommand{\PW}{\mathswitchr W}
\newcommand{\PZ}{\mathswitchr Z}

\newcommand{\Pe}{\mathswitchr e}
\newcommand{\Pne}{\mathswitch \nu_{\mathrm{e}}}

\newcommand{\Pnebar}{\mathswitch \bar\nu_{\mathrm{e}}}

\newcommand{\Pd}{\mathswitchr d}
\newcommand{\PD}{\mathswitchr D}

\newcommand{\Pu}{\mathswitchr u}
\newcommand{\PU}{\mathswitchr U}
\newcommand{\Ps}{\mathswitchr s}
\newcommand{\Pb}{\mathswitchr b}
\newcommand{\Pc}{\mathswitchr c}
\newcommand{\Pt}{\mathswitchr t}
\newcommand{\Pq}{\mathswitchr q}
\newcommand{\Pep}{\mathswitchr {e^+}}
\newcommand{\Pem}{\mathswitchr {e^-}}

\newcommand{\PWp}{\mathswitchr {W^+}}
\newcommand{\PWm}{\mathswitchr {W^-}}

\newcommand{\Pp}{\mathswitchr {p}}

\def\mathswitch#1{\relax\ifmmode#1\else$#1$\fi}

\newcommand{\MW}{\mathswitch {M_\PW}}

\newcommand{\MZ}{\mathswitch {M_\PZ}}

\newcommand{\Mb}{\mathswitch {m_\Pb}}
\newcommand{\Mt}{\mathswitch {m_\Pt}}
\newcommand{\GW}{\mathswitch {\Gamma_\PW}}
\newcommand{\GZ}{\Gamma_{\PZ}}
\newcommand{\PL}{\mathswitch {P_\PL}}

\newcommand{\NCf}{\mathswitch {N_{\mathrm{C}}^f}}

\newcommand{\scrs}{\scriptscriptstyle}
\newcommand{\sw}{\mathswitch {s_{\scrs\PW}}}
\newcommand{\cw}{\mathswitch {c_{\scrs\PW}}}

\newcommand{\gs}{g_{\mathrm{s}}}
\newcommand{\als}{\alpha_{\mathrm{s}}}

\def\ie{i.e.\ }
\def\eg{e.g.\ }
\def\cf{cf.\ }

\newcommand{\ord}{{\cal O}}
\newcommand{\Oa}{\mathswitch{{\cal{O}}(\alpha)}}

\newcommand{\rT}{{\mathrm{T}}}
\newcommand{\rL}{{\mathrm{L}}}

\newcommand{\rd}{{\mathrm{d}}}

\newcommand{\ri}{{\mathrm{i}}}

\newcommand{\M}{{\cal{M}}}

\newcommand{\EW}{\mathrm{EW}}
\newcommand{\CM}{\mathrm{CM}}

\newcommand{\Born}{\mathrm{Born}}
\newcommand{\born}{\mathrm{Born}}
\newcommand{\fact}{{\mathrm{fact}}}
\newcommand{\nonfact}{{\mathrm{nf}}}
\newcommand{\sing}{{\mathrm{sing}}}
\newcommand{\finite}{{\mathrm{finite}}}
\newcommand{\SU}{\mathrm{SU}}
\newcommand{\U}{\mathrm{U}}

\newcommand{\virt}{\mathrm{virt}}

\newcommand{\LL}{\mathrm{LL}}

\newcommand{\TT}{\mathrm{TT}}

\newcommand{\ieps}{\ri\epsilon}

\newcommand{\ew}{{\mathrm{ew}}}

\newcommand{\DPA}{{\mathrm{DPA}}}
\newcommand{\LPA}{{\mathrm{LPA}}}

\def\Li{\mathop{\mathrm{Li}_2}\nolimits}
\def\cLi{\mathop{{\cal L}i_2}\nolimits}
\def\Re{\mathop{\mathrm{Re}}\nolimits}

\def\arc{\mathop{\mathrm{arc}}\nolimits}

\newcommand{\qqVV}{{\bar q_1 q_2\to V_1 V_2}}
\newcommand{\Vff}{{V \to f\bar f'}}
\newcommand{\Vaff}{{V_1 \to f_3 \bar f_4}}
\newcommand{\Vbff}{{V_2 \to f_5 \bar f_6}}
\newcommand{\qqVVffff}{\bar q_1 q_2\to V_1 V_2 \to 4f}
\newcommand{\qqffff}{\bar q_1 q_2\to 4f}
\newcommand{\qqffffg}{\qqffff\ga}
\newcommand{\sparton}{\hat s}
\newcommand{\uparton}{\hat u}
\newcommand{\tparton}{\hat t}

\newcommand{\nr}{N}
\newcommand{\Mbar}{\overline{M}}
\newcommand{\scpr}[2]{(#1#2)}
\newcommand{\ttwo}{\tilde t}
\newcommand{\stwo}{\tilde s}
\newcommand{\stwotwo}{\bar s}
\newcommand{\ffp}{\mathswitch{\mathrm{f\/f}'}}
\newcommand{\mfp}{\mathswitch{\mathrm{mf}'}}
\newcommand{\mmp}{\mathswitch{\mathrm{mm}'}}
\newcommand{\mf}{\mathswitch{\mathrm{mf}}}
\newcommand{\mm}{\mathswitch{\mathrm{mm}}}
\newcommand{\im}{\mathswitch{\mathrm{im}}}
\newcommand{\iif}{\mathswitch{\mathrm{if}}}
\newcommand{\nf}{{\mathrm{nf}}}

\newcommand{\cew}{C^{\ew}}

\newcommand{\PT}{P_{\mathrm{T}}}
\newcommand{\ET}{E_{\mathrm{T}}}
\newcommand{\PTmiss}{P_{\mathrm{T}}^{\mathrm{miss}}}
\newcommand{\PTmax}{P_{\mathrm{T}}^{\mathrm{max}}}
\newcommand{\PTcut}{P_{\mathrm{T}}^{\mathrm{cut}}}
\newcommand{\MT}{M_{\mathrm{T}}}

\newcommand{\Minv}{M_{\mathrm{inv}}}
\newcommand{\Minvcut}{M_{\mathrm{inv}}^{\mathrm{cut}}}
\newcommand{\AEWS}{\mathrm{AEWS}}

\newcommand{\Log}[1]{\log \left( #1\right) }
\newcommand{\Logq}[1]{\log^2 \left( #1\right) }
\newcommand{\mr}[1]{\mathrm{#1}}

\makeatletter
\newcount\@tempcntc
\def\@citex[#1]#2{\if@filesw\immediate\write\@auxout{\string\citation{#2}}\fi
  \@tempcnta\z@\@tempcntb\m@ne\def\@citea{}\@cite{\@for\@citeb:=#2\do
    {\@ifundefined
       {b@\@citeb}{\@citeo\@tempcntb\m@ne\@citea
        \def\@citea{,\penalty\@m\ }{\bf ?}\@warning
       {Citation `\@citeb' on page \thepage \space undefined}}%
    {\setbox\z@\hbox{\global\@tempcntc0\csname
b@\@citeb\endcsname\relax}%
     \ifnum\@tempcntc=\z@ \@citeo\@tempcntb\m@ne
       \@citea\def\@citea{,\penalty\@m}
       \hbox{\csname b@\@citeb\endcsname}%
     \else
      \advance\@tempcntb\@ne
      \ifnum\@tempcntb=\@tempcntc
      \else\advance\@tempcntb\m@ne\@citeo
      \@tempcnta\@tempcntc\@tempcntb\@tempcntc\fi\fi}}\@citeo}{#1}}

\def\@citeo{\ifnum\@tempcnta>\@tempcntb\else\@citea
  \def\@citea{,\penalty\@m}%
  \ifnum\@tempcnta=\@tempcntb\the\@tempcnta\else
   {\advance\@tempcnta\@ne\ifnum\@tempcnta=\@tempcntb \else
\def\@citea{--}\fi
    \advance\@tempcnta\m@ne\the\@tempcnta\@citea\the\@tempcntb}\fi\fi}
\makeatother

\def\draftdate{\relax}
\def\mpar#1{\relax}
\def\mda{\relax}
\def\mua{\relax}
\def\mla{\relax}
\def\draft{
\def\thtystars{******************************}
\def\sixtystars{\thtystars\thtystars}
\typeout{}
\typeout{\sixtystars**}
\typeout{* Draft mode!
         For final version remove \protect\draft\space in source file *}
\typeout{\sixtystars**}
\typeout{}
\def\draftdate{\today}
\def\mua{\marginpar[\boldmath\hfil$\uparrow$]%
                   {\boldmath$\uparrow$\hfil}%
                    \typeout{marginpar: $\uparrow$}\ignorespaces}
\def\mda{\marginpar[\boldmath\hfil$\downarrow$]%
                   {\boldmath$\downarrow$\hfil}%
                    \typeout{marginpar: $\downarrow$}\ignorespaces}
\def\mla{\marginpar[\boldmath\hfil$\rightarrow$]%
                   {\boldmath$\leftarrow $\hfil}%
                    \typeout{marginpar: $\leftrightarrow$}\ignorespaces}
\def\Mua{\marginpar[\boldmath\hfil$\Uparrow$]%
                   {\boldmath$\Uparrow$\hfil}%
                    \typeout{marginpar: $\Uparrow$}\ignorespaces}
\def\Mda{\marginpar[\boldmath\hfil$\Downarrow$]%
                   {\boldmath$\Downarrow$\hfil}%
                    \typeout{marginpar: $\Downarrow$}\ignorespaces}
\def\Mla{\marginpar[\boldmath\hfil$\Rightarrow$]%
                   {\boldmath$\Leftarrow $\hfil}%
                    \typeout{marginpar: $\Leftrightarrow$}\ignorespaces}
\def\mpar##1{\marginpar{\hbadness10000%
                      \sloppy\hfuzz10pt\boldmath\bf##1}%
                      \typeout{marginpar: ##1}\ignorespaces}
\overfullrule 5pt
\oddsidemargin -15mm
\marginparwidth 29mm
}

\newcommand{\thismonth}{\ifcase\month\or January\or February\or March \or April
\or May \or June \or July \or August \or September \or \November \or 
\December\fi}

\makeatletter

\def\eqnarray{\stepcounter{equation}\let\@currentlabel=\theequation
\global\@eqnswtrue
\global\@eqcnt\z@\tabskip\@centering\let\\=\@eqncr
$$\halign to \displaywidth\bgroup\hskip\@centering
  $\displaystyle\tabskip\z@{##}$\@eqnsel&\global\@eqcnt\@ne
  \hskip 2\arraycolsep \hfil${##}$\hfil
  &\global\@eqcnt\tw@ \hskip 2\arraycolsep $\displaystyle\tabskip\z@{##}$\hfil
   \tabskip\@centering&\llap{##}\tabskip\z@\cr}
\def\appendix{\par
 \setcounter{section}{0} \setcounter{subsection}{0}
 \def\thesection{\Alph{section}}}

\makeatother

\newcommand{\lsim}
{\;\raisebox{-.3em}{$\stackrel{\displaystyle <}{\sim}$}\;}
\newcommand{\gsim}
{\;\raisebox{-.3em}{$\stackrel{\displaystyle >}{\sim}$}\;}

\begin{document}
\tolerance=100000
\thispagestyle{empty}
\setcounter{page}{0}

\thispagestyle{empty}
\def\thefootnote{\fnsymbol{footnote}}
\setcounter{footnote}{1}
\null
\draftdate\hfill  PSI-PR-04-08
\\
\strut\hfill ZU-TH 13/04 \\
\strut\hfill DFTT 19/04\\
\strut\hfill hep-ph/0409247
\vskip 0cm
\vfill
\begin{center}
  {\Large \bf Logarithmic electroweak corrections to\\
    gauge-boson pair production at the LHC
\par} \vskip 2.5em
{\large
{\sc E. Accomando$^1$, A.~Denner$^2$ and A. Kaiser$^{2,3}$}}%
\\[.5cm]
$^1$ {\it Dipartimento di Fisica Teorica, Universit\`a di Torino,\\
Via P. Giuria 1, 10125 Torino, Italy}
\\[0.3cm]
$^2$ {\it Paul Scherrer Institut\\
CH-5232 Villigen PSI, Switzerland}
\\[0.3cm]
$^3$ {\it Institute of Theoretical Physics\\ University of Z\"urich, CH-8057 
Z\"urich, Switzerland}
\par
\end{center}\par
\vskip 2.0cm \vfill {\bf Abstract:} \par We have studied the effects
of the complete logarithmic electroweak $\Oa$ corrections on the
production of vector-boson pairs $\PW\PZ$, $\PZ\PZ$, and $\PW\PW$ at
the LHC.  These corrections are implemented into a Monte Carlo program
for $\Pp\Pp\to 4f (+\gamma )$ with final states involving four or two
leptons using the double-pole approximation.  We numerically
investigate purely leptonic final states and find that electroweak
corrections lower the predictions by 5--30\% in the physically
interesting region of large di-boson invariant mass and large angle of
the produced vector bosons.
\par
\vskip 1cm
\noindent
September 2004
\par
\null
\setcounter{page}{0}
\clearpage
\def\thefootnote{\arabic{footnote}}
\setcounter{footnote}{0}

\section{ Introduction }
\label{sec:intro}

The production of gauge-boson pairs provides an excellent opportunity
to test the non-abelian structure of the Standard Model (SM).
Gauge-boson-pair-production amplitudes involve trilinear gauge-boson
couplings.  Therefore, the corresponding cross sections depend very
sensitively on the non-abelian structure of the underlying theory. For
this reason, vector-boson pair production has found continuous
interest in the literature. In the last few years, gauge-boson
self-interactions were directly measured at LEP2 and Tevatron. Still,
up to now these couplings have not been determined with the same
precision as other boson properties, such as their masses and
couplings to fermions.  Despite of the high statistics reached at LEP2
in producing $\PWp\PWm$ pairs, the resulting limits on possible
anomalous couplings, which parametrize deviations from SM predictions
due to new physics occurring at energy scales of order of tens of TeV,
are not very stringent. The weakness of the LEP2 measurement is 
the rather modest centre-of-mass (CM) energy of the produced W-boson
pairs.  On the other hand, anomalous gauge-boson couplings cause
strong enhancements in the gauge-boson-pair-production cross section
especially at large values of the di-boson invariant mass
$M_{VV^\prime}$ ($V,V^\prime =\PW,\PZ$).  A significant improvement in
the bounds on triple-gauge-boson couplings is expected from
measurements at future colliders operating at high energies such as
the Large Hadron Collider (LHC).  Therefore, in order to achieve a
better precision in the determination of these couplings, it will be
useful to analyse di-boson production at hadron colliders at the
highest possible CM energies.

Vector-boson pairs also constitute a background to other kinds of
new-physics searches. One of the gold-plated signals for supersymmetry
at hadron colliders is chargino--neutralino pair production, which
would give rise to final states with three charged leptons and missing
transverse momentum; the primary background to this signature is given
by $\PW\PZ$ production.  Also final states coming from $\PZ\PZ$
production could fake that supersymmetry signature if one of the
leptons is lost in the beam pipe. Finally, $\PW^\pm\PW^\mp$ can dirty
the measurements of chargino and slepton pair production, which both
give rise to two leptons and missing energy.  Leptonic final states,
coming from $\Pp\Pp\to VV^\prime$ ($V,V'=\PW,\PZ$), could also fake
$\PZ\PZ$, $\PW\PZ$, and $\PW\PW$ vector-boson scattering signals which
are again expected to be enhanced at high CM energies.

In the near future, the LHC will be the main source of vector-boson pairs 
with large invariant mass $M_{VV^\prime}$. The machine will collect thousands 
of events, the exact statistics depending on the particular process and
luminosity \cite{Haywood:1999qg}.  With LHC approaching its goal of an
integrated luminosity of $100\fba^{-1}$, a large data sample will be
available to start a detailed investigation of the trilinear vertices.

In order to match the experimental precision, theoretical predictions
need to have an accuracy of the order of a few per cent to allow for a
decent analysis of the data. At lowest order, this demands taking into
account all spin correlations and finite-width effects. The easiest
way to fulfil this requirement is to go beyond the
{\it{production$\times$decay}} approach by computing the full
processes $\Pp\Pp\to 4f$. The next step consists in a full
understanding and control of higher-order QCD and electroweak (EW)
corrections.  In the past years, a large effort has gone into accurate
calculations of hadronic di-boson production (for a review on the
subject see \citere{Haywood:1999qg}).  The $\ord(\alpha_s)$ QCD
corrections to massive gauge-boson pair production and decay have been 
extensively analysed by many authors
\cite{Mele:1990bq,Frixione:1992pj,%
Ohnemus:1994ff,Baur:1995uv,%
Baur:1995aj,Dixon:1999di,Campbell:1999ah}.
Several NLO Monte Carlo programs have been constructed and cross
checked so that complete $\ord(\alpha_s)$ corrections are now
available \cite{Dixon:1999di,Campbell:1999ah}. QCD corrections turn
out to be quite significant at LHC energies. They can increase the
lowest-order cross section by a factor of two if no cuts are applied
and by one order of magnitude for large transverse momentum or large
invariant mass of the vector bosons
\cite{Mele:1990bq,Frixione:1992pj}.  By including a jet veto, their
effects can be drastically reduced to the order of tens of per cent
\cite{Baur:1995aj,Dixon:1999di}, but in any case they have to be
considered to get realistic and reliable estimates of total cross
sections and distributions.

In view of the envisaged precision of a few per cent at the LHC, also
a discussion of EW corrections is in order. For single $\PW$- and
$\PZ$-boson production, $\Oa$ corrections have been computed taking
into account the full electromagnetic and weak contributions
\cite{Baur:1999kt}. One-loop weak corrections have been also
investigated for $\Pt\bar\Pt$ production \cite{Beenakker:1993yr},
$\Pb\bar \Pb$ production \cite{Maina:2003is}, $\ga/\PZ+\mr{jet}$
hadro-production \cite{Maina:2004rb}, WH and ZH production
\cite{Ciccolini:2003jy}, as well as for $\ga Z$ production
\cite{Hollik:2004tm}.  By contrast, gauge-boson pair production at
hadron colliders is commonly treated by including only universal
radiative corrections such as the running of the electromagnetic
coupling, and corrections to the $\rho$ parameter.  This approach is
based on the belief that the remaining EW corrections (dominated by
double-logarithmic contributions) are not relevant at the LHC just
because physical cross sections decrease strongly with increasing
invariant mass of the gauge-boson pairs, \ie where EW corrections can
be non-negligible.  However, a first analysis of the effect of
one-loop logarithmic EW corrections on $\PW\PZ$ and $\PW\ga$
production processes at the LHC \cite{Accomando:2001fn} has instead
demonstrated that $\Oa$ corrections are of the same order or bigger
than the statistical error, when exploring the large di-boson
invariant-mass and small rapidity region.

The fact that $\Oa$ EW corrections grow with increasing energy is well
known since long time. EW corrections are in fact dominated by double
and single logarithms of the ratio of the energy to the EW scale.
Analyses of the general high-energy behaviour of EW corrections have
been extensively performed (see for instance
\citeres{Beenakker:1993tt,ewee}). A process-independent recipe for the
calculation of logarithmic EW corrections is given in
\citeres{Denner:2001jv,Denner:2001gw,Pozzorini:2001rs}, where it has
been shown that the logarithmic one-loop corrections to arbitrary EW
processes factorize into tree-level amplitudes times universal
correction factors.

Using the method of
\citeres{Denner:2001jv,Denner:2001gw,Pozzorini:2001rs}, we investigate
in this paper the effect of logarithmic EW corrections to the hadronic
production of $\PW^\pm \PZ$, $\PZ \PZ$, and $\PW^\pm \PW^\mp$ pairs in
the large-invariant-mass region of the hard process at the LHC.  Going
beyond the analysis of \citere{Accomando:2001fn}, which addressed only
logarithmic contributions originating from above the EW scale, here we
consider also the effect of the complete logarithmic electromagnetic
corrections.  Since the aim of the paper is to analyse the structure
of the $\Oa$ EW corrections and to give an estimate of their size, we
have not included QCD corrections.

The simplest experimental analyses of gauge-boson pair production will
rely on purely leptonic final states. Semi-leptonic channels, where
one of the vector bosons decays hadronically, have been analysed at
the Tevatron \cite{semitev} showing that these events suffer from the
background due to the production of one vector boson plus jets via
gluon exchange.  For this reason, we study only di-boson production where 
both gauge bosons decay leptonically into $\Pe$ or $\mu$.

The paper is organized as follows: the strategy of our calculation is
presented in \refse{sec:strategy-calculation}.  In
\refse{sec:matrix_elements} we describe the calculation of the
lowest-order matrix elements, and in \refse{sec:ewrc} the analytical
results for the virtual logarithmic EW one-loop corrections
are summarized.  The treatment of soft and collinear singularities is
discussed in \refse{sec:treatm-soft-coll}.  While the general setup of
our numerical calculation is given in \refse{sec:processes},
\refse{sec:results} contains a numerical discussion for processes
mediated by $\PW\PZ$, $\PZ\PZ$, and $\PW\PW$ production.  Our findings
are summarized in \refse{sec:concl}. Appendices
\ref{sec:non-fact-phot} and \ref{app:scalints} contain results for
non-factorizable corrections to a general class of processes and the
corresponding integrals. Some coupling factors are listed in
\refapp{sec:couplings}.

\section{Strategy of the calculation}
\label{sec:strategy-calculation}

We consider the production of massive gauge-boson pairs in
hadron--hadron collisions. The generic process can be written as 
\beq
h_1 + h_2 \to V_1 + V_ 2 + X \to f_3 + \bar f_4 + f_5 + \bar f_6 + X,
\eeq 
where $h_1$ and $h_2$ denote the incoming hadrons,
$V_1$ and $V_2$ two arbitrary massive gauge bosons, \eg W or Z bosons,
$f_3,f_5$ the outgoing fermions, $\bar f_4,\bar f_6$ the outgoing
antifermions, and $X$ the remnants of the hadrons.

In the parton model the corresponding cross sections are obtained from
a convolution as
\beqar\label{eq:convolution}
\rd\si^{h_1 h_2}(P_1,P_2,p_f) = \sum_{i,j}\int_0^1\rd x_1 \int_0^1\rd x_2\,
\Phi_{i,h_1}(x_1,Q^2)\Phi_{j,h_2}(x_2,Q^2) \,
\rd\hat\si^{ij}(x_1P_1,x_2P_2,p_f),\nln
\eeqar
where $p_f$ summarizes the final-state momenta, $\Phi_{i,h_1}$ and
$\Phi_{j,h_2}$ are the distribution functions of the partons $i$ and
$j$ in the incoming hadrons $h_1$ and $h_2$ with momenta $P_1$ and
$P_2$, respectively, $Q$ is the factorization scale, and
$\rd\hat\si^{ij}$ represent the differential cross sections for the
partonic processes averaged over colours and spins of the partons. The
sum $\sum_{i,j}$ runs over all possible quarks and antiquarks of
flavour $\Pu,\Pd,\Pc$, and $\Ps$.

The relevant parton processes are of the form
\begin{eqnarray}\label{eq:parton_process_dr}
\bar q_1(p_1,\sigma_1) + q_2(p_2,\sigma_2) &\to&
 V_1(k_1,\la_1) + V_2(k_2,\la_2) \nl
 &\to& f_3(p_3,\sigma_3) + \bar f_4(p_4,\sigma_4) + f_5(p_5,\sigma_5) +
 \bar f_6(p_6,\sigma_6).
\end{eqnarray}
The arguments label the momenta $p_i,k_l$ and helicities 
$\si_i=\pm 1/2,\la_l=0,\pm1$ of the corresponding incoming partons,
outgoing fermions, and virtual gauge bosons.  We often use only the
signs to denote the helicities.  The momenta of the incoming partons
are related to the momenta of the hadrons by $p_1= x_1 P_{1}$ and
$p_2= x_2 P_{2}$
if $i$ is an antiquark and $j$ a quark and by $p_2= x_1 P_{1}$ and
$p_1= x_2 P_{2}$ in the opposite case.

The corresponding lowest-order partonic cross sections are calculated
using the complete matrix elements. This means that we include the full
set of Feynman diagrams, in this way accounting for all irreducible
background coming from non-doubly resonant contributions. 
The calculation of the matrix elements for the complete process
\begin{eqnarray}\label{eq:parton_process}
\bar q_1(p_1,\sigma_1) + q_2(p_2,\sigma_2) &\to&
f_3(p_3,\sigma_3) + \bar f_4(p_4,\sigma_4) + f_5(p_5,\sigma_5) +
 \bar f_6(p_6,\sigma_6)
\end{eqnarray}
is described in \refse{sec:matrix_elements}.

The electroweak radiative corrections to \refeq{eq:parton_process}
consist of virtual corrections, resulting from loop diagrams, as well
as of real corrections, originating from the processes
\begin{eqnarray}\label{eq:realprocess}
\bar q_1(p_1,\sigma_1) + q_2(p_2,\sigma_2) \to
 f_3(p_3,\sigma_3) + \bar f_4(p_4,\sigma_4) + f_5(p_5,\sigma_5) +
 \bar f_6(p_6,\sigma_6) 
+ \gamma(k,\la_\ga)
\nln
\end{eqnarray}
with an additional photon with momentum $k$ and helicity
$\la_\ga=\pm1$.  Both have to be combined properly in order to ensure
the cancellations of soft and collinear singularities (\cf
\refse{sec:treatm-soft-coll}).

For the calculation of the radiative corrections we follow the
approach used for the process $\Pep\Pem\to\PWp\PWm\to4f$ in
\citere{Denner:2000bj}.  The virtual corrections are calculated in the
double-pole approximation (DPA), \ie we take only those terms into
account that are enhanced by two resonant massive gauge-boson
propagators. The real corrections are calculated from the full matrix
elements for the processes \refeq{eq:realprocess}.

\subsection{Double-pole approximation for virtual corrections}
\label{sec:DPA}

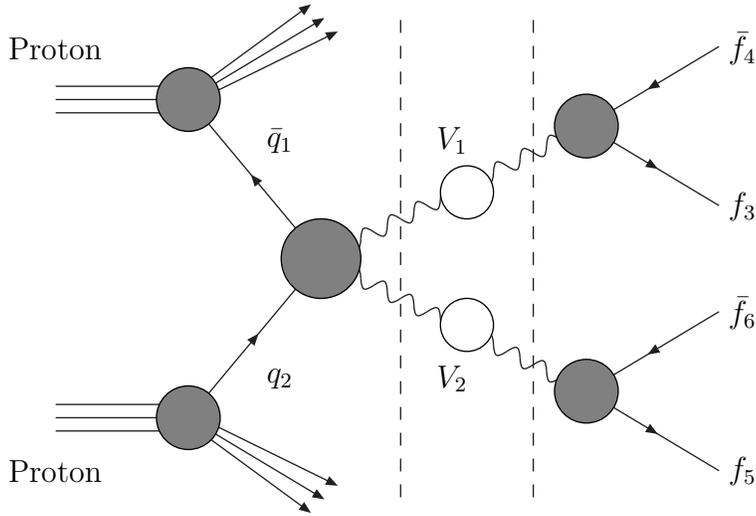
\begin{figure}
\begin{center}
\begin{picture}(400,200)(0,0)

\SetOffset(0,0)

\Line(50,160)(100,160)
\Line(50,165)(100,165)
\Line(50,155)(100,155)
\ArrowLine(150,100)(100,160)
\LongArrow(100,160)(150,190)
\LongArrow(100,162)(145,195)
\LongArrow(100,158)(155,185)

\Line(50,40)(100,40)
\Line(50,45)(100,45)
\Line(50,35)(100,35)
\ArrowLine(100,40)(150,100)
\LongArrow(100,40)(150,10)
\LongArrow(100,42)(155,15)
\LongArrow(100,38)(145,5)

\GCirc(100,40){12}{0.5}
\GCirc(100,160){12}{0.5}

\Photon(150,100)(250,50){3}{10}
\Photon(150,100)(250,150){-3}{10}
\GCirc(150,100){15}{0.5}

\DashLine(180,10)(180,190){5}
\DashLine(230,10)(230,190){5}
\GCirc(205,125){10}{1.0}
\GCirc(205,75){10}{1.0}

\ArrowLine(300,180)(250,150)
\ArrowLine(250,150)(300,120)
\ArrowLine(300,80)(250,50)
\ArrowLine(250,50)(300,20)

\GCirc(250,150){12}{0.5}
\GCirc(250,50){12}{0.5}

\Text(50,180)[]{Proton}
\Text(50,20)[]{Proton}

\Text(135,145)[]{$\bar q_1$}
\Text(135,55)[]{$q_2$}

\Text(310,180)[]{$\bar f_4$}
\Text(310,120)[]{$f_3$}
\Text(310,80)[]{$\bar f_6$}
\Text(310,20)[]{$f_5$}

\Text(200,145)[]{$V_1$}
\Text(200,55)[]{$V_2$}

\end{picture}
\end{center}

\caption{Structure of the  process $\Pp\Pp\to V_1 V_2+X \to 4f+X$}
\label{Full_Amplitude_DPA}
\end{figure}

In DPA the processes $\qqVVffff$ are divided into the production of
on-shell gauge bosons and their decay into fermion--antifermion pairs
(see \reffi{Full_Amplitude_DPA}).

At tree level, the matrix elements in DPA for the partonic processes
$\qqVVffff$ factorize into those for the production of two on-shell
bosons, $\M_{\Born}^{\bar q_1 q_2\to V_{1,\la_1}V_{2,\la_2}}$, the
propagators of these bosons, and the matrix elements for their
on-shell decays, $\M_\Born^{V_{1,\la_1}\to f_3\bar f_4}$ and
$\M_\Born^{V_{2,\la_2}\to f_5\bar f_6}$,
\beqar\label{eq:BornVV} \M_{\Born,\DPA}^{\qqVVffff} &=&
P_{V_1}(k_1^2)\;P_{V_2}(k_2^2)
  \sum_{\la_1,\la_2}\M_{\Born}^{\bar q_1q_2\to
  V_{1,\la_1}V_{2,\la_2}} \M_\Born^{V_{1,\la_1}\to
  f_3\bar f_4}\M_\Born^{V_{2,\la_2}\to f_5\bar f_6}.
\eeqar 
The sum runs over the physical helicities $\la_1,\la_2=0,\pm1$ of the
on-shell projected gauge bosons $V_1$ and $V_2$ with momenta $k_1$ and
$k_2$, respectively.  The propagators of the massive gauge bosons 
\begin{equation}\label{eq:PV}
 P_V(p^2) = \frac{1}{p^2 - M_V^2+\theta(p^2)\ri M_V\Ga_V}, 
\quad V=\PW,\PZ,
\end{equation} 
involve besides the masses of the gauge bosons also their widths, which we
consider as constant and finite for time-like momenta. 
The on-shell matrix elements are calculated using on-shell projected
momenta as defined in Appendix~A of \citere{Denner:2000bj}.
Of course, the momenta in the resonant propagators are not projected
on shell.

In DPA there are two types of corrections, factorizable and
non-factorizable ones.  The former are those that can be associated
to one of the production or decay subprocesses, the latter are those
that connect these subprocesses.

The factorizable corrections can be expressed in terms of the
corrections to the on-shell gauge-boson-pair-production and -decay
subprocesses. The matrix elements for the virtual factorizable
corrections to the processes $\qqVVffff$ can be written as
\beqar\label{eq:factcorrVV}
\de\M_{\virt,\DPA,\fact}^{\qqVVffff} &=&
P_{V_1}(k_1^2)\;P_{V_2}(k_2^2)
\sum_{\la_1,\la_2}\biggl\{\de\M_{\virt}^{\bar q_1q_2\to V_{1,\la_1}V_{2,\la_2}}
\M_{\Born}^{V_{1,\la_1}\to f_3\bar f_4}
\M_{\Born}^{V_{2,\la_2}\to f_5\bar f_6}\nl
&&{}+\M_{\Born}^{\bar q_1q_2\to V_{1,\la_1}V_{2,\la_2}}
\de\M_{\virt}^{V_{1,\la_1}\to f_3\bar f_4}
\M_{\Born}^{V_{2,\la_2}\to f_5\bar f_6}\nl
&&{}+\M_{\Born}^{\bar q_1q_2\to V_{1,\la_1}V_{2,\la_2}}
\M_{\Born}^{V_{1,\la_1}\to f_3\bar f_4}
\de\M_{\virt}^{V_{2,\la_2}\to f_5\bar f_6}\biggr\},
\eeqar
where $\delta\M_{\virt}^{\bar q_1q_2\to V_{1,\la_1}V_{2,\la_2}}$,
$\delta\M_\virt^{V_{1,\la_1}\to  f_3\bar f_4}$, and
$\delta\M_\virt^{V_{2,\la_2}\to  f_5\bar f_6}$ denote the virtual
corrections to the on-shell matrix elements for the gauge-boson production
and decay processes.  

The non-factorizable corrections yield a simple correction factor
$\de^{\virt}_{\nonfact,\DPA}$ to the lowest-order cross section.
Its explicit form is given in \refse{sec:nfRC}.

The contribution of the complete virtual corrections in DPA to the
cross section reads
\begin{eqnarray}\label{eq:virt_corr}
\int \rd\sigma_{\virt,\DPA}^{\qqVVffff}&=&
\frac{1}{2 \sparton}\int \rd\Phi_{4f} \Bigl[
2\Re \Bigl(\Bigl(\M^{\qqVVffff}_{\born,\DPA}\Bigr)^*
\de\M^{\qqVVffff}_{\virt,\DPA,\fact}\Bigr) 
\nonumber\\
&& {}
+\Bigl|\M^{\qqVVffff}_{\born,\DPA}\Bigr|^2 
\de^{\virt}_{\nonfact,\DPA} \Bigr],
\end{eqnarray}
where $\rd\Phi_{4f} $ denotes the four-particle phase-space element
and $\sparton=(p_1+p_2)^2$ the square of the CM energy in the partonic
system.  For some four-fermion final states resonant massive gauge
bosons can either be formed from the pairs $(f_3,\bar f_4)$ and
$(f_5,\bar f_6)$ or from the pairs $(f_3,\bar f_6)$ and $(f_5,\bar
f_4)$. Denoting the isospin partner of $f$ by $f'$, this is the case
for $f_3=f_4=f_5=f_6$ and $f_3=f_4=f_5=f'_6$, which allow for two
different sets of ZZ and WZ pairs, respectively.  In all these
cases a DPA has to be defined for each of the two sets of resonant
gauge bosons separately, and the cross sections from
these two cases have to be summed.%
\footnote{In our numerical calculation we actually do not add the
  cross sections but the matrix elements for the different resonant
  sets. Since the interference between these different contributions
  is non-doubly resonant, this is equivalent within DPA accuracy.}%

Finally, we have to take care of the proper matching of the infrared
(IR) and collinear singularities.  In general the total cross section is
composed as
\begin{equation}\label{eq:crosssection0}
\int \rd \sigma =
\int\nolimits_{\displaystyle \Phi_{4f}} \rd \sigma_{\Born}^{\qqffff}+
\int\nolimits_{\displaystyle \Phi_{4f}} \rd \sigma_{\virt}^{\qqffff}
+\int\nolimits_{\displaystyle \Phi_{4f\ga}} \rd\sigma^{\qqffffg}.
\end{equation}
Here $\rd\sigma_{\Born}^{\qqffff}$ is the full differential
lowest-order cross section for $\qqffff$ that is to be integrated over
the four-particle phase space $\Phi_{4f}$, \ie
\begin{eqnarray}\label{eq:sigma_Born}
\rd\sigma_{\Born}^{\qqffff}&=&
\frac{1}{2 \sparton}
\Bigl|\M^{\qqffff}_{\born}\Bigr|^2 \rd\Phi_{4f} .
\end{eqnarray}
Similarly, $\rd\sigma^{\qqffffg}$, which describes the real
corrections, is the full lowest-order cross section for $\qqffffg$ to
be integrated over the five-particle phase space $\Phi_{4f\ga}$, and
$\rd \sigma_{\virt}^{\qqffff}$ denotes the virtual one-loop
corrections.

Both $\rd\sigma^{\qqffffg}$ and $\rd \sigma_{\virt}^{\qqffff}$ involve
soft and collinear singularities that cancel in their sum.  Taking the
DPA for $\rd \sigma_{\virt}^{\qqffff}$ but not for
$\rd\sigma^{\qqffffg}$ spoils this cancellation. Therefore, we
subtract the singular contributions
$\rd\sigma_{\virt,\sing}^{\qqffff}$ before we impose the DPA and
replace \refeq{eq:crosssection0} by
\begin{eqnarray}\label{eq:crosssection}
\int \rd \sigma &=&
\int\nolimits_{\displaystyle \Phi_{4f}} \rd \sigma_{\Born}^{\qqffff}+
\int\nolimits_{\displaystyle \Phi_{4f}} 
(\rd \sigma_{\virt,\DPA}^{\qqffff}-\rd \sigma_{\virt,\sing,\DPA}^{\qqffff})
\nl&&{}
+\int\nolimits_{\displaystyle \Phi_{4f}} \rd \sigma_{\virt,\sing}^{\qqffff}
+\int\nolimits_{\displaystyle \Phi_{4f\ga}} \rd\sigma^{\qqffffg}.
\end{eqnarray}
Since the finite, non-logarithmic terms of
$\rd\sigma_{\virt,\sing}^{\qqffff}$ are not uniquely defined, this
procedure leads to an ambiguity, which is, however, of the order of
the uncertainty of the DPA.  Since the IR- and fermion-mass-singular
part is not treated in DPA, the logarithmic photonic corrections are
not affected by this ambiguity. We use the definition of
$\rd\sigma_{\virt,\sing}^{\qqffff}$ as given in
\refse{sec:defin-finite-virt}.

\subsection{High-energy approximation}
\label{sec:hea}

In contrast to \citere{Denner:2000bj}, we do not calculate the
EW $\Oa$ corrections completely, but we only calculate the
logarithmic corrections in the high-energy limit.  To this end, we
consider the limit where all kinematical invariants
$s_{ij}=(p_i+p_j)^2$ are large compared with the weak-boson mass scale,
$|s_{ij}|\gg\MW^2$, and take into account all contributions
proportional to $\al\log^2(|s_{ij}|/\MW^2)$ or
$\al\log(|s_{ij}|/\MW^2)$. Note that this approximation is not
applicable to the full processes $\qqVVffff$ because of the presence
of the resonances with invariant masses of the order of $\MW$. But
it is perfectly applicable to the subprocesses $\qqVV$, $\Vaff$, and
$\Vbff$ appearing in the DPA.

Since we assume that all kinematical invariants are of the same order
of magnitude, we can write all large energy-dependent EW logarithms in
terms of $\log(\hat s/\MW^2)$, where $\hat s$ is the CM energy squared
of the partonic process.  In \citere{Accomando:2001fn} we have taken
into account all contributions involving logarithms of the form
$\log(\hat s/\MW^2)$.  These arise from scales larger than $\MW$ and
can be written in an $\SU(2)\times\U(1)$ symmetric form.  We did not
include logarithmic corrections of purely electromagnetic origin
arising from scales smaller than $\MW$.  These involve logarithms of
the form $\log(\MW^2/m_f^2)$ or $\log(\MW^2/\la^2)$, where $\la$ is
the photon mass regulator. In the present paper all large logarithms
of electromagnetic origin are included as well.  Since the decay
processes involve no large-energy variable, the corresponding virtual
corrections involve no large EW logarithms. However, they give rise to
large electromagnetic logarithms.

In the high-energy approximation we omit all mass-suppressed terms,
\ie terms of order $\MW^2/\hat s$. Therefore we can omit the channels
with one longitudinal and one transverse gauge boson that are mass
suppressed for the di-boson production processes $\qqVVffff$ and take
into account only the corrections to the dominating channels involving
two transverse ($\TT$) or two longitudinal ($\LL$) gauge bosons.
On the other hand, we take into account the exact kinematics by
evaluating the complete four-fermion phase space and use the exact
values of the kinematical invariants in all formulas.

The logarithmic virtual EW corrections to the dominating channels 
of $\qqVV$
are calculated using the general results for a high-energy
approximation given in
\citeres{Denner:2001jv,Denner:2001gw,Pozzorini:2001rs}.  The validity
of these results relies on the assumption that all kinematical
variables ${\sparton}$, $|{\tparton}|$, and $|{\uparton}|$ are large
compared with $\MW^2$ and approximately of the same size,
\beq\label{HEA}
{\sparton}\sim |{\tparton}|\sim |{\uparton}|\gg \MW^2.
\eeq
This implies that the produced gauge bosons have to be emitted at
sufficiently large angles with respect to the beam. Hence, the
validity range of the high-energy logarithmic approximation for the
radiative corrections corresponds to the central region of the boson
scattering angle in the di-boson rest frame.  The $t$-channel pole in
the Born matrix element gives rise to additional enhanced logarithms
when integrated over the full kinematical range.  Since these terms
are not included in our $\Oa$ analysis, we have to take care that we
do not get sizeable contributions from small scattering angles with
respect to the beam. On the other hand, our formulas do not fake
spurious contributions as long as
${\sparton},|{\tparton}|,|{\uparton}|\gsim\MW^2$, since the large
logarithms become small for
${\sparton},|{\tparton}|,|{\uparton}|\sim\MW^2$.

The logarithmic approximation yields the dominant corrections for
large kinematical invariants $|s_{ij}|\gg\MW^2$, but neglects finite,
non-logarithmic, process-dependent $\Oa$ contributions. For
$\Pep\Pem\to\PWp\PWm$, where complete $\Oa$ corrections and
their high-energy limit are available \cite{Beenakker:1993tt}, the
latter turn out to be of order of a few per cent. We assume that this
holds as well for similar processes like hadronic di-boson production.
Neglecting non-logarithmic terms can therefore be considered a
reasonable approximation at the LHC, where the experimental accuracy
in the high-energy regime is at the few-per-cent level.

\section{The matrix elements for $\qqffff$}
\label{sec:matrix_elements}

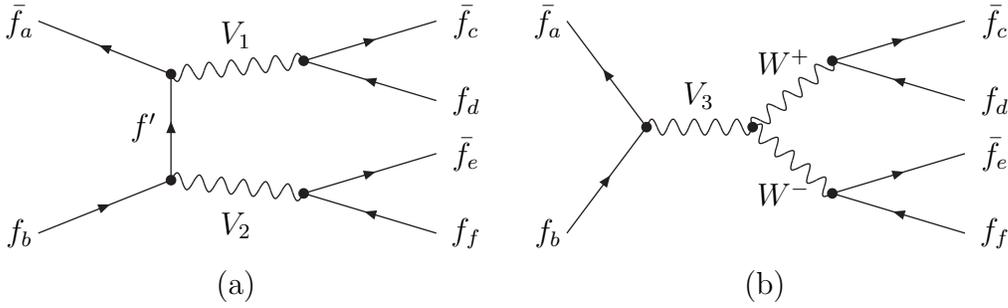
\begin{figure}
\begin{center}
\begin{picture}(400,110)(0,-10)
%
\SetOffset(-35,0)
\ArrowLine(50,10)(100,30)
\ArrowLine(100,70)(50,90)
\ArrowLine(100,30)(100,70)
\Photon(100,30)(150,25){3}{6}
\Photon(100,70)(150,75){-3}{6}
\ArrowLine(150,75)(200,90)
\ArrowLine(200,60)(150,75)
\ArrowLine(150,25)(200,40)
\ArrowLine(200,10)(150,25)
\Vertex(100,30){2}
\Vertex(100,70){2}
\Vertex(150,75){2}
\Vertex(150,25){2}
\Text(43,90)[]{$\bar f_a$}
\Text(43,10)[]{$f_b$}
\Text(212,90)[]{$\bar f_c$}
\Text(212,60)[]{$f_d$}
\Text(212,40)[]{$\bar f_e$}
\Text(212,10)[]{$f_f$}
\Text(125,85)[]{$V_1$}
\Text(125,13)[]{$V_2$}
\Text(90,50)[]{$f'$}
\Text(125,-10)[]{(a)}
%
\SetOffset(165,0)
\ArrowLine(50,10)(80,50)
\ArrowLine(80,50)(50,90)
\Photon(80,50)(120,50){3}{5}
\Photon(120,50)(150,25){3}{5}
\Photon(120,50)(150,75){3}{5}
\ArrowLine(150,75)(200,90)
\ArrowLine(200,60)(150,75)
\ArrowLine(150,25)(200,40)
\ArrowLine(200,10)(150,25)
\Vertex(80,50){2}
\Vertex(120,50){2}
\Vertex(150,75){2}
\Vertex(150,25){2}
\Text(43,90)[]{$\bar f_a$}
\Text(43,10)[]{$f_b$}
\Text(212,90)[]{$\bar f_c$}
\Text(212,60)[]{$f_d$}
\Text(212,40)[]{$\bar f_e$}
\Text(212,10)[]{$f_f$}
\Text(100,62)[]{$V_3$}
\Text(132,75)[]{$W^+$}
\Text(132,25)[]{$W^-$}
\Text(125,-10)[]{(b)}
\end{picture}
\end{center}
\caption{The fundamental topologies for processes with six external fermions}
\label{fundamentaltopologies}
\end{figure}

Our calculation involves two independent sets of matrix elements. The
first set consists of the complete lowest-order matrix elements for the
processes $\qqffff$ and $\qqffffg$. This set is based on the generic
matrix elements for $\mr{e}^+ \mr{e}^- \to 4 \, \mr{fermions} +
\gamma$ given in \citere{Denner:1999gp}. All Feynman diagrams
contributing to a $2 f\to 4 f$ process can be constructed from only
two fundamental topologies (see \reffi{fundamentaltopologies}) by
permuting the external particles $f_a, \dots, f_f$.

The second set of matrix elements is used for the calculation of the
corrections in the DPA. It includes the matrix elements $\M^{\qqVV}$ for the
production of a pair of transverse or longitudinal gauge bosons and
the matrix elements $\M^{\Vff}$ for their decay.

\subsection{Matrix elements for four-fermion production}

We need the amplitudes for the parton processes
\refeq{eq:parton_process} and \refeq{eq:realprocess}.  To this end we
consider a generic process with three incoming antifermions
$\bar f_1,\bar f_3,\bar f_5$ and three incoming fermions $f_2,f_4,f_6$:
\begin{equation} \label{eq:6fprocess}
\bar f_1(p_1,\sigma_1) + f_2(p_2,\sigma_2) +
\bar f_3(-p_3,-\sigma_3) + f_4(-p_4,-\sigma_4) +
\bar f_5(-p_5,-\sigma_5) + f_6(-p_6,-\sigma_6) \to 0 .
\end{equation}
The arguments denote the (incoming) momenta and helicities
of the incoming fermions and antifermions.
 
Each Feynman diagram for the process \refeq{eq:6fprocess} corresponds
to one of the two generic diagrams in
\reffi{fundamentaltopologies}.  These generic diagrams are given by
\beqar \label{eq:genericMa}
\lefteqn{
\mathcal{M}^{\mr{a},V_1,V_2}(\bar f_a,f_b,\bar f_c,f_d,\bar f_e,f_f)
= -4 e^4\left( \sum_{f^\prime} C_{V_1 \bar{f}_a f^\prime}^{\sigma_a,
    \sigma^\prime} C_{V_2 \bar{f}^\prime f_b}^{-\sigma^\prime,
    \sigma_b} \right) C_{V_1 \bar{f}_c f_d}^{\sigma_c, \sigma_d}
C_{V_2 \bar{f}_e f_f}^{\sigma_e, \sigma_f} } \quad 
 \nl && \times
\frac{P_{V_1}\bigl((p_c+p_d)^2\bigr) P_{V_2}\bigl((p_e+p_f)^2\bigr)}
{(p_b+p_e+p_f)^2}
A_2^{\sigma_a,\sigma_c,\sigma_e}(p_a,p_b,p_c,p_d,p_e,p_f), \nl[1ex]
\lefteqn{\mathcal{M}^{\mr{b},V_3}(\bar f_a,f_b,\bar f_c,f_d,\bar f_e,f_f) = 
-4 e^4 C_{V_3 W^+ W^-} C_{V_3 \bar{f}_a f_b}^{\sigma_a, \sigma_b}
C_{W^+ \bar{f}_c f_d}^{\sigma_c, \sigma_d} C_{W^- \bar{f}_e
  f_f}^{\sigma_e, \sigma_f} }\quad
&&\phantom{\mathcal{M}^{\mr{b},V_3}(\bar f_a,f_b,\bar f_c,f_d,\bar f_e,f_f) = 
-4 e^4 C_{V_3
  \mr{W}^+\mr{W}^-} C_{V_3 \bar{f}_a f_b}^{\sigma_a, \sigma_b}
C_{\mr{W}^+ \bar{f}_c f_d}^{\sigma_c, \sigma_d} C_{\mr{W}^- \bar{f}_e
  f_f}^{\sigma_e, \sigma_f} P_{V_3}(\bigl(p_a)}
\nl && \times
 P_{V_3}\bigl((p_a+p_b)^2\bigr)
P_{\mr{W}}\bigl((p_c+p_d)^2\bigr) P_{\mr{W}}\bigl((p_e+p_f)^2\bigr)
A_3^{\sigma_a}(p_a,p_b,p_c,p_d,p_e,p_f),
\eeqar
and the auxiliary functions $A_2$ and $A_3$ can be found in
\citere{Denner:1999gp}.  The gauge-boson propagators $P_V$ are defined
by \refeq{eq:PV}, and $e$ is the electric charge of the positron.
For the photon ($V=A$), the Z boson, and the W boson, the generic couplings
$C^{\sigma_a, \sigma_b}_{V \bar{f}_a f_b}$ are listed in
\refeq{eq:rel_C_I} with \refeq{eq:F_couplings}.  The coupling $C_{V_3
  W^+ W^-}$ is given by
\beq
C_{A W^+ W^-} = 1,\qquad C_{Z W^+ W^-} = -\frac{\cw}{\sw}.
\eeq
where $\cw=\MW/\MZ$ and $\sw$ are the cosine and sine of the
electroweak mixing angle, respectively.
If gluons are present, \refeq{eq:genericMa} corresponds to the matrix
element with colour matrices omitted, and the generic fermionic
couplings read
\beq
 C^{\sigma_a, \sigma_b}_{g \bar{f}_a f_b} =  \delta_{\sigma_a, -\sigma_b}
\frac{\gs}{e}
\eeq
with the strong gauge coupling $g_{s}=\sqrt{4\pi\als}$.  The matrix
elements for outgoing particles are simply obtained by inverting the
helicities and momenta.  It is convenient to define the objects
\beqar 
\mathcal{M}^\mr{weak}_\pm=
\sum_{\{i_1,i_3,i_5\}} \sum_{\{i_2,i_4,i_6\}}
\frac{\mr{sign}(\{i_1,i_3,i_5\})\,\mr{sign}(\{i_2,i_4,i_6\}) \pm 1}{2}
\nl \times \Biggl[ \sum_{V_1=W^\pm,Z,\gamma} \sum_{V_2=W^\pm,Z,\gamma}
\mathcal{M}^{\mr{a},V_1,V_2}(\bar f_{i_1},f_{i_2},\bar f_{i_3},f_{i_4},\bar f_{i_5},f_{i_6}) \nl +
\sum_{V_3=Z,\gamma} 
\mathcal{M}^{\mr{b},V_3}(\bar f_{i_1},f_{i_2},\bar f_{i_3},f_{i_4},\bar f_{i_5},f_{i_6})
\Biggr]
\eeqar 
and 
\beqar \mathcal{M}^\mr{gluon}_\pm =
\sum_{\{i_1,i_3,i_5\}} \sum_{\{i_2,i_4,i_6\}}
\frac{\mr{sign}(\{i_1,i_3,i_5\})\,\mr{sign}(\{i_2,i_4,i_6\}) \pm 1}{2}
\nl \times \Biggl[ \sum_{V=W^\pm,Z,\gamma}
\mathcal{M}^{\mr{a},g,V}(\bar f_{i_1},f_{i_2},\bar f_{i_3},f_{i_4},\bar f_{i_5},f_{i_6}) \nl
+\sum_{V=W^\pm,Z,\gamma}
\mathcal{M}^{\mr{a},V,g}(\bar f_{i_1},f_{i_2},\bar f_{i_3},f_{i_4},\bar f_{i_5},f_{i_6}) \Biggr], 
\eeqar 
where the two sums run over the permutations of the fermions and
antifermions and $\mr{sign}(\{i_1,i_3,i_5\})$ and
$\mr{sign}(\{i_2,i_4,i_6\})$ give the signs of these permutations.
Note that $\mathcal{M}^\mr{weak}_+$ is the sum of all diagrams with a
positive signature of all permutations and $\mathcal{M}^\mr{weak}_-$
is the sum of all diagrams with a negative signature.  All diagrams
that are not present in the SM, \eg diagrams including
a $Z \bar{u} e^-$ coupling or a $W^+ \bar{d} u$ coupling, drop out because the
corresponding values of the generic couplings vanish.

For a process that involves just one quark--antiquark pair, the matrix
element squared and summed over colours and spins of the fermions and
antifermions reads
\begin{equation} \label{Studentenschnitte}
|\mathcal{M}_{2 \, \mr{quarks}}|^2 = 
N_\mr{colour} \sum_{\sigma_1,\dots,\sigma_6}
   |\mathcal{M}^\mr{weak}_+ + \mathcal{M}^\mr{weak}_-|^2
\end{equation}
with the colour factor $N_\mr{colour}=3$.  For a specific $2f\to4f$
process this has to be divided by $4 N_\mr{av} N_\mr{sym}$ in order to
average over the polarizations and colours of the initial state and to
take into account identical particles in the final state.  For initial
state quarks we have $N_\mr{av}=9$, and the symmetry factor is given
by $N_\mr{sym}=2^{N_\mr{id}}$ for $N_\mr{id}$ pairs of identical
particles in the final state.

If there are four quarks, two additional complications
must be taken into account. First the exchange of gluons between two
pairs of quarks becomes possible giving rise to additional Feynman
diagrams with gluon exchange, and secondly we get different colour
structures.  For the squared matrix element with four quarks summed
over polarizations and colours we find
\beqar \left| \mathcal{M}_{4 \, \mr{quarks}} \right| ^2 &=& 
\sum_{\sigma_1,\dots,\sigma_6} \biggl[ 
9\left| \mathcal{M}^\mr{weak}_+ \right| ^2 
+ 9 \left|\mathcal{M}^\mr{weak}_- \right| ^2 
+ 2 \left|\mathcal{M}^\mr{gluon}_+ \right| ^2 
+ 2 \left|  \mathcal{M}^\mr{gluon}_- \right| ^2 
\nl&&{}
+ 6 \Re \left(
  \mathcal{M}^\mr{weak}_+ (\mathcal{M}^\mr{weak}_-)^* \right)  
+ 8\Re \left( \mathcal{M}^\mr{weak}_+ (\mathcal{M}^\mr{gluon}_-)^*
\right)
\nl&&{} 
+ 8 \Re \left( \mathcal{M}^\mr{weak}_-
  (\mathcal{M}^\mr{gluon}_+)^* \right) 
- \frac{4}{3} \Re \left(
  \mathcal{M}^\mr{gluon}_+ (\mathcal{M}^\mr{gluon}_-)^* \right)
\biggr].  
\eeqar 
We do not consider processes with six quarks.

The matrix elements for four-fermion-plus-photon production are constructed 
in complete analogy to the matrix elements for four-fermion production
from the generic diagrams given in \citere{Denner:1999gp}.

\subsection{Matrix elements for $\qqVV$ and $\Vff$}

In DPA, we need matrix elements for the processes
\beqar 
\bar q_1(p_1,\sigma_1) + q_2(p_2,\sigma_2) &\to& V_1(k_1,\la_1) +
V_2(k_2,\la_2), \nl 
V_1(k_1,\la_1) &\to& f_3(p_3,\sigma_3) + \bar f_4(p_4,\sigma_4), \nl 
V_2(k_2,\la_2) &\to& f_5(p_5,\sigma_5) + \bar f_6(p_6,\sigma_6).
\eeqar 
We take all pairs of massive gauge bosons, \ie $\mr{W}^+ \mr{W}^-$,
$\mr{W}^\pm \mr{Z}$, and $\mr{Z} \mr{Z}$, into account.  Owing to the
mixing of the $\mr{Z}$ boson with the photon also matrix elements for
$\mr{W}^\pm \gamma$ and $\mr{Z} \gamma$ production occur in the
results for the logarithmic EW radiative corrections. The
logarithmic corrections for matrix elements involving
longitudinal gauge bosons are calculated with the Goldstone-boson
equivalence theorem. As a consequence of the mixing of the would-be
Goldstone bosons with the Higgs boson, also the matrix elements for
the production of a gauge boson and a Higgs boson appear.  All these
matrix elements have been calculated with the Weyl--van der Waerden
spinor formalism. The results can be found in
\citere{Kaiserdissertation}.

\section{Logarithmic EW corrections}
\label{sec:ewrc}

In DPA, the $\Oa$ contributions consist of factorizable corrections to
gauge-boson production and decay as well as non-factorizable
corrections as summarized in \refeq{eq:virt_corr}. In the following we
list these corrections in the high-energy approximation.

\subsection{Corrections to gauge-boson production}
\label{sec:qqVVcorr}

In this section, we present the analytical formulas for the
logarithmic EW corrections to the polarized partonic
subprocesses
\beq\label{duprocess}
\bar q_1(p_1,\sigma_1) + q_2(p_2,\sigma_2) \to V_1(k_1,\la_1) +
V_2(k_2,\la_2) ,
\eeq 
which can be derived from the general results given in
\citere{Denner:2001jv}. The photon field is denoted by $A$. The
Mandelstam variables read
\beq\label{eq:mandelstam}
{\sparton}=(p_{1}+p_2)^2,\ \ \ {\tparton}=(p_{1}-k_1)^2, \ \ \ 
{\uparton}=(p_{1}-k_2)^2 ,
\eeq
where the momenta of the initial and final states are incoming and
outgoing, respectively.

The one-loop corrections are evaluated in the limit \refeq{HEA}, and
we neglect combinations of gauge-boson helicities that are
mass-suppressed compared with $\sqrt{{\sparton}}$ in this limit.
Thus, we do not consider corrections to the case of mixed longitudinal
and transversely polarized gauge bosons.  We calculate corrections to
the non-suppressed purely longitudinal final state
$(\la_1,\la_2)=(0,0)$, which we denote by $(\la_1,\la_2)=(\rL,\rL)$,
and to the purely transverse final states denoted by
$(\la_1,\la_2)=(\rT,\rT)$, which includes the non-suppressed
opposite-helicity final states $(\la_1,\la_2)=(\pm,\mp)$ and the
suppressed equal-helicity final states $(\la_1,\la_2)=(\pm,\pm)$.

The leading and next-to-leading logarithms depend only on tree-level
amplitudes and quantum numbers of the external particles, and thus are
universal.  Following \citere{Denner:2001jv}, the logarithmic
EW corrections can be written in the form
\begin{eqnarray}
\delta \M^{\bar{q}_1 q_2 \to V_{1,\la_1} V_{2,\la_2}} &=& 
 \delta^{\mathrm{LSC}} \mathcal{M} + \delta^{\mathrm{SSC}}\mathcal{M} 
 + \delta^{\mathrm{C}}\mathcal{M} + \delta^{\mathrm{PR}}\mathcal{M}.
\end{eqnarray}
Here $\delta^{\mathrm{LSC}}\M$ denotes the contribution of leading
soft-collinear corrections and $ \delta^{\mathrm{SSC}}\mathcal{M}$
the contribution of the next-to-leading soft-collinear
corrections, which are angular dependent. The term
$\delta^{\mathrm{C}}\mathcal{M}$ contains the collinear logarithms and
the logarithms related to the renormalization of the incoming and
outgoing fields.  Finally, $\delta^{\mathrm{PR}}\mathcal{M}$ are the
logarithmic corrections that arise from parameter renormalization.

The following results are directly based on the formulas of
\citere{Denner:2001jv} and written in a generic way for all processes
$\qqVV$. In this way they have been directly implemented in our Monte
Carlo program.  The amplitudes involving longitudinal gauge bosons are
evaluated using the Goldstone-boson equivalence theorem.  Therefore, it
is convenient to write the results in terms of Born matrix elements
with external would-be Goldstone bosons.%
\footnote{We denote the would-be Goldstone bosons corresponding to the
  Z and W bosons by $\chi$ and $\phi$, respectively.}  These matrix
elements have to be understood as shorthands for matrix elements with
external longitudinal gauge bosons according to
\beqar
  \M_{\born}^{\bar{q}_1 q_2 \to \Phi_1 \Phi_2} &=&  
  (-\ri)^{(Q_{V_1}+1)} (-\ri)^{(Q_{V_2}+1)} 
  \M_{\born}^{\bar{q}_1 q_2 \to V_{1,\rL} V_{2,\rL} } , \nl
  \M_{\born}^{\bar{q}_1 q_2 \to H \Phi}&=&  
  (-\ri)^{(Q_{V}+1)} \M_{\born}^{\bar{q}_1 q_2 \to H V_\mr{L} },
\eeqar
where $Q_{V}$ are the charges of the outgoing gauge bosons.

\subsubsection{Leading soft-collinear corrections}

According to formulas (3.6) and (3.7) of \citere{Denner:2001jv}, the
leading soft-collinear corrections read
\begin{eqnarray}
 \delta^{\mr{LSC}} \M^{\bar{q}_1 q_2 \to V_{1,\rT} V_{2,\rT}}
&=&     \Bigl( \delta^{\mr{LSC}}_{\bar q_1 \bar q_1} 
         +\delta^{\mr{LSC}}_{q_2 q_2} 
         +\delta^{\mr{LSC}}_{V_1 V_1} 
         +\delta^{\mr{LSC}}_{V_2 V_2} \Bigr) 
            \M_{\born}^{\bar{q}_1 q_2 \to V_{1,\rT} V_{2,\rT}} 
\nl &&
      { }+ \delta_{V_1 Z} \delta^{\mr{LSC}}_{A V_1} 
            \M_{\born}^{\bar{q}_1 q_2 \to A V_{2,\rT}} 
      { }+ \delta_{V_2 Z} \delta^{\mr{LSC}}_{A V_2} 
            \M_{\born}^{\bar{q}_1 q_2 \to V_{1,\rT} A} 
\end{eqnarray}
 for transverse gauge bosons and
\begin{eqnarray}
 \delta^{\mr{LSC}} \M^{\bar{q}_1 q_2 \to V_{1,\rL} V_{2,\rL}}
  &=& 
    \Bigl( \delta^{\mr{LSC}}_{\bar q_1 \bar q_1} 
         +\delta^{\mr{LSC}}_{q_2 q_2} 
         +\delta^{\mr{LSC}}_{\Phi_1 \Phi_1} 
         +\delta^{\mr{LSC}}_{\Phi_2 \Phi_2} \Bigr) 
            \M_{\born}^{\bar{q}_1 q_2 \to \Phi_1 \Phi_2} 
\end{eqnarray}
for longitudinal gauge bosons, where $\Phi_1$ and $\Phi_2$ denote the
would-be Goldstone bosons corresponding to $V_{1,\rL}$ and
$V_{2,\rL}$, respectively.  The factors
$\delta^{\mr{LSC}}_{\varphi^\prime \varphi} $ are defined as
\begin{eqnarray}
  \delta^\mathrm{LSC}_{\varphi^\prime \varphi} &=&
  - \frac{\alpha}{8 \pi} C^{\mathrm{ew}}_{\varphi^\prime \varphi} 
   \log^2 \left( \frac{\sparton}{\MW^2} \right)
  +\delta_{\varphi^\prime \varphi} 
     \Biggl\{ \frac{\alpha}{4 \pi} (I^Z_{\varphi} )^2
     \log \left( \frac{\sparton}{\MW^2} \right) 
     \log \left( \frac{\MZ^2}{\MW^2} \right) \nl && {}
    -\frac{1}{2} Q^2_{\varphi} 
  L^{\mathrm{em}}(\sparton,\lambda^2,M^2_{\varphi})
     \Biggr\},
\end{eqnarray}
where $M_{\varphi}$ and $Q_{\varphi}$  are the mass and relative
charge, respectively, of the field $\varphi =q, \bar{q}, W^\pm , Z,
\phi^\pm , \chi$, and $\lambda$ is the photon mass regulator.

The term $L^{\mathrm{em}}$ contains all leading soft-collinear 
logarithms of pure electromagnetic origin:
\begin{eqnarray}
  L^{\mathrm{em}}(\sparton,\lambda^2,M^2_{\varphi}) = 
    \frac{\alpha}{4 \pi} \Biggl\{ 
  2 \Log{\frac{\sparton}{\MW^2}} \Log{\frac{\MW^2}{\lambda^2}}
                 +\Logq{\frac{\MW^2}{\lambda^2}} 
                 -\Logq{\frac{M^2_{\varphi}}{\lambda^2}}
                         \Biggr\} .
\end{eqnarray}
The relevant non-vanishing components of the EW Casimir
operator $\cew$ read 
\beqar \cew_{qq}&=& \cew_{\bar q\bar q} =
\frac{(1+2\cw^2)+4(Q_q^2-2 I^3_q Q_q)\sw^2}{4\sw^2\cw^2}, \nl
\cew_{\phi^\pm\phi^\pm}&=&\cew_{\chi\chi}=\frac{1+2\cw^2}{4\sw^2\cw^2},\qquad
\cew_{W^\pm W^\pm}=\frac{2}{\sw^2}, \nl 
\cew_{AA} &=& 2,\qquad \cew_{AZ} =
\cew_{ZA} =-2\frac{\cw}{\sw},\qquad \cew_{ZZ} =2\frac{\cw^2}{\sw^2},
\eeqar 
and the relevant squared \PZ-boson couplings are given by
\beqar
(I^Z_{q})^2&=&(I^Z_{\bar q})^2 = \frac{(Q_q\sw^2-I^3_q)^2}{\sw^2\cw^2},\qquad
(I^Z_{W})^2 = \frac{\cw^2}{\sw^2},\nl 
(I^Z_{\phi})^2 &=&\frac{(\cw^2-\sw^2)^2}{4\sw^2\cw^2},\qquad (I^Z_{\chi})^2=
\frac{1}{4\sw^2\cw^2}.
\eeqar
Finally, $Q_q$ and $I^3_q$ denote the relative charge and the third
component of the weak isospin of the quark $q$.

\subsubsection{Subleading soft-collinear corrections}

The angular-dependent subleading soft-collinear corrections are
obtained from formula (3.12) of \citere{Denner:2001jv}.
For the production of transverse gauge bosons we get
\begin{eqnarray}
\lefteqn{\delta_{\mathrm{neutral}}^{\mathrm{SSC}} 
   \M^{\bar{q}_1 q_2 \to V_{1,\rT} V_{2,\rT}} =
   \frac{\alpha}{2 \pi} \sum_{V=A,Z}
   \left[ \Log{\frac{\sparton}{\MW^2}}+
     \de_{VA}\Log{\frac{\MW^2}{\lambda^2}} 
   \right]}\quad \nl && \times
   \biggl{\{} \Log{\frac{|\tparton|}{\sparton}} \left[ 
                I^{\bar{V}}_{\bar q_1 \bar q_1} I^{V}_{\bar V_1 \bar{V}_1}
              + I^{\bar{V}}_{q_2 q_2} I^{V}_{\bar V_2 \bar{V}_2} 
                               \right]  
\nl && {}\quad
              + \Log{\frac{|\uparton|}{\sparton}} \left[ 
                 I^{\bar{V}}_{\bar q_1 \bar q_1} I^{V}_{\bar V_2 \bar{V}_2} 
              +  I^{\bar{V}}_{q_2 q_2} I^{V}_{\bar V_1 \bar{V}_1}
            \right] \biggl{\}} 
\,\M_{\born}^{\bar{q}_1 q_2 \to V_{1,\rT} V_{2,\rT}} 
\end{eqnarray}
from the exchange of neutral virtual gauge bosons and 
\beqar\label{eq:delta_ssc_charged}
\lefteqn{\delta_{\mathrm{charged}}^{\mathrm{SSC}} 
\M^{\bar{q}_1 q_2 \to V_{1,\rT} V_{2,\rT}} =
   \frac{\alpha}{2 \pi} \sum_{V=W^\pm}\sum_{V'=A,Z,W^\pm}\sum_{q'}\,
   \Log{\frac{\sparton}{\MW^2}} } \qquad 
 \nl && {} \times
   \biggl{\{} \Log{\frac{|\tparton|}{\sparton}} \left[ 
   I^{\bar{V}}_{ \bar q^\prime\bar q_1} I^{V}_{\bar V^\prime \bar{V}_1} 
  \M_{\born}^{\bar q^\prime q_2 \to V_{\rT}^\prime V_{2,\rT}}
  + I^{\bar{V}}_{{q}^\prime q_2} I^{V}_{\bar V^\prime \bar{V}_2} 
  \M_{\born}^{\bar{q}_1 q^\prime  \to V_{1,\rT} V_{\rT}^\prime }
                               \right] \nl && \quad {} 
  + \Log{\frac{|\uparton|}{\sparton}} \left[ 
   I^{\bar{V}}_{ \bar q^\prime\bar q_1} I^{V}_{\bar V^\prime \bar{V}_2} 
  \M_{\born}^{\bar q^\prime q_2 \to V_{1,\rT}  V_{\rT}^\prime}
  + I^{\bar{V}}_{{q}^\prime q_2} I^{V}_{\bar  V^\prime \bar{V}_1} 
  \M_{\born}^{\bar{q}_1 q^\prime \to V_{\rT}^\prime V_{2,\rT}}
                               \right] \biggl{\}} 
\eeqar
from the exchange of charged virtual gauge bosons.  The charge
conjugated of the gauge boson $V$ is denoted by $\bar V$.  The
couplings $I^{V_1}_{ \bar V_2 \bar V_3}$ are defined in
\refeq{eq:V_couplings}.  The couplings $I^{\bar{V}}_{{q}^\prime q}$,
given in \refeq{eq:F_couplings}, involve the quark-mixing matrix,
and quark mixing requires the sum over $q^\prime$
in \refeq{eq:delta_ssc_charged}.  After using the unitarity of the
quark-mixing matrix, the EW logarithmic corrections have
exactly the same dependence on its matrix elements as the lowest
order.

For the production of longitudinal gauge bosons we find
\begin{eqnarray}
\lefteqn{ \delta_{\mathrm{neutral}}^{\mathrm{SSC}} 
\M^{\bar{q}_1 q_2 \to V_{1,\rL} V_{2,\rL}} =
  \ri^{((1+Q_{V_1})+(1+Q_{V_2}))}\,
  \delta_{\mathrm{neutral}}^{\mathrm{SSC}} \M^{\bar{q}_1 q_2
    \to S_1 S_2} }\quad \nl
   &=&\ri^{((1+Q_{V_1})+(1+Q_{V_2}))} \frac{\alpha}{2 \pi} 
  \sum_{V=A,Z} \sum_{S^\prime =\chi,H,\phi^\pm}
   \left[ \Log{\frac{\sparton}{\MW^2}}+
     \de_{VA}\Log{\frac{\MW^2}{\la^2}} \right] \nl && \times
   \biggl{\{} \Log{\frac{|\tparton|}{\sparton}} \left[ 
               I^{\bar{V}}_{\bar q_1 \bar q_1} I^{V}_{\bar  S^\prime \bar{S}_1}
                 \M_{\born}^{\bar{q}_1 q_2 \to S^\prime S_2}
             +  I^{\bar{V}}_{q_2 q_2} I^{V}_{\bar  S^\prime \bar{S}_2} 
                 \M_{\born}^{\bar{q}_1 q_2 \to S_1 S^\prime }
                              \right] \nl &&\quad
              + \Log{\frac{|\uparton|}{\sparton}} \left[ 
               I^{\bar{V}}_{\bar q_1 \bar q_1} I^{V}_{\bar S^\prime \bar{S}_2} 
                 \M_{\born}^{\bar{q}_1 q_2 \to S_1  S^\prime}
              +  I^{\bar{V}}_{q_2 q_2} I^{V}_{\bar  S^\prime \bar{S}_1}   
                 \M_{\born}^{\bar{q}_1 q_2 \to S^\prime S_2}              
                               \right] \biggl{\}}
\end{eqnarray}
from the exchange of neutral virtual gauge bosons and 
\begin{eqnarray}
\lefteqn{ \delta_{\mathrm{charged}}^{\mathrm{SSC}} 
\M^{\bar{q}_1 q_2 \to V_{1,\rL} V_{2,\rL}} =
  \ri^{((1+Q_{V_1})+(1+Q_{V_2}))}  \,
  \delta_{\mathrm{charged}}^{\mathrm{SSC}} 
  \M^{\bar{q}_1 q_2 \to S_1 S_2}}\quad \nl
  &=& \ri^{((1+Q_{V_1})+(1+Q_{V_2}))} \frac{\alpha}{2 \pi} 
  \sum_{V=W^\pm} \sum_{S^\prime =\chi,H,\phi^\pm}\sum_{q'}\,
   \Log{\frac{\sparton}{\MW^2}}  
\nl && \times
   \biggl{\{} \Log{\frac{|\tparton|}{\sparton}} \left[ 
    I^{\bar{V}}_{\bar q^\prime\bar q_1 } I^{V}_{\bar  S^\prime \bar{S}_1} 
                 \M_{\born}^{\bar q^\prime q_2 \to S^\prime S_2}
    + I^{\bar{V}}_{{q}^\prime q_2} I^{V}_{\bar  S^\prime \bar{S}_2} 
                 \M_{\born}^{\bar{q}_1 q^\prime  \to S_1 S^\prime }
                               \right] \nl &&\quad
              + \Log{\frac{|\uparton|}{\sparton}} \left[ 
     I^{\bar{V}}_{ \bar q^\prime\bar q_1} I^{V}_{\bar  S^\prime \bar{S}_2} 
                 \M_{\born}^{\bar q^\prime q_2 \to S_1  S^\prime}
    + I^{\bar{V}}_{{q}^\prime q_2} I^{V}_{\bar  S^\prime \bar{S}_1} 
                 \M_{\born}^{\bar{q}_1 q^\prime \to S^\prime S_2} 
                               \right] \biggl{\}} 
\end{eqnarray}
from the exchange of charged virtual gauge bosons, where the couplings
$I^{V}_{\bar S^\prime \bar{S}_2} $ are defined in
\refeq{eq:S_couplings}.

\subsubsection{Collinear logarithms}

The single collinear logarithms can be read off from formulas (4.2),
(4.6), (4.10), (4.22), and (4.33) of \citere{Denner:2001jv}.
Their contribution to the gauge-boson-production 
matrix element reads
\begin{eqnarray}
 \delta^{\mr{C}} \M^{\bar{q}_1 q_2 \to V_{1,\rT} V_{2,\rT}} &=& 
   \left( \delta^{\mr{C}}_{\bar q_1 \bar q_1}
         +\delta^{\mr{C}}_{q_2 q_2} 
         +\delta^{\mr{C}}_{V_{1,\rT} V_{1,\rT}} 
         +\delta^{\mr{C}}_{V_{2,\rT} V_{2,\rT}} \right) 
    \M_{\born}^{\bar{q}_1 q_2 \to V_{1,\rT} V_{2,\rT}} \nl &&{}
  + \delta_{V_1 Z} \delta^{\mr{C}}_{A V_{1,\rT}} 
     \M_{\born}^{\bar{q}_1 q_2 \to A V_{2,\rT}} 
  + \delta_{V_2 Z} \delta^{\mr{C}}_{A V_{2,\rT}} 
     \M_{\born}^{\bar{q}_1 q_2 \to V_{1,\rT} A} 
\end{eqnarray}  
 in the case of transverse gauge bosons and 
\begin{eqnarray}
 \delta^{\mr{C}} \M^{\bar{q}_1 q_2 \to V_{1,\rL} V_{2,\rL}} &=& 
   \left( \delta^{\mr{C}}_{\bar q_1 \bar q_1}
         +\delta^{\mr{C}}_{q_2 q_2} 
         +\delta^{\mr{C}}_{V_{1,\rL} V_{1,\rL}} 
         +\delta^{\mr{C}}_{V_{2,\rL} V_{2,\rL}} \right) 
    \M_{\born}^{\bar{q}_1 q_2 \to V_{1,\rL} V_{2,\rL}} 
\end{eqnarray} 
in the case of longitudinal gauge bosons. The collinear correction
factors for the different particles read
\begin{eqnarray}\label{eq:coll-logs}
 \delta^\mathrm{C}_{q_\si q_\si} &=& 
  \delta^\mathrm{C}_{\bar q_{-\sigma} \bar q_{-\sigma}} =
\frac{\alpha}{4 \pi} 
   \left[ \frac{3}{2} C^\mathrm{ew}_{q_\sigma q_\si} \Log{ \frac{\sparton}{\MW^2}} 
          + Q_{q_\si}^2 \left( \frac{1}{2} \Log{\frac{\MW^2}{m_f^2}} +
            \Log{\frac{\MW^2}{\lambda^2}} \right) \right],  \nl
\delta^\mathrm{C}_{W^\pm_\mr{T} W^\pm_\mr{T}} &=& \frac{\alpha}{4 \pi} 
  \biggl[ \frac{19}{12 \sw^2} \Log{\frac{\sparton}{\MW^2}} 
  + \Log{\frac{\MW^2}{\lambda^2}} 
           \biggr],\nl
\delta^\mathrm{C}_{Z_\mr{T} Z_\mr{T}} &=& \frac{\alpha}{4 \pi} 
\frac{19-38 \sw^2-22 \sw^4}{12 \sw^2 \cw^2} \Log{\frac{\sparton}{\MW^2}} 
, \nl
\delta^\mathrm{C}_{A Z_\mr{T}} &=& -\frac{\alpha}{4 \pi} 
        \frac{19+22 \sw^2}{6 \sw \cw} \Log{\frac{\sparton}{\MW^2}} 
,\nl
\delta^\mathrm{C}_{W^\pm_\mr{L} W^\pm_\mr{L}} &=& 
   \frac{\alpha}{4 \pi} 
   \biggl[ \frac{1+2 \cw^2}{2 \sw^2 \cw^2} \Log{\frac{\sparton}{\MW^2}} 
   - \frac{3}{4 \sw^2} \frac{\Mt^2}{\MW^2} \Log{\frac{\sparton}{\Mt^2}}
   + \Log{\frac{\MW^2}{\lambda^2}}   
   \biggr], \nl
\delta^\mathrm{C}_{Z_\mr{L} Z_\mr{L}} &=& 
      \frac{\alpha}{4 \pi} 
      \biggl[ \frac{1+2 \cw^2}{2 \sw^2 \cw^2} \Log{\frac{\sparton}{\MW^2}} 
      - \frac{3}{4 \sw^2} \frac{\Mt^2}{\MW^2} \Log{\frac{\sparton}{\Mt^2}}
        \biggr].
\end{eqnarray}
While in \citere{Denner:2001jv} all masses of the order of the
EW scale were replaced by $\MW$ in the arguments of the large
logarithms, we keep $\Mt$ in the logarithm resulting from top-quark
loops.

\subsubsection{Logarithms from parameter renormalization}

The parameter renormalization gives rise to the so-called
counter-term contributions which result from
\begin{equation}\label{eq:para-ren}
\delta^\mathrm{PR} \M^{\bar{q}_1 q_2 \to V_1 V_2} =
  \left. 
  \frac{ \partial \M_{\born}^{\bar{q}_1 q_2 \to V_1 V_2}}{\partial e}
       \delta e 
+  \frac{ \partial \M_{\born}^{\bar{q}_1 q_2 \to V_1 V_2}}{\partial \cw}
       \delta \cw \right|_{\mu^2=\sparton},
\end{equation}
where $\de e$ and $\de\cw$ are the counter terms to the electric
charge $\Pe$ and the cosine of the weak mixing angle $\cw
={M_\mr{W}}/{M_\mr{Z}}$, respectively.
The mass parameter $\mu^2$ of dimensional regularization is set to $\sparton$
in order not to introduce spurious large logarithms.
The counter terms depend on the explicit renormalization conditions.
We fix them in the on-shell scheme and obtain in logarithmic
approximation:
\begin{eqnarray}\label{eq:charge-ren}
\frac{\delta e}{e} &=& \frac{1}{2} \left[ \frac{\alpha}{4 \pi} 
       \frac{11}{3} \Log{\frac{\sparton}{\MW^2}} 
   +\Delta \alpha(\MW^2) \right] 
, \\
 \frac{\delta \cw}{\cw} &=& -\frac{\alpha}{4 \pi} 
   \frac{19+22 \sw^2}{12\cw^2} 
   \Log{\frac{\sparton}{\MW^2}}
  ,
\end{eqnarray}
where 
\beq
\Delta\alpha(\MW^2) = \frac{\al}{3\pi}\sum_{f\neq t}
\NCf Q_{f}^2 \ln\left(\frac{\MW^2}{m_f^2}\right)
\eeq
describes the running of $\al$ from zero to the EW scale.

\subsection{Corrections to gauge-boson decay}
\label{one_loop_decay}
\label{sec:Vffcorr}

Since the decay matrix elements are independent of a large energy
scale, no energy-dependent logarithms appear in the corresponding corrections.
The only large logarithms result from electromagnetic corrections, \ie
from diagrams with photon exchange. For massless fermions these
corrections turn out to be proportional to the lowest-order matrix element
\begin{eqnarray}
 \delta \M^{V_\la \to f \bar{f}'} &=& 
\delta_{V f \bar{f}'} \M_{\born}^{V_\la \to f \bar{f}'}.
\end{eqnarray} 
In the logarithmic approximation the correction factors
for $V=\mr{Z}$ and $V= \mr{W}^\pm$ read
\beqar\label{eq:decay_corr}
\delta_{\mr{Z} f \bar{f}} &=& \frac{\alpha}{4 \pi} Q_f^2 
 \biggl[  \ln^2 \frac{m_f^2}{\lambda^2}           
-\ln^2 \frac{\MZ^2}{\lambda^2}
 +\ln \frac{\MZ^2}{m_f^2} 
+2 \ln \frac{\MZ^2}{\lambda^2}\biggr]
+\frac{1}{2}\De\alpha(\MZ^2),
\nl[1ex]                                 
\delta_{\mr{W} f \bar f'} &=& \frac{\alpha}{4 \pi}        
\biggl\{\frac{1}{2}Q_{f}^2 \biggl[
 \ln^2 \frac{m_{f}^2}{\lambda^2}           
-\ln^2 \frac{\MW^2}{\lambda^2}
 +\ln \frac{\MW^2}{m_{f}^2} 
+2 \ln \frac{\MW^2}{\lambda^2}\biggr]     
\nl&&{}
+\frac{1}{2}Q_{f'}^2 \biggl[
 \ln^2 \frac{m_{f'}^2}{\lambda^2}           
-\ln^2 \frac{\MW^2}{\lambda^2}
 +\ln \frac{\MW^2}{m_{f'}^2} 
+2 \ln \frac{\MW^2}{\lambda^2}\biggr]  \nl 
&&{} + (Q_{f'}-Q_{f})^2 \ln \frac{\MW^2}{\lambda^2} 
\biggr\}                                                 
+\frac{1}{2}\De\alpha(\MW^2).
\eeqar

\subsection{Non-factorizable corrections}
\label{sec:nfRC}

The non-factorizable corrections for a general class of processes are
evaluated in \refapp{sec:non-fact-phot}.  For the processes
\refeq{eq:parton_process} the corresponding correction factor to the
lowest-order cross section reads
\begin{eqnarray}
\de^{\virt}_{\nonfact,\DPA}&=& \frac{\alpha}{\pi} \biggl(
   - \sum_{i=3}^{4} \sum_{j=5}^{6} Q_i Q_j \theta_d(i) \theta_d(j) 
      \Re \{ \Delta_1(k_1,p_i;k_2,p_j) \} \nl
&& + \sum_{k=1}^{2} \sum_{i=3}^{4} Q_k Q_i \theta_d(k) \theta_d(i)
      \Re \{ \Delta_2(p_k;k_1,p_i) \} \nl
&& + \sum_{k=1}^{2} \sum_{j=5}^{6} Q_k Q_j \theta_d(k) \theta_d(j)
      \Re \{ \Delta_2(p_k;k_2,p_j) \} \biggr) 
\end{eqnarray}
in DPA where $Q_i$ are the relative charges of the fermions
corresponding to the external legs and $\theta_d$ is defined 
by
\beq\label{def_theta_d}
\theta_d(i)=\left\{ \barr{ll} 
 +1 & \mbox{for incoming fermions and outgoing antifermions} \\
 -1 & \mbox{for incoming antifermions and outgoing fermions}
\earr \right.\,
\eeq
and accounts for the sign difference of the charges of fermions and
antifermions.  In the high-energy limit we assume $\MW^2\ll|r|$ for
all kinematical invariants $r$ that are not fixed to a certain mass
value like $s_{34} = (p_3 + p_4)^2 = M_V^2$ after on-shell projection
and keep only logarithmic terms. If we apply this approximation to the
non-factorizable corrections we find rather simple expressions for the
quantities $\Delta_1$ and $\Delta_2$:
\begin{eqnarray}\label{eq:delta1}
\Delta_1(k_1,p_i;k_2,p_j) &=&  
\frac{1}{2} ( s_{ij} \bar{s} - \stwo_{1j} \stwo_{2i}) 
D_0^{\mr{he}}( - k_2 +p_j, k_1 +p_j, p_i+p_j, m_j, M_2, M_1, m_i)  \nl
  &&{}  + \log \left( \frac{ (k_2^2-\overline{M}_2^2) {M}_1}{(k_1^2-\overline{M}_1^2) {M}_2} \right) 
      \log \frac{ \stwo_{2i}}{\stwo_{1j}} \nl
  &&{}  + \left[ 2 + \log \frac{ s_{ij}}{\bar{s}} \right] 
      \left[ \log \frac{\lambda {M}_2}{\overline{M}_2^2 - k_2^2} + \log \frac{\lambda {M}_1}{\overline{M}_1^2 - k_1^2} \right],
 \\
\Delta_2(p_k;k_l,p_j) &=&  2 \log \frac{\lambda {M}_l}{\overline{M}_l^2 - k_l^2} 
  \left[ \log \frac{ {\ttwo}_{kl}}{t_{kj}} -1 \right]. 
\end{eqnarray}
The invariants are defined as $\stwo_{lj}= (k_l + p_j)^2$, $\ttwo_{kl}
= (p_k-k_l)^2 $, $s_{ij}= (p_i + p_j)^2$, $t_{ij} = (p_i - p_j)^2$ and
$\bar{s} = s_{12} = (p_1+ p_2 )^2 $ and have to be calculated using
the appropriate on-shell-projected momenta.  Note that the invariants
$\stwo_{lj}$, $\ttwo_{kl}$ and $\bar{s}$ differ from the corresponding
invariants defined in \refeq{eq:shorthands} appearing in
\refeq{eq:Deltaoreshe} and \refeq{eq:Deltatreshe}. It it crucial for
the cancellation of the IR singularities that we use here the same
definitions as in \refeq{eq:mandelstam}.  In the high-energy limit we
are interested in, the differences between the definitions disappear.
In the $D_0^\mr{he}$ function, which is given in \refeq{D00he}, also
the on-shell-projected momenta enter. The original set of momenta
enters the non-factorizable corrections only in the terms
$k_l^2-\overline{M}_l^2$, where
$\overline{M_l} = \sqrt{M_l^2 - \ri M_l \Gamma_l}$ are the complex
masses of the gauge bosons. We note that we could omit the first two
lines of \refeq{eq:delta1} in the logarithmic approximation since they
do not contain large logarithms.

\section{Treatment of soft and collinear photon emission}
\label{sec:treatm-soft-coll}

In this section we describe the treatment of soft and collinear photon
emission. Soft and collinear singularities are regularized by an
infinitesimal photon mass $\la$ and small fermion masses,
respectively. The masses of the external fermions are denoted by $m_i$
($p_i^2=m_i^2\to0$). 

\subsection{Phase-space slicing}
\label{sec:phase-space-slicing}
\newcommand{\delsoft}{\de_{\mathrm{s}}}
\newcommand{\delcoll}{\de_{\mathrm{c}}}

For the evaluation of the real corrections we use the 
phase-space slicing method, where the phase space is divided into singular
and non-singular regions. The singular regions are integrated
analytically thus allowing the explicit cancellation of the
singularities against their counterparts in the virtual
corrections. The finite reminder is evaluated by using Monte Carlo
techniques.

For the actual implementation of this well-known procedure (see \eg
\citere{Berends:1981uq}) we closely follow the approach of
\citere{Denner:2000bj}.  We divide the five-particle phase space into
soft, collinear, and finite regions by introducing the cut parameters
$\delsoft$ and $\delcoll$, respectively. The soft region contains
photons with energies $E_{\gamma} < \delsoft \sqrt{\sparton}/2= \Delta
E$ in the CM frame of the incoming partons.  The collinear region
contains all photons with $E_{\gamma}>\Delta E$ but collinear to any
charged fermion, \ie with $1-\delcoll < \cos\theta_{\ga f} < 1$, where
$\theta_{\ga f}$ is the angle between the charged fermion and the
emitted photon in the partonic CM frame.  The finite region contains
all photons with $E_{\gamma}>\Delta E$ and $-1 < \cos\theta_{\ga f} <
1-\delcoll$ for all charged fermions.
 
In the soft and collinear regions, the squared matrix element
$|\M^{\qqffffg}|^2$ factorizes into the leading-order squared matrix
element $|\M^{\qqffff}_{\Born}|^2$ and a soft or collinear factor as
long as $\delsoft$ and $\delcoll$ are sufficiently small.  Also the
five-particle phase space factorizes into a four-particle phase space,
and a soft or collinear part.  As a consequence the contribution of
the real corrections can be written as
\begin{equation}\label{eq:si_4fg}
  \int \nolimits_{\displaystyle{\Phi_{4f\ga}}} 
  \mr{d} \sigma_{\mr{real}}^{\qqffffg} =
  \int \nolimits_{\displaystyle{\Phi_{4f\ga}}} 
  \mr{d} \sigma_{\mr{finite}}^{\qqffffg} +
  \int \nolimits_{\displaystyle{\Phi_{4f}}} \mr{d} \sigma_{\mr{soft}} +
  \int \nolimits_{\displaystyle{\Phi_{4f}}} \mr{d} \sigma_{\mr{coll}} .
\end{equation}

In the soft-photon region, we use the soft-photon approximation, \ie
the photon four-momentum $k$ is omitted everywhere but in the
IR-singular propagators. Since we neglect $k$ also in the resonant
gauge-boson propagators we have to assume $E_{\gamma}< \Delta E \ll
\Gamma_V$, $V=\PW,\PZ$.  In this region, $\rd\sigma^{\qqffffg}$ can be
written as \cite{Yennie:1961ad,Denner:kt}
\begin{equation}
\mr{d}\sigma_\mathrm{soft} = 
   -\mr{d}\sigma_{\mathrm{Born}}^{\bar q_1 q_2\to 4f} 
   \frac{\alpha}{2 \pi} 
   \sum_{i=1}^6 \sum_{j=1}^6 
   Q_i Q_j \theta_d(i) \theta_d(j) I_{ij},
\end{equation}
where
\begin{equation}
I_{ij}=\frac{1}{2\pi} 
\int_{E_{\gamma}< \Delta E \atop |{\bf k}|^2=E_{\gamma}^2-\lambda^2} 
\frac{\rd^3 {\bf k}}{E_{\gamma}} \frac{p_i p_j}{(kp_i) (kp_j)}.
\end{equation}
The explicit expression for the integrals $I_{ij}$ can be found in
\citeres{Denner:kt,'tHooft:1978xw}.  Since we only investigate high
energies, we can assume that the energies $ E_i$ of the external
fermions in the CM frame are large compared with their masses, 
$ E_i \gg m_i$, and keep the fermion masses $m_i$ only as regulators
so that $\mr{d} \sigma_{\mathrm{soft}}$ can be written as
\begin{eqnarray}\label{eq:si_soft}
 \mr{d}\sigma_\mathrm{soft} &=& \mr{d}\sigma_{\mathrm{Born}}^{\bar q_1 q_2
    \to 4f}
    \frac{\alpha}{2 \pi} \sum_{i=1}^5 \sum_{j=i+1}^6 
     Q_i Q_j \theta_d(i)  \theta_d(j)
    \Biggl\{ 2 \log \left( \frac{2 \Delta E}{\lambda}\right) \nl && {}
    \times \biggl[ 2-\log\biggl( \frac{s_{ij}^2}{m_i^2 m_j^2} \biggr) \biggr]
          -2 \log\left( \frac{ 4 E_i E_j}{m_i m_j} \right)
          +\frac{1}{2} \log^2\left( \frac{4 E_i^2}{m_i^2} \right)  \nl && {}
          +\frac{1}{2} \log^2\left( \frac{4 E_j^2}{m_j^2} \right)  
          +\frac{ 2 \pi^2}{3} +2 \, \mathrm{Li}_2\left( 1- \frac{ 4 E_i E_j}{s_{ij}} \right) 
    \Biggl\} ,
\end{eqnarray}
where $s_{ij}=(p_i+p_j)^2$.

In the collinear region, we use the collinear limit, \ie the
components of the photon four-momentum $k$ perpendicular to the
momentum of the collinear fermion are omitted everywhere but in the
singular propagators.  The collinear cross section is divided into a
part $\mr{d}\sigma^{\mr{initial}}_{\mr{coll}}$ originating from
initial-state radiation and a part
$\mr{d}\sigma^{\mr{final}}_{\mr{coll}}$ originating from final-state
radiation,
\begin{equation}\label{eq:si_collin}
 \mr{d} \sigma_{\mr{coll}} = \mr{d}\sigma_{\mr{coll}}^{\mr{initial}} 
       + \mr{d}\sigma_{\mr{coll}}^{\mr{final}}.
\end{equation}
While the emission of photons from the final state does not change the
kinematics of the subprocess, the initial-state radiation causes a
loss of energy of the incoming partons.  In the latter case  assuming
unpolarized incoming partons the cross section reads
\begin{equation} 
  \mr{d}\sigma^{\mr{initial}}_{\mr{coll}} =
  \sum_{i=1,2} \frac{\alpha}{2 \pi} Q_i^2
  \int_0^{1-\delta_s} \mr{d} z_i  
\left( \frac{1+z_i^2}{1-z_i} 
      \Log{\frac{\hat{s}}{m_i^2} \frac{\delta_c}{2}} 
          -\frac{2 z_i}{1-z_i} \right)
  \mr{d}\sigma_{\mr{Born}}^{\bar q_1 q_2\to 4f}(z_i p_i) ,
\end{equation}
where $z_i$ denotes the fraction of energy of the incoming parton $i$
that is left after emission of the collinear photon, and $\hat{s}$ is
defined in \refeq{eq:mandelstam}.  The differential cross section for
final-state radiation reads
\begin{eqnarray}
 \mr{d}\sigma_{\mr{coll}}^{\mr{final}} &=&
  \sum_{i=3}^6 \frac{\alpha}{2 \pi} Q_i^2 
  \left( \left[ \frac{3}{2} + 2 \Log{\frac{\Delta E}{E_i}} \right] 
         \left[ 1- \Log{\frac{4 E_i^2}{m_i^2} \frac{\delta_c}{2}} \right]
         + 3  - \frac{2 \pi^2}{3} \right) 
   \mr{d}\sigma_{\mr{Born}}^{\bar q_1 q_2\to 4f}.
\end{eqnarray}
Note that this procedure implicitly assumes that photons within small
cones collinear to charged final-state fermions will never be
separated from those collinear fermions.

\begin{figure}
  \unitlength 1cm
  \begin{center}
  \begin{picture}(16.0,7.5)
  \put(0,0){\epsfig{file=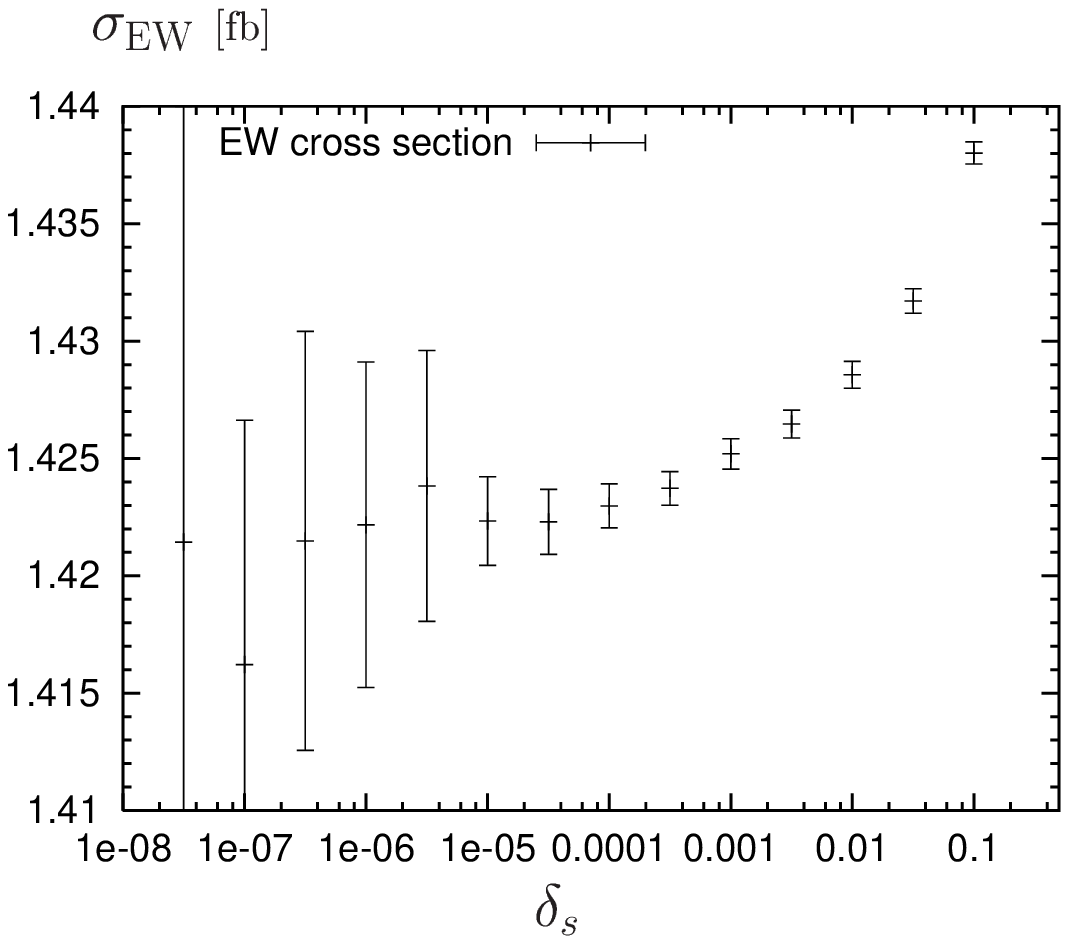,width=8cm}}
  \put(8,0){\epsfig{file=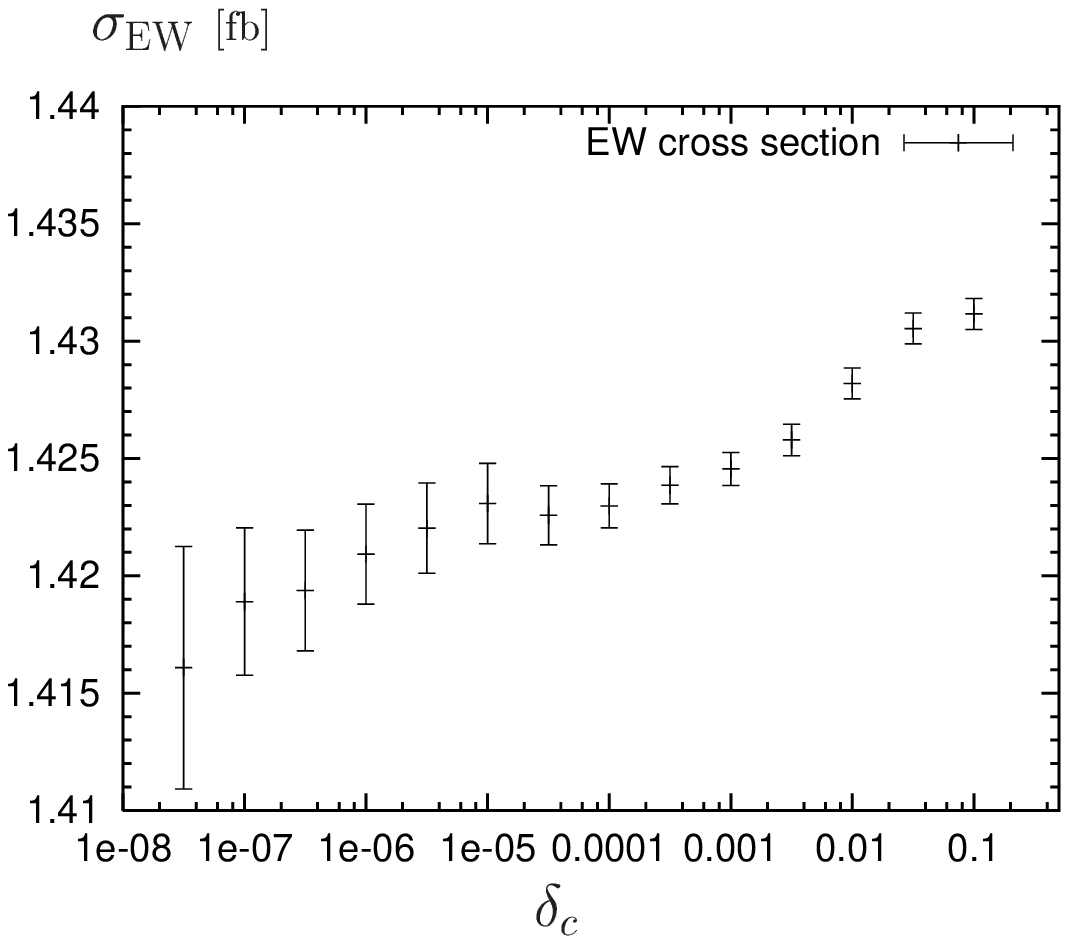,width=8cm}}
  \end{picture}
  \end{center}
\caption{Dependence of the cross section for
  $\nu_e\Pep\mu^-\bar\nu_\mu$ production in the scenario
  \refeq{eq:WWscenarioII} on the phase-space slicing cuts.  Left:
  Dependence on $\delta_s$ for $\delta_c=10^{-4}$.  Right: Dependence
  on $\delta_c$ for $\delta_s=10^{-4}$.}
\label{phasespaceslicingcuts}
\end{figure}

Subtracting the soft and collinear cross sections \refeq{eq:si_soft}
and \refeq{eq:si_collin} from the cross section of the process
$\qqffffg$ \refeq{eq:si_4fg} yields the finite part
$\rd\sigma_{\finite}^{\qqffffg}$.  The different contributions depend
on the cut parameters $\delsoft$ and $\delcoll$.  The dependence on
these technical cuts cancels in the sum when the cut parameters are
chosen to be small enough so that the soft-photon and collinear
approximations apply.  The variation of the cross section for
$\nu_e\Pep\mu^-\bar\nu_\mu$ production in the scenario
\refeq{eq:WWscenarioII} with the parameters $\delta_s$ and $\delta_c$
is shown in \reffi{phasespaceslicingcuts}.  While the numerical
integration becomes unstable for very small cuts, and the soft and
collinear approximations fail for too large cuts, the cross section is
independent of the cuts within integration errors for
$10^{-5}\lsim\delsoft\lsim10^{-3}$ and
$10^{-6}\lsim\delcoll\lsim10^{-3}$. For the numerical analysis we have
chosen $\delsoft=10^{-4}$ and $\delcoll=10^{-4}$.

\subsection{Definition of finite virtual corrections}
\label{sec:defin-finite-virt}

We fix the finite parts of $\rd \sigma_{\virt,\sing}^{\qqffff}$
entering \refeq{eq:crosssection} by adopting the convention of
\citere{Denner:2000bj},
\begin{eqnarray}\label{eq:virt-sing}
 \mr{d} \sigma_\mr{virt,sing(,DPA)}^{\bar q_1 q_2 \to 4f} &=& 
   \mr{d} \sigma_\mr{Born (,DPA)}^{\bar q_1 q_2 \to 4f} \frac{\alpha}{2 \pi} 
   \sum_{i=1}^5 \sum_{j=i+1}^6 Q_i Q_j \theta_d(i) \theta_d(j) \nl && \times
   \left[ \mathcal{L}(s_{ij},m_{i}^2)+  \mathcal{L}(s_{ij},m_{j}^2)
         + C_{ij} + C_{ji} \right]
\end{eqnarray}
with the invariants $s_{ij} = (p_i + p_j)^2$, the masses $m_i$ of the
external fermions, their relative charges $Q_i$, $\theta_d$ given by
\refeq{def_theta_d},
\begin{eqnarray}
 \mathcal{L}(s_{ij},m_i^2) &=& 
   \log \frac{m_i^2}{s_{ij}} \log \frac{\lambda^2}{s_{ij}}
  +\log \frac{\lambda^2}{s_{ij}} -\frac{1}{2} \log^2 \frac{m_i^2}{s_{ij}} 
  + \frac{1}{2} \log \frac{m_i^2}{s_{ij}},
\end{eqnarray}
and the constant terms $C_{ij}$ defined as
\begin{eqnarray} 
 C_{ij} &=& -\frac{\pi^2}{3}+2, \quad \mbox{if $i$ and $j$ are incoming}, \nl 
 C_{ij} &=& \,\,\,\,\, \frac{\pi^2}{6}-1, \quad \mbox{if $i$ is incoming and $j$ is outgoing}, \nl 
 C_{ij} &=& -\frac{\pi^2}{2}+\frac{3}{2}, \quad \mbox{if $i$ is outgoing and $j$ is incoming}, \nl 
 C_{ij} &=& -\frac{\pi^2}{3}+\frac{3}{2}, \quad \mbox{if $i$ and $j$ are outgoing}.
\end{eqnarray}
Of course, in logarithmic approximation the constants $C_{ij}$ could
be omitted.

\subsection{Absorption of mass singularities in parton distributions}

After combining real and virtual corrections, the $\Oa$-corrected
partonic cross section still contains mass-singular terms of the form
$\alpha \log m_{q_k}$ involving the masses $m_{q_k}$ of the incoming
partons.  These terms arise from collinear emission of photons in the
initial state. In analogy to the $\overline{\mr{MS}}$ factorization scheme
for next-to-leading-order QCD corrections, we absorb these collinear
singularities into the quark distributions. To this end, we replace
the parton-distribution functions in \refeq{eq:convolution} as
\begin{eqnarray}\label{eq:parton_ren}
 \Phi_{q,h}(x,Q^2) &\to& \Phi_{q,h}(x,Q^2) - \frac{\alpha}{2 \pi} Q_q^2 
  \int_x^1 \frac{\mr{d}z}{z} \Phi_{q,h}\left(\frac{x}{z},Q^2\right)  \nl
 &&  \times \left[ 
  \frac{1+z^2}{1-z}
  \left( \log \frac{Q^2}{m_q^2} - 2  \log(1-z)-1 \right) \right]_+ ,
\end{eqnarray}
where the usual $[\dots]_+$ prescription is defined by
\begin{equation}\label{renormalizedpdfequ}
 \int_x^1 \mr{d} z \left[ f(z) \right]_+ g(z) 
= \int_x^1 \mr{d}z f(z) g(z)- \int_0^1 \mr{d}z f(z) g(1) .
\end{equation} 
The replacement \refeq{eq:parton_ren} amounts to a contribution
\begin{eqnarray}
  \mr{d} \sigma_\mr{pdf} &=& -\frac{\alpha}{2 \pi} \sum_{i=1}^2 
     Q_i^2 \int_0^1 \mr{d}z 
     \left[ \frac{1+z^2}{1-z} 
     \left( \log \frac{Q^2}{m_i^2} - 2 \log(1-z) -1 \right) \right]_+
     \mr{d}\sigma_\mr{Born}^{\bar q_1 q_2 \to 4f}(z_ip_i) 
\nln
\end{eqnarray}
that has to be added to the partonic cross section. When adding
$\mr{d} \sigma_\mr{pdf}$ to $\mr{d} \sigma_{\mr{soft}}$, $\mr{d}
\sigma_{\mr{coll}}$, and  $\rd\sigma_{\virt,\sing}^{\qqffff}$ all 
IR and collinear singularities, \ie all $\ln(\la^2)$ and 
$\ln(m_i^2)$ terms, cancel. 

The absorption of the collinear $\Oa$ singularities into the parton
distributions requires also the inclusion of the corresponding
corrections into the DGLAP evolution of these distributions and into
their fit to experimental data. At present time there exist no
published PDFs in which the photonic $\mathcal{O}(\alpha)$ corrections
are consistently included.
An approximative inclusion of the $\Oa$ corrections to the DGLAP
evolution shows \cite{Kripfganz:1988bd} that the impact of these
corrections is below about $1\%$. Therefore, these effects are below our
aimed accuracy of a few per cent and can be neglected.

\section{Setup of the numerical analysis}
\label{sec:processes}

We consider three classes of processes:
\renewcommand{\labelenumi}{(\roman{enumi})}
\begin{enumerate}
\item $\Pp\Pp\to l\nu_ll^\prime\bar{l^\prime}($+$\gamma )$,
\qquad 
\item $\Pp\Pp\to l\bar{l}l^\prime\bar{l^\prime}(+\gamma )$,
\qquad 
\item $\Pp\Pp\to l\bar\nu_l\nu_{l^\prime}\bar{l^\prime}(+\gamma )$, 
\end{enumerate}
where $l,l^\prime=\Pe$ or $\mu$. In our notation, $l\nu_l$ indicates
both $l^-\bar\nu_l$ and $l^+\nu_l$.  The first class is characterized
by three isolated charged leptons plus missing energy in the final
state.  This channel includes $\PW\PZ$ production as intermediate
state.  The second class is purely mediated by $\PZ\PZ$ production,
while the third class is related to $\PW^\pm\PW^\mp$ production.  When
there is a unique flavor in the final state, $l=l'$, the third
process receives also a $\PZ\PZ$ contribution.
 
All above-mentioned processes are described by \refeq{eq:convolution}.
Since the two incoming hadrons are protons and we sum over final
states with opposite charges, we find
\beqar\label{eq:convolWZ}
\rd\si^{\Pp\Pp}(P_1,P_2,p_f) = 
\int_0^1\rd x_1 \rd x_2 &&\sum_{U=\Pu,\Pc}\sum_{D=\Pd,\Ps}
\Bigl[\Phi_{\bar\PD,\Pp}(x_1,Q^2)\Phi_{\PU,\Pp}(x_2,Q^2)\,\rd\hat\si^{\bar\PD\PU}
(x_1P_1,x_2P_2,p_f)
\nl&&{}
+\Phi_{\bar\PU,\Pp}(x_1,Q^2)\Phi_{\PD,\Pp}(x_2,Q^2)\,\rd\hat\si^{\bar\PU\PD}
(x_1P_1,x_2P_2,p_f)
\nl&&{}
+\Phi_{\bar\PD,\Pp}(x_2,Q^2)\Phi_{\PU,\Pp}(x_1,Q^2)\,\rd\hat\si^{\bar\PD\PU}
(x_2P_2,x_1P_1,p_f)
\nl&&{}
+\Phi_{\bar\PU,\Pp}(x_2,Q^2)\Phi_{\PD,\Pp}(x_1,Q^2)\,\rd\hat\si^{\bar\PU\PD}
(x_2P_2,x_1P_1,p_f)
\Bigr]
\eeqar
for $\PW\PZ$ production and 
\beqar\label{eq:convolVV}
\rd\si^{\Pp\Pp}(P_1,P_2,p_f) = 
\int_0^1\rd x_1 \rd x_2 &&\sum_{q=\Pu,\Pd,\Pc,\Ps}
\Bigl[\Phi_{\bar\Pq,\Pp}(x_1,Q^2)\Phi_{\Pq,\Pp}(x_2,Q^2)\,\rd\hat\si^{\bar\Pq\Pq}
(x_1P_1,x_2P_2,p_f)
\nl&&{}
+\Phi_{\bar\Pq,\Pp}(x_2,Q^2)\Phi_{\Pq,\Pp}(x_1,Q^2)\,\rd\hat\si^{\bar\Pq\Pq}
(x_2P_2,x_1P_1,p_f)
\Bigr]
\eeqar
for $\PZ\PZ$ and $\PW\PW$ production in leading order of QCD.  Since
the initial state is forward--backward symmetric for two incoming
protons, this cross section is forward--backward symmetric.

For the free parameters we use the input values \cite{Hagiwara:pw,mtop}:
\beq\label{eq:SMpar}
\begin{array}[b]{lcllcllcl}
G_\mu &= & 1.16637 \times 10^{-5} \GeV^{-2}, \qquad &
\MW & = & 80.425\GeV, \qquad &
\MZ & = & 91.1876\GeV, \qquad  \\
\Mt & = & 178.0 \GeV,  \quad \\
\end{array}
\eeq
The weak mixing angle is fixed by $\sw^2=1-\MW^2/\MZ^2$.  Moreover, we
adopted the so-called $G_{\mu}$-scheme, which effectively includes
higher-order contributions associated with the running of the
electromagnetic coupling and the leading universal two-loop
$\Mt$-dependent corrections. To this end we parametrize the
lowest-order matrix element in terms of the effective coupling
$\alpha_{G_{\mu}}=\sqrt{2}G_{\mu}\MW^2\sw^2/\pi 
= 7.543596\ldots \times 10^{-3}$  
and omit the explicit contributions proportional to
$\De\al(\MW^2)$ and $\De\al(\MZ^2)$ in \refeq{eq:charge-ren} and
\refeq{eq:decay_corr}.
In this setup our results are independent of the masses of the
(internal) light quarks.
Additional input parameters are the quark-mixing matrix elements whose
values have been taken to be $V_{\Pu\Pd}=0.974$ \cite{Hocker:2001xe},
$V_{\Pc\Ps}=V_{\Pu\Pd}$, $V_{\Pu\Ps}= -V_{\Pc\Pd}= \sqrt{1-
  |V_{\Pu\Pd}|^2} = 0.226548\ldots$, 
$V_{\Pt\Pb}=1$, and zero for all other matrix elements.

For the numerical results presented here, we have used the fixed-width scheme 
with $\GZ$ and $\GW$ from standard formulas 
\beqar
\GZ &=& {\alpha\MZ\over{24\sw^2\cw^2}} 
\Bigl[21-40\sw^2+{160\over 3}\sw^4-9{\Mb^2\over\MZ^2}
+{\Mb^4\over\MZ^4}(24\sw^2-16\sw^4)
\nl&&\qquad\qquad{}
+{\alpha_s\over\pi}\Bigl(15-28\sw^2+{88\over 3}\sw^4\Bigr)\Bigr]
\eeqar
and 
\beq
\GW ={\alpha\MW\over{2\sw^2}}\left [{3\over 2}+{\alpha_s\over\pi}
\right ] .
\eeq
Using $\alpha_s=0.117$ for the strong coupling and $\Mb=4.9\GeV$
we obtain 
\beq
\GZ = 2.505044\ldots \GeV,\qquad 
\GW = 2.099360\ldots \GeV.
\eeq
As to parton distributions, we have used CTEQ6M \cite{cteq} at the
following factorization scales for the three classes of processes.
We have chosen
\begin{equation}\label{eq:scaleWZ}
Q^2={1\over 2}\left (\MW^2+\MZ^2+\PT^2(l\nu_l)+
\PT^2(l^\prime\bar{l^\prime})\right )
\end{equation}
for $\PW\PZ$ production and 
\begin{equation}\label{eq:scaleZZ}
Q^2={1\over 2}\left (2\MZ^2+\PT^2(l\bar l)+
\PT^2(l^\prime\bar{l^\prime})\right )
\end{equation}
for $\PZ\PZ$ production, \ie we use the transverse momentum  $\PT$ of the
produced gauge bosons to fix $Q^2$. For $\PW\PW$ production, where the 
gauge bosons cannot be reconstructed, we take 
\begin{equation}\label{eq:scaleWW}
Q^2={1\over 2}\left (2\MW^2+\PT^2(l)+\PT^2(l^\prime )+
\PT^2(\nu\nu^\prime )\right ) .
\end{equation}
For final states that allow for two different sets of reconstructed
gauge bosons, we choose the average of the corresponding scales from 
\refeq{eq:scaleWZ}--\refeq{eq:scaleWW} if both reconstructed sets pass
the cuts.
This scale choice appears to be appropriate for the calculation of
differential cross sections, in particular for vector-boson
transverse-momentum distributions. It generalizes the scale of
\citeres{Frixione:1992pj,Dixon:1999di} to final states with identical
particles.

We have, moreover, implemented a general set of cuts, proper for LHC 
analyses, defined as follows:
\begin{itemize}
\item {charged lepton transverse momentum $\PT(l)>20\GeV$},
  
\item {missing transverse momentum $\PTmiss> 20\GeV$
for final states with one neutrino and  $\PTmiss> 25\GeV$
for final states with two neutrinos},
  
\item {charged lepton pseudo-rapidity $|\eta_l |< 3$}, where
  $\eta_l=-\log\left (\tan(\theta_l/2)\right )$, and $\theta_l$ is the
  polar angle of particle $l$ with respect to the beam.
\end{itemize}
These cuts approximately simulate the detector acceptance.  In
addition to the above-mentioned cuts, we impose requirements on the
separation of the charged lepton and the photon.  We consider the
following photon recombination procedure:
\begin{itemize}
\item {Photons with a rapidity $|\eta_{\gamma} |> 3$ are treated as
invisible}.

\item {If the photon is central enough ($|\eta_{\gamma} |< 3$) and 
the rapidity--azimuthal-angle separation between charged 
lepton and photon $\Delta R_{l\gamma}=
\sqrt{(\eta_l-\eta_\gamma )^2+(\phi_l -\phi_\ga)^2}<0.1$, then the photon 
and lepton momentum four-vectors are combined into an effective lepton
momentum}.

\item {If the photon is central enough ($|\eta_{\gamma} |< 3$), 
the rapidity--azimuthal-angle separation $\Delta R_{l\gamma}>0.1$,
and the photon energy $E_\gamma < 2\GeV$, then the momenta of the photon 
and of the nearest charged lepton are recombined}.

\end{itemize}
The effective lepton momentum must pass the acceptance cuts given
above, and we use effective lepton momenta to define the above
mentioned factorization scales.
 
For the processes considered, we have also implemented further 
cuts which are described in due time. In the following sections, we present 
results for the LHC at $\CM$ energy $\sqrt s=14\TeV$ and an integrated 
luminosity $L=100\fba^{-1}$.

We have performed several consistency checks of our program.  We
showed that the cross sections do neither depend on the photon mass
$\lambda$ nor on the fermion masses $m_f$, which have been used as
regulators. The independence of the phase-space-slicing parameters
$\delta_s$ and $\delta_c$ has been checked as discussed in
\refse{sec:phase-space-slicing}.  Moreover, all results for WZ
production have been verified by a second independent program at the
level below $1\%$.

\section{Phenomenological results}
\label{sec:results}

In the following, we discuss the phenomenological implications of the
$\Oa$ EW corrections to vector-boson pair production at the LHC. We
examine the impact of EW corrections on observables popularly employed
to study anomalous gauge-boson couplings and vector-boson scattering.
Systematic studies of the effect of anomalous couplings on the
production of gauge-boson pairs have pointed out that in general
deviations from the SM predictions should be particularly enhanced
when gauge bosons are produced at high energies and large scattering
angles in the di-boson rest frame \cite{Haywood:1999qg}.  The same
kinematical region is also proper to search for the scattering of
strongly interacting vector bosons.  On the other hand, EW corrections
are expected to be maximally pronounced in precisely these same
regions. It is therefore interesting to discuss their effect in the
aforesaid kinematical configuration.

In order to illustrate the behaviour and the size of the $\Oa$
contributions, we analyse different distributions according to the
chosen final states.  We work under the setup described in
\refse{sec:processes}.

\subsection{$\PW\PZ$ production}
\label{sec:WZ}

First, we study the leptonic processes 
$\Pp\Pp\to l\nu_l l^\prime\bar{l^\prime}$ with $l,l^\prime = \Pe$ or
$\mu$.  These final states can be mediated by $\PW\PZ$ production and
allow to test the trilinear $\PW\PW\PZ$ coupling.

For this case we have chosen to investigate four distributions, two
momentum distributions:
\begin{description}
\item[\qquad$\PTmax(l)$:] maximal transverse momentum of the three
  charged leptons,

\item[\qquad$\PTmiss$:] missing transverse momentum,
\end{description}
and two angular distributions:
\begin{description}
\item[\qquad$\Delta y(\PZ l)= y(\PZ) - y(l)$:] difference of the
  rapidities of the reconstructed $\PZ$~boson and the charged lepton
  from $\PW$-boson decay,
  
\item[\qquad$y(l^- )$:] rapidity of the negatively charged lepton
  coming from the reconstructed $\PZ$~boson.
\end{description}
The rapidity is defined from the energy $E$ and the longitudinal
momentum $P_{\rL}$ by $y=0.5\log ((E+P_\rL)/(E-P_\rL))$.

In addition to the standard cuts and the photon recombination recipe given
in \refse{sec:processes}, we reconstruct the $\PZ$ boson by imposing
\beq\label{addcutsZ}
|M(l^\prime\bar{l^\prime} )-\MZ|< 20\GeV
\end{equation}
if $l\ne l'$.  If identical particles are present in the final state,
\ie $l=l'$, we choose as reconstructed $\PZ$~boson the lepton pair
with invariant mass closer to $\MZ$.  Since the neutrino produced in
the leptonic $\PW$-boson decay cannot be directly detected, the
longitudinal component of its momentum is not measurable. Therefore,
there is not enough information to reconstruct the invariant mass of
the $\PW$ boson.  Instead, the transverse mass defined as
$\MT(l\nu_l)=\sqrt{\ET^2(l\nu_l)-\PT^2(l\nu_l)}$ is a physical
quantity and can be restricted in order to isolate the doubly-resonant
signal over the irreducible background. In the following, we require
\beq\label{addcutsW}
\MT(l\nu_l)< \MW +20\GeV.
\end{equation}
We note that under these kinematical cuts the exact result is well
approximated by the DPA. The difference, which is about $15\%$ without
gauge-boson reconstruction, goes down to per-cent level (for a
detailed discussion see \citere{Accomando:2001fn}). Thus, we can
safely adopt the DPA for calculating EW radiative corrections.

As an illustration of the role played by $\Oa$ corrections, we
study the above-mentioned distributions in two different kinematical
regions both characterized by large energies and scattering angles in
the di-boson rest frame. As a first scenario, we restrict the
transverse momentum of the reconstructed $\PZ$ boson, $\PT(\PZ)$, by
\begin{equation}\label{eq:WZscenarioI}
\PT(\PZ)> 300\GeV.
\end{equation}
As a second scenario, we impose cuts on the invariant mass of the
three charged leptons, $\Minv(ll^\prime\bar{l^\prime})$, and the
difference $\Delta y(\PZ l)=y(\PZ)-y(l)$ between the rapidity of the 
reconstructed $\PZ$ boson and of the charged lepton coming from the
$\PW$-boson decay:
\begin{equation}\label{eq:WZscenarioII}
\Minv(ll^\prime\bar{l^\prime})> 500\GeV,\qquad 
|\Delta y(\PZ l)|< 3.
\end{equation}
Under these cuts, all invariants $\hat{s}, {\tparton}, {\uparton}$ are
large compared with the $\PW$-boson mass.  In particular,
$\sqrt{\hat{s}} \gsim 600\GeV$ and $\sqrt{{\tparton}},
\sqrt{{\uparton}} \gsim 200\GeV$ for most events and $\sqrt{\hat{s}} >
500\GeV$ and $\sqrt{{\tparton}}, \sqrt{{\uparton}} > 100\GeV$ for all.%
Thus, the conditions \refeq{HEA}, under which the logarithmic
high-energy approximation is valid, are fulfilled in these two
kinematical regions.

\begin{figure}
  \unitlength 1cm
  \begin{center}
  \begin{picture}(16.,15.)
  \put(0,7){\epsfig{file=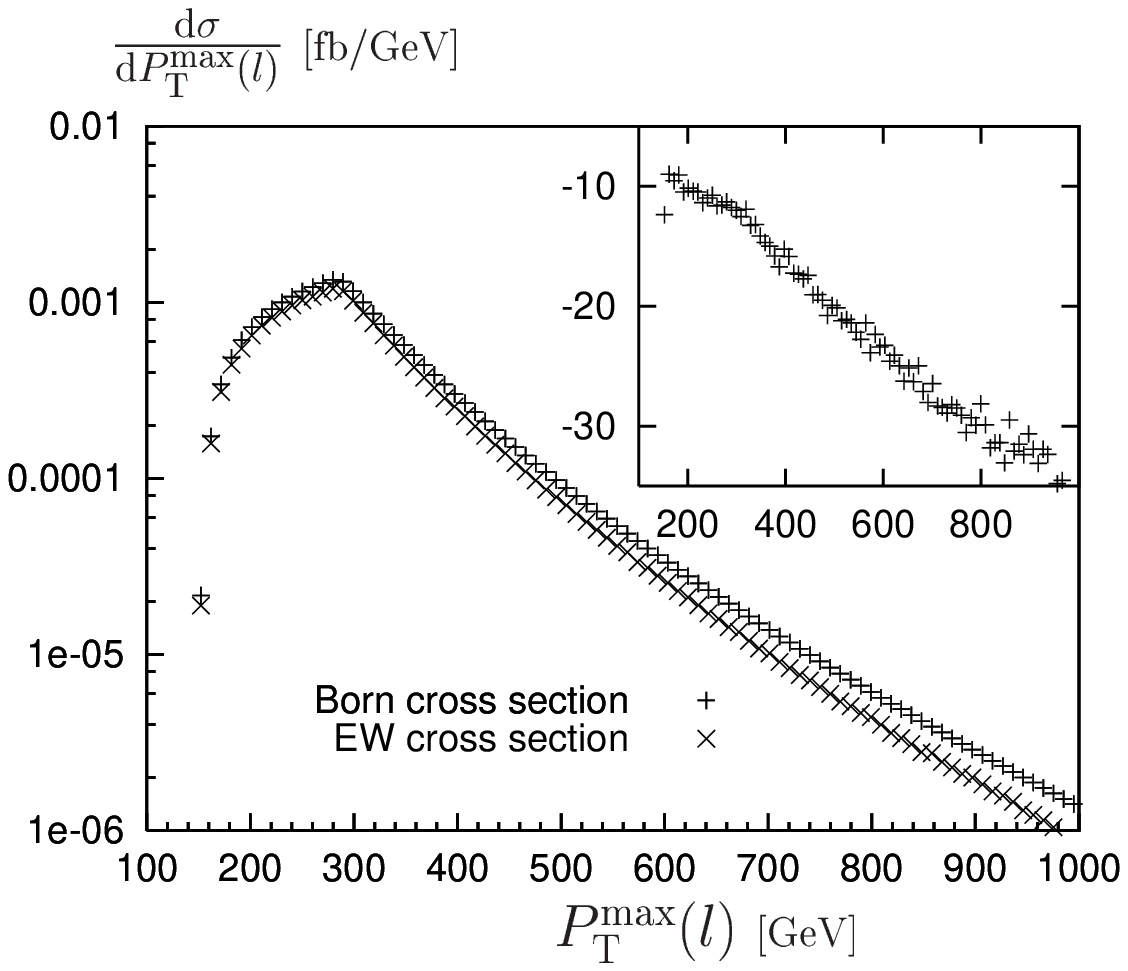,width=8cm}}
  \put(8,7){\epsfig{file=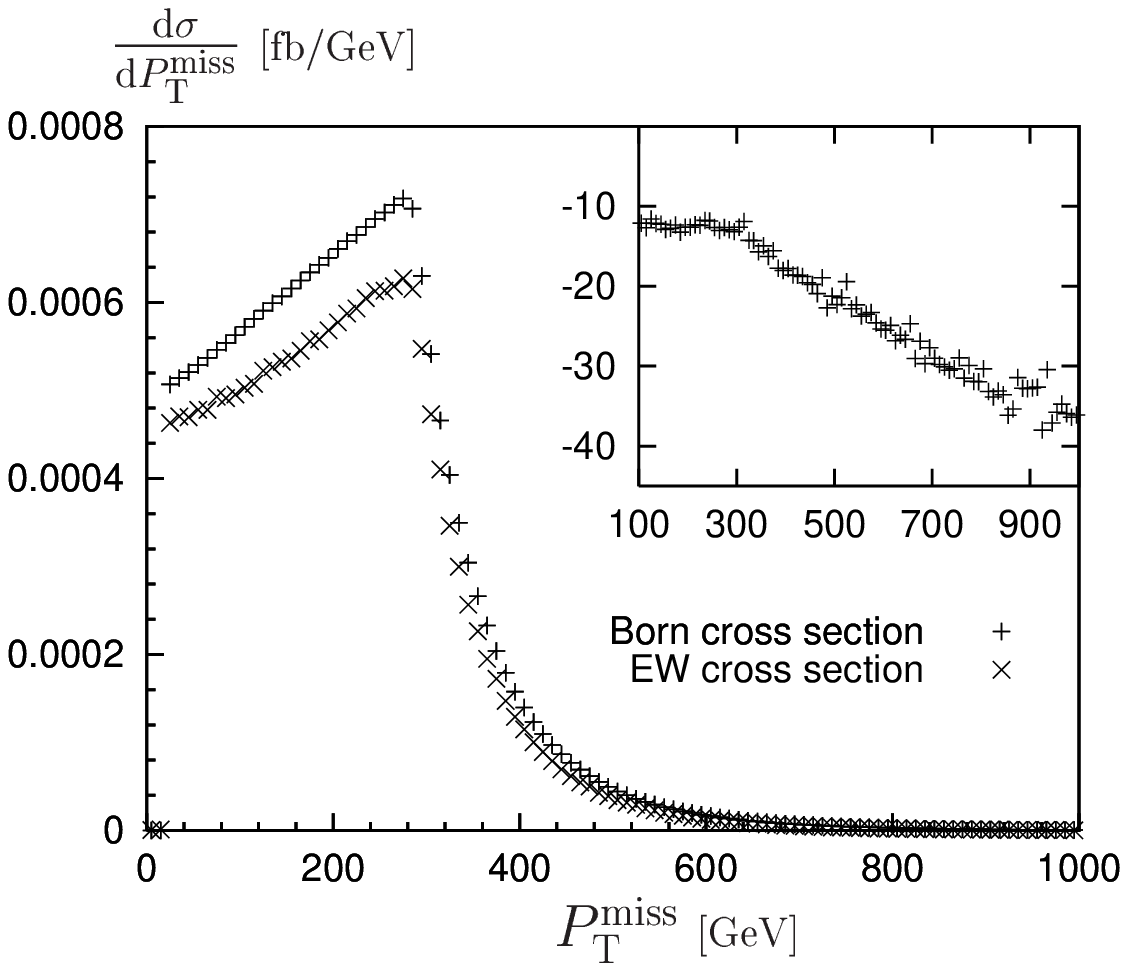,width=8cm}}
  \put(0,0){\epsfig{file=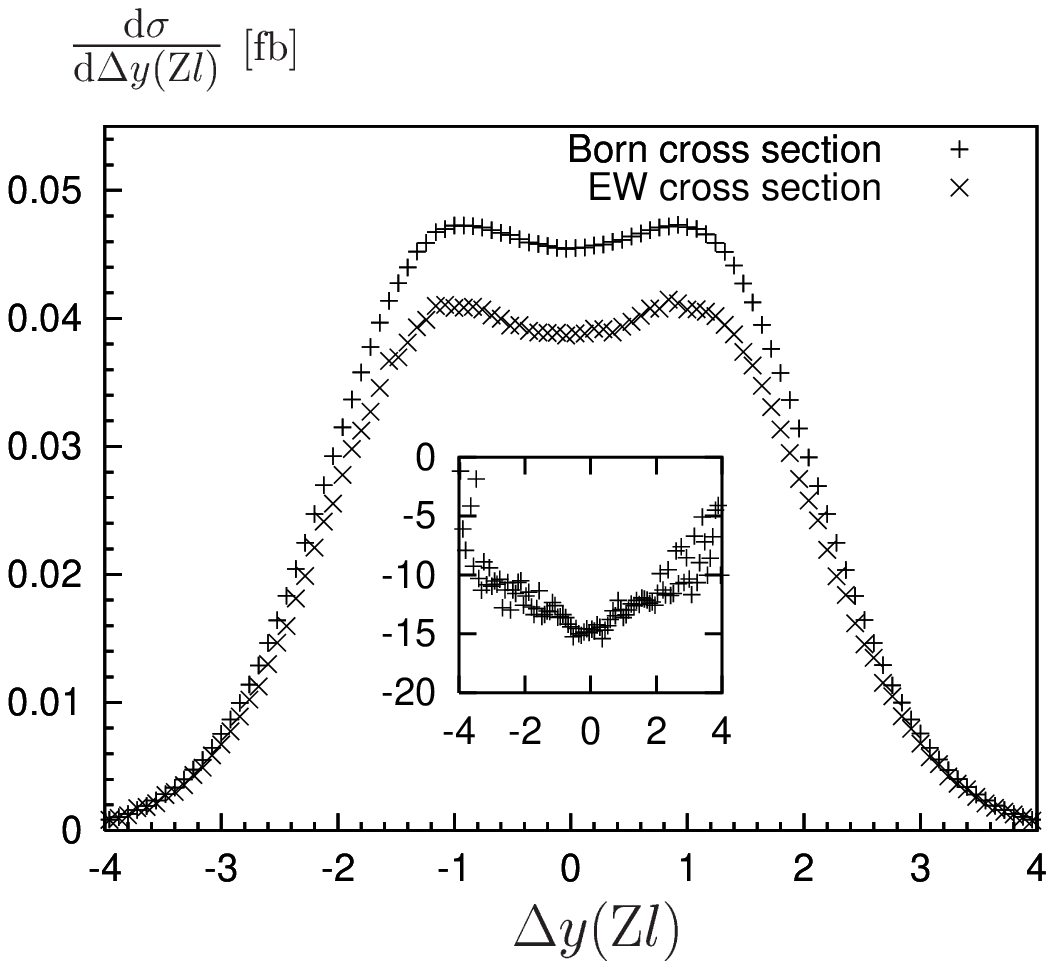,width=8cm}}
  \put(8,0){\epsfig{file=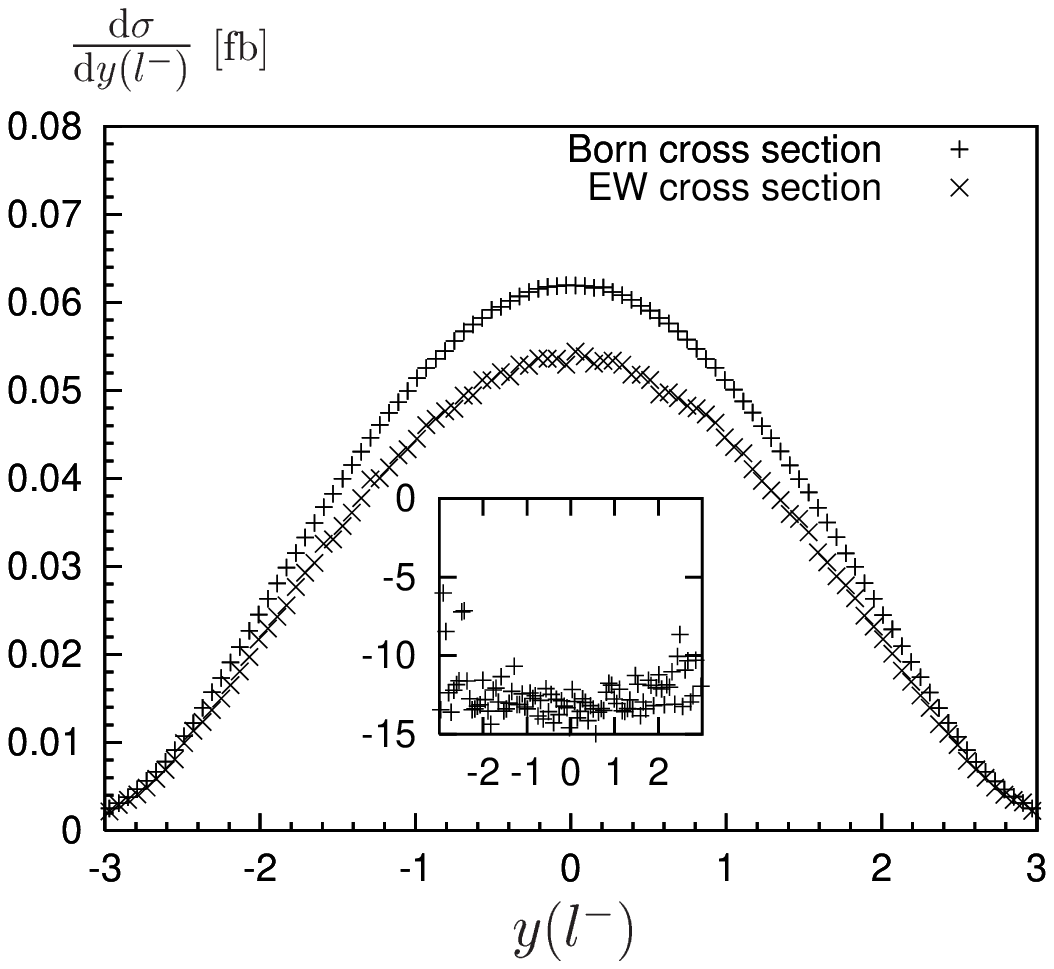,width=8cm}}
  \end{picture}
  \end{center}
\caption{Distributions for $\PW\PZ$ production:
  (a) Maximal transverse momentum of the charged leptons, (b) Missing
  transverse momentum.  (c) Difference in rapidity between the
  reconstructed $\PZ$ boson and the charged lepton coming from the
  $\PW$-boson decay.  (d) Rapidity of the $\mu^-$.  The contributions
  of the final states $\Pe^-\Pnebar\mu^-\mu^+$ and
  $\Pne\Pe^+\mu^-\mu^+$ are summed, and standard cuts as well as $\PT(\PZ)>
  300\GeV$ are applied.  The inset plots show the $\Oa$
  corrections relative to the Born results in per cent.}
\label{fi:WZ_s1}
\end{figure}
We start discussing the scenario \refeq{eq:WZscenarioI}.  In
\reffi{fi:WZ_s1} we have plotted the four distributions for the
complete process $\Pp\Pp\to \Pe^-\Pnebar\mu^-\mu^+,
\Pne\Pe^+\mu^-\mu^+ $, \ie we sum over the two charge-conjugate final
states.  As a general feature the EW corrections are negative and
lower the Born cross section by more than 10$\%$. For the individual
distributions we observe the following.  EW corrections reduce the
distribution in $\PTmax(l)$ by the order of 10$\%$ at low to modest
$\PTmax(l)$ values. This effect grows with increasing $\PTmax(l)$ as
shown by the long tail where the contribution of EW corrections can
amount to more than $30\%$.  This is of course the result of enhanced
EW logarithms at large energies, which are enforced by the large
$\PTmax(l)$.  The missing-transverse-momentum distribution shows the
same qualitative behaviour. At low values the correction amounts to
about $-12\%$, while at high $\PTmiss$ it increases up to $-40\%$.  As
stated in the literature, the large $\PT$ region is an ideal place to
look for new physics. As an example, the $\PT(\PZ)$ distribution has
been found to be much more sensitive to new-physics effects than the
$\PW\PZ$ invariant-mass distribution, which in principle should give a
more direct access to the energy scale \cite{Baur:1995aj}.  This
feature is shared by $\PTmax(l)$ and $\PTmiss$ we just discussed.

As to angular distributions, EW corrections are maximal at low
rapidity values in both cases, where once again effects due to new
physics could be more enhanced. A low rapidity corresponds in fact to
large scattering angles of the produced vector bosons in their rest
frame. As shown in the lower left plot of \reffi{fi:WZ_s1}, the
distribution in the rapidity difference $\Delta y(\PZ l)$ exhibits a
characteristic dip, relic of an approximate radiation zero at high
energy \cite{Baur:1994prl}.  Of course, new physics could have
observable consequences on the shape of this variable
\cite{Baur:1995aj}. The general tendency is to fill in the dip, but in
certain models the approximate zero may even become more pronounced.
It is thus important to consider the impact of radiative corrections
to this relevant signal.  In the last decade, the effect of NLO QCD
corrections has been extensively analysed
\cite{Baur:1995aj,Dixon:1999di}. It can completely spoil the
significance of the dip, if one measures the inclusive $\PW\PZ+X$
production.  By imposing a jet veto, the QCD corrections get
drastically reduced to about 20$\%$ of the Born result, at the same
time diminuishing the dependence of the NLO cross section on the
factorization scale.  As shown in the third plot of \reffi{fi:WZ_s1},
EW corrections can be of the same order as QCD effects but with
opposite sign. So, they slightly increase the dip.

\begin{figure}
  \unitlength 1cm
  \begin{center}
  \begin{picture}(16.,15.)
  \put(0,7){\epsfig{file=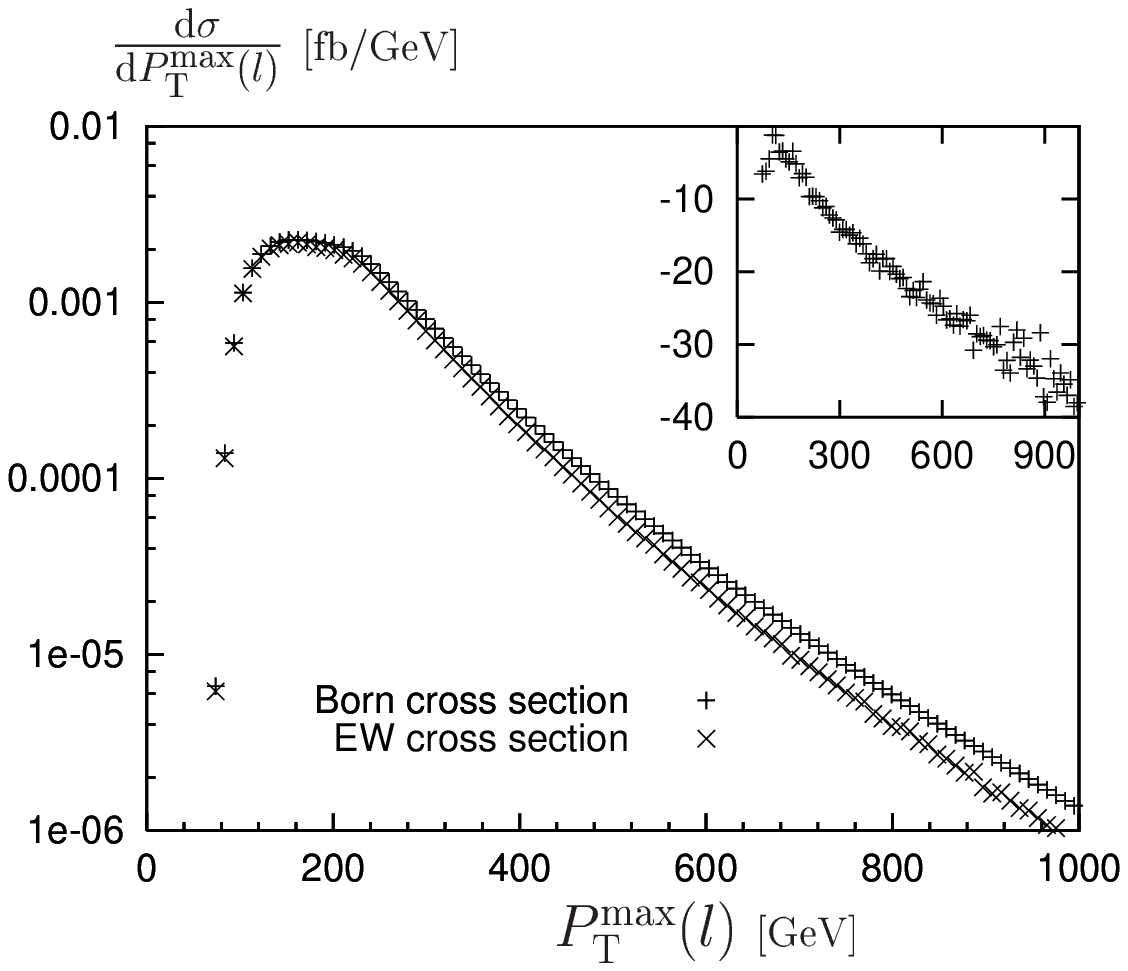,width=8cm}}
  \put(8,7){\epsfig{file=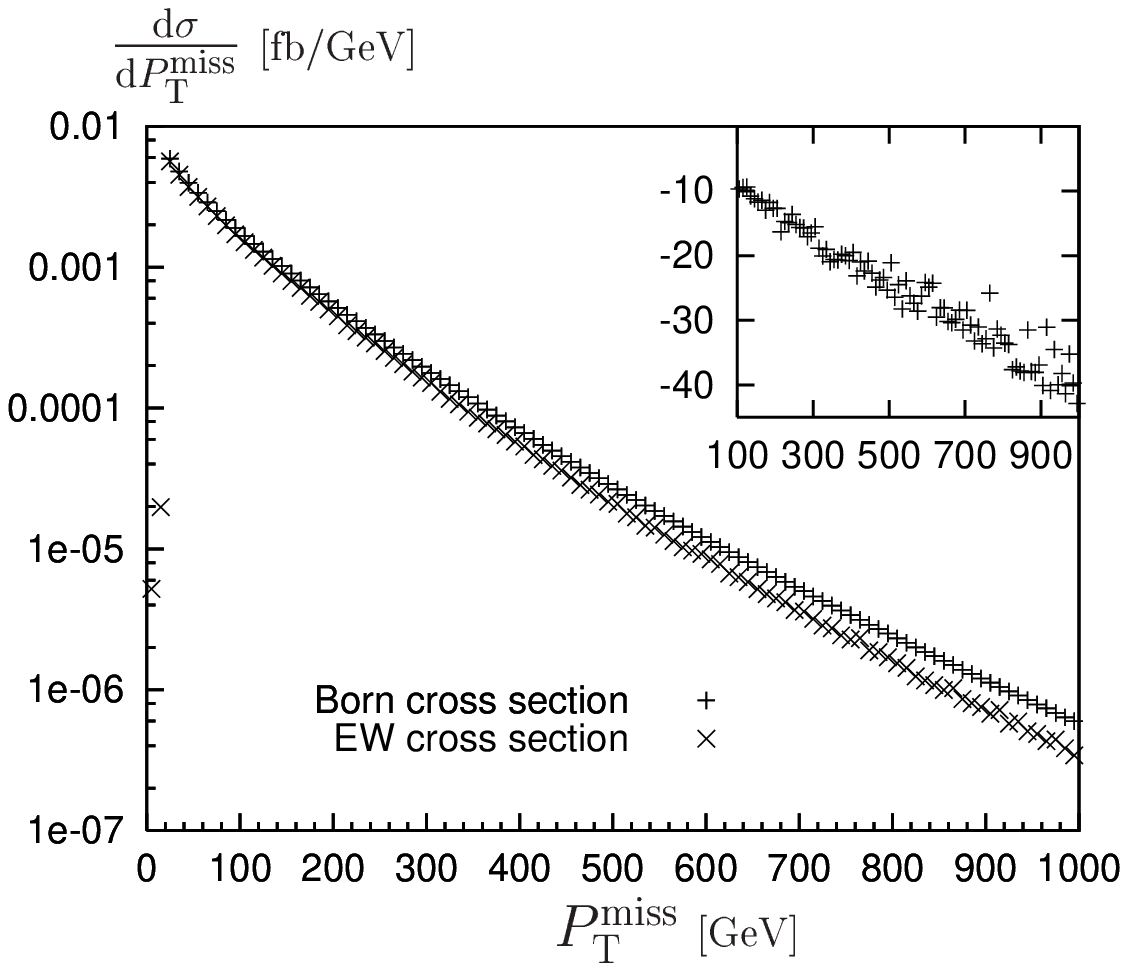,width=8cm}}
  \put(0,0){\epsfig{file=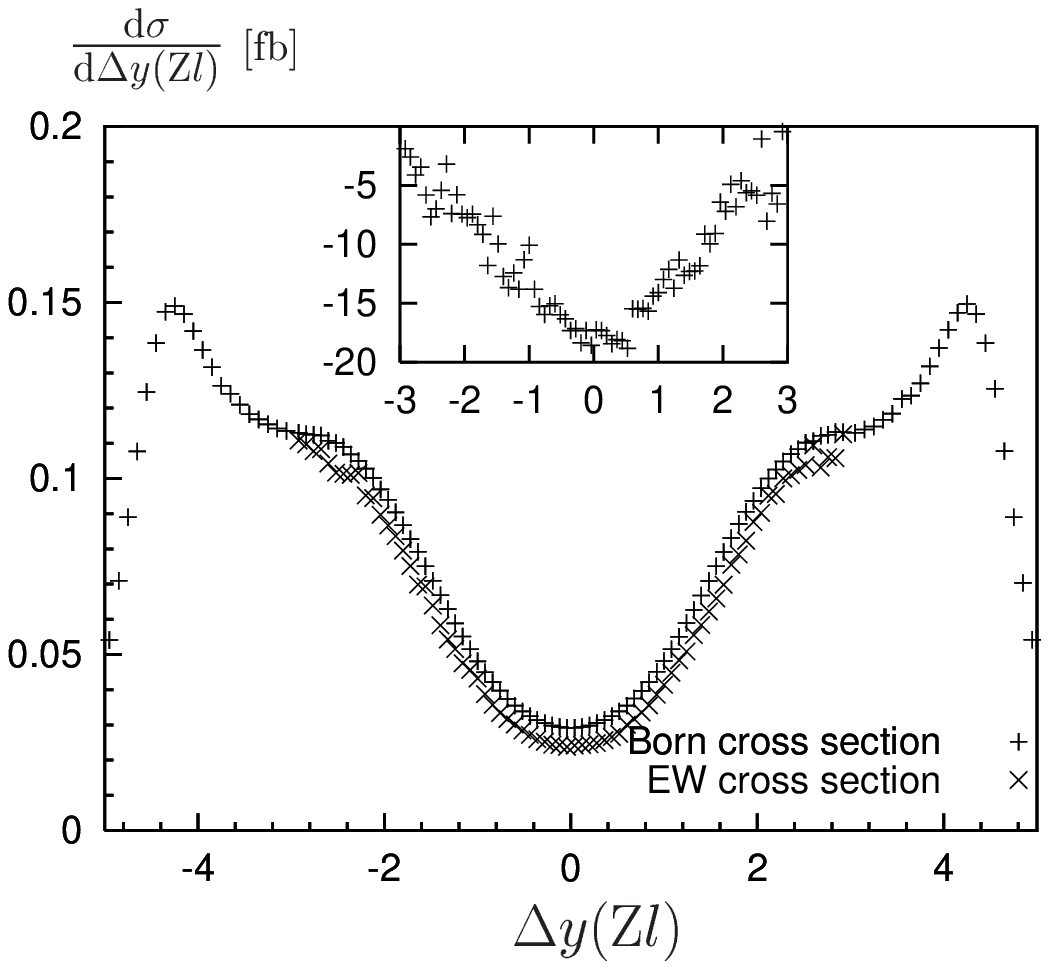,width=8cm}}
  \put(8,0){\epsfig{file=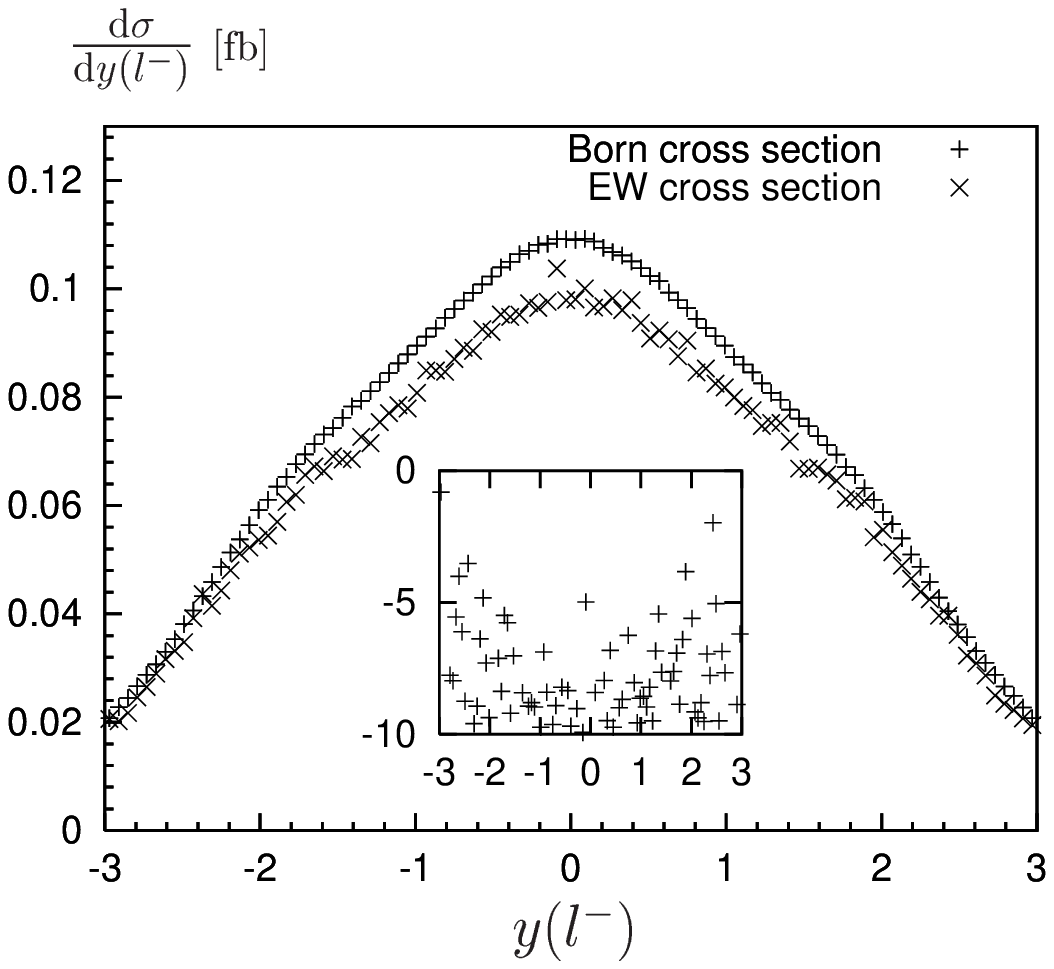,width=8cm}}
  \end{picture}
  \end{center}
\caption{Distributions for $\PW\PZ$ production:
  (a) Maximal transverse momentum of the charged leptons. (b) Missing
  transverse momentum.  (c) Difference in rapidity between the
  reconstructed $\PZ$ boson and the charged lepton coming from the
  $\PW$-boson decay.  (d) Rapidity of the $\mu^-$.  The contributions
  of the final states $\Pe^-\nu_\Pe\mu^-\mu^+$ and $\Pe^+\bar
  \nu_\Pe\mu^-\mu^+$ are summed, and standard cuts as well as
  $\Minv(ll^\prime\bar{l^\prime})> 500\GeV$ and $|\Delta y(\PZ l)|<
  3$ are applied. The last cut is omitted for the $\De y(\PZ l)$
  distribution in lowest order. The inset plots show the $\Oa$
  corrections relative to the Born results in per cent.}
\label{fi:WZ_s2}
\end{figure}
Of course, radiative corrections do not only depend on the considered
distribution but also on the selected cuts.  Figure~\ref{fi:WZ_s2}
shows a second set of plots for the same set of distributions as above
but in the scenario \refeq{eq:WZscenarioII}.  The influence of the
radiative corrections on the two 
momentum-like variables is analogous to the one observed in the
previous case.  The main difference between the two selected
kinematical regions is in the shape of the $\Delta y(\PZ l)$
distribution.  Here, the radiation-zero dip strongly increases. This
is due to the fact that the requirement
$\Minv(ll^\prime\bar{l^\prime})> 500\GeV$ forces the reconstructed
$\PZ$ boson and the charged lepton from the $\PW$-boson decay to be
produced at large separation angle. This effect translates into a
depletion of events in the central region of low rapidity difference.
Radiative corrections are more pronounced in this suppressed region.

To measure the significance of the EW corrections, a naive but direct
way is to compare their magnitude with the expected statistical error.
In \refta{ta:WZ_s1} we have listed the relative deviation $\Delta$ and
the statistical accuracy, estimated by taking as a luminosity
$L=100\fba^{-1}$ for two experiments, in the scenario
\refeq{eq:WZscenarioI} for some values of the cut on the transverse
momentum of the reconstructed $\PZ$ boson. To this purpose, we sum
over all eight final states $\Pem\Pnebar\mu^-\mu^+$,
$\Pne\Pep\mu^-\mu^+$, $\mu^-\bar\nu_\mu\Pem\Pep$,
$\nu_\mu\mu^+\Pem\Pep$, $\mu^-\bar\nu_\mu\mu^-\mu^+$,
$\nu_\mu\mu^+\mu^-\mu^+$, $\Pem\Pnebar\Pem\Pep$, and
$\Pne\Pep\Pem\Pep$.  In \refta{ta:WZ_s2}, we give the same entries but
for the scenario \refeq{eq:WZscenarioII} and for different values of
the cut on the invariant mass of the three charged leptons.
\begin{table}\centering
$$
\begin{array}{|c|c|c|c|c|c|c|}
\hline 
\multicolumn{7}{|c|}{\Pp\Pp\to
  l\nu_ll^\prime\bar{l^\prime}\rule[-2ex]{0ex}{5ex}}\\
\hline
\PTcut(\PZ)~[\mathrm{GeV}] & 
\sigma_{\Born}~[\mathrm{fb}] & \sigma_{\AEWS}~[\mathrm{fb}] &
\sigma_{\virt}^{\finite}~[\mathrm{fb}] & 
~\sigma_{\EW} ~[\mathrm{fb}]~ &~\Delta~[\%]~~ & 1/\sqrt{2L\sigma_{\Born}}~[\%] \\
\hline
\hline
250 & 1.672 & 1.563 & 1.553 & 1.489 & -10.9 & 5.5  \\ \hline
300 & 0.876 & 0.794 & 0.789 & 0.761 & -13.1 & 7.6  \\ \hline
350 & 0.489 & 0.431 & 0.428 & 0.413 & -15.5 & 10.1 \\ \hline
400 & 0.287 & 0.246 & 0.244 & 0.236 & -17.8 & 13.2 \\ \hline
450 & 0.175 & 0.146 & 0.145 & 0.141 & -19.7 & 16.9 \\ \hline
500 & 0.111 & 0.090 & 0.089 & 0.087 & -21.2 & 21.2 \\ \hline
\end{array}$$
\caption {Cross section for $\Pp\Pp\to l\nu_l l^\prime \bar{l^\prime}$ 
for various values of $\PTcut(\PZ)$. Here we
have summed over all eight final states with $l,l'=\Pe$ or $\mu$.} 
\label{ta:WZ_s1}
\end{table}%
\begin{table}\centering
$$
\begin{array}{|c|c|c|c|c|c|c|}
\hline 
\multicolumn{7}{|c|}{\Pp\Pp\to
  l\nu_ll^\prime\bar{l^\prime}\rule[-2ex]{0ex}{5ex}}\\
\hline
\Minvcut(ll^\prime\bar{l^\prime})~[\mathrm{GeV}] & 
\sigma_{\Born}~[\mathrm{fb}] & \sigma_{\AEWS}~[\mathrm{fb}] &
\sigma_{\virt}^{\finite}~[\mathrm{fb}] & 
~\sigma_{\EW} ~[\mathrm{fb}]~ &~\Delta~[\%]~~ & 1/\sqrt{2L\sigma_{\Born}}~[\%] \\
\hline
\hline
500 & 1.729 & 1.689 & 1.692 & 1.601 & -7.4 & 5.4 \\
\hline
600 & 0.899 & 0.858 & 0.860 & 0.814  & -9.5  & 7.5 \\
\hline
700 & 0.508 & 0.474 & 0.476 & 0.452 & -10.9 & 9.9 \\
\hline
800 & 0.304 & 0.278 & 0.279 & 0.264 & -13.3 & 12.8 \\
\hline
900 & 0.190 & 0.170 & 0.171 & 0.161 & -15.1 & 16.2 \\
\hline
1000 & 0.123 & 0.108 & 0.109 & 0.102 & -16.7 & 20.2 \\
\hline
\end{array}$$
\caption {Cross section for $\Pp\Pp\to l\nu_l l^\prime \bar{l^\prime}$ 
for $|\Delta y(\PZ l)|< 3$ and various values of
$\Minvcut(ll^\prime\bar{l^\prime})$. Here we have summed over all
eight final states with $l,l'=\Pe$ or $\mu$.}  
\label{ta:WZ_s2}
\end{table}%
The integration errors in these and the following tables are at the
level below 1\%.
The comparison of the expected statistical error with the EW
corrections indicates that these are non-negligible and
can be comparable with the experimental precision up to about
$\PTcut(\PZ)=500\GeV$ or $\Minvcut(ll^\prime\bar{l^\prime})=1\TeV$. In
these regions the corrections range between $-7$ and $-22\%$, being
slightly more enhanced in the first scenario.  Of course, their
significance depends on the available luminosity.  This kind of
accuracy is needed only in a high-luminosity run.

Besides the lowest-order cross section $\sigma_{\Born}$ and the cross
section $\sigma_{\EW}$ including the complete logarithmic EW
corrections, we have also inserted two entries representing partial
results in \reftas{ta:WZ_s1} and \ref{ta:WZ_s2} in order to give an
idea of the individual contributions.  The cross section including
only the EW logarithms originating from above the EW scale,
$\MW$, is denoted by $\sigma_{\AEWS}$. This term neglects all
IR- and mass-singular terms coming from the mass gap between the
photon and the weak gauge bosons and is exactly the part computed in
\citere{Accomando:2001fn} for the same process.  The column
$\sigma_{\virt}^{\finite}$ contains instead the full finite virtual
correction, \ie the full logarithmic EW corrections with the
IR- and mass-singular contribution \refeq{eq:virt-sing} subtracted.
The difference between $\sigma_{\AEWS}$ and $\sigma_{\virt}^{\finite}$
is numerically small, despite of the fact that it contains
logarithmic contributions.  The dominant contribution to this
difference is in fact proportional to
$\al/(2\pi)\log(s/M^2)[\log(s/M^2)-3]$ which is suppressed for
energies between $500\GeV$ and $1\TeV$ owing to cancellations in the
bracket.

\subsection{$\PZ\PZ$ production}
\label{sec:ZZ}

In this section we extend our analysis to the processes $\Pp\Pp\to
l\bar{l}l^\prime\bar{l^\prime}$ ($l,l'=\Pe$ or $\mu$). This channel is
proper for studying the impact of trilinear neutral gauge-boson
vertices, $\PZ\PZ\PZ$ and $\PZ\PZ\gamma$, on physical observables.
While these couplings are absent in the SM Lagrangian, one-loop
corrections induce small but not-vanishing values for them.
Significantly larger couplings are predicted by non-standard models,
where new physics appearing at energy scales much larger than those
which can be directly probed at forthcoming experiments can be
parametrized in terms of anomalous neutral self-interactions.
At LEP2 and Tevatron, the $\PZ\PZ\gamma$ vertex has been measured
through $\PZ\gamma$ production. LEP2 has been able to produce also
$\PZ\PZ$ pairs but with poor statistics.  At the LHC several
thousands of such $\PZ\PZ$ pairs will be produced, allowing for more
stringent bounds on $\PZ\PZ\PZ$ and $\PZ\PZ\gamma$ vertices.  The
envisioned increase in statistics, and the possibility to observe
significant deviations due to new physics interactions have gathered a
renewed interest in the literature \cite{NAC}.

$\PZ\PZ\PZ$ and $\PZ\PZ\gamma$ couplings affect the production of
longitudinal or transverse $\PZ$ bosons in a different way. Therefore,
the helicity of the decay products coming from $\PZ\PZ$ production
constitutes a valuable information.  Up to now, on one side the
aforementioned studies have been performed in the {\it
  production$\times$decay} approach, neglecting all spin correlations
and irreducible background contributions.  On the other side, accurate
calculations of QCD corrections have been carried out in
\citere{Dixon:1999di}.  In this section, we illustrate the results of
a complete calculation of four-fermion production mediated by
$\PZ\PZ$ production including logarithmic EW corrections. We
focus, in particular, on the effect of the EW corrections on
the distributions mostly discussed in the literature \cite{NACbaur}.

We consider the same kind of observables as in the previous section,
with the only difference that we replace the distribution in
the missing transverse momentum by the distribution in the
maximal transverse momentum of the reconstructed $\PZ$ bosons. To
be precise we plot distributions in:
\begin{description}
\item[\qquad$\PTmax(l)$:] maximal transverse momentum of the four
  charged leptons,
  
\item[\qquad$\PT(\PZ)$:] maximal transverse momentum of the two
  reconstructed $\PZ$ bosons,

\item[\qquad$\Delta y(\PZ\PZ)$:] rapidity difference between the two
  reconstructed $\PZ$ bosons,

\item[\qquad$y(\mu^- )$:] rapidity of the $\mu^-$.
\end{description}
The $\PZ$~bosons are reconstructed by imposing \refeq{addcutsZ}, and
for identical particles in the final state we choose the possibility
where the reconstructed \PZ~bosons are closer to their mass shell.  We
have checked that the accuracy of the DPA is at the level of a few per
cent for this case.

We show results for the specific process
$\Pp\Pp\to\Pe^-\Pe^+\mu^-\mu^+$ and only for the scenario
characterized by the requirement
\begin{equation}\label{eq:ZZscenarioII}
\Minv(l\bar{l}l^\prime\bar{l^\prime})> 500\GeV,\qquad 
|\Delta y(\PZ\PZ)|< 3.
\end{equation}
An analogous behaviour holds for the scenario with $\PT(Z)>300\GeV$
for both reconstructed Z~bosons. We have verified that for both these
scenarios the conditions \refeq{HEA} for the validity of the
logarithmic high-energy approximation are fulfilled.

\begin{figure}
  \unitlength 1cm
  \begin{center}
  \begin{picture}(16.,15.)
  \put(0,7){\epsfig{file=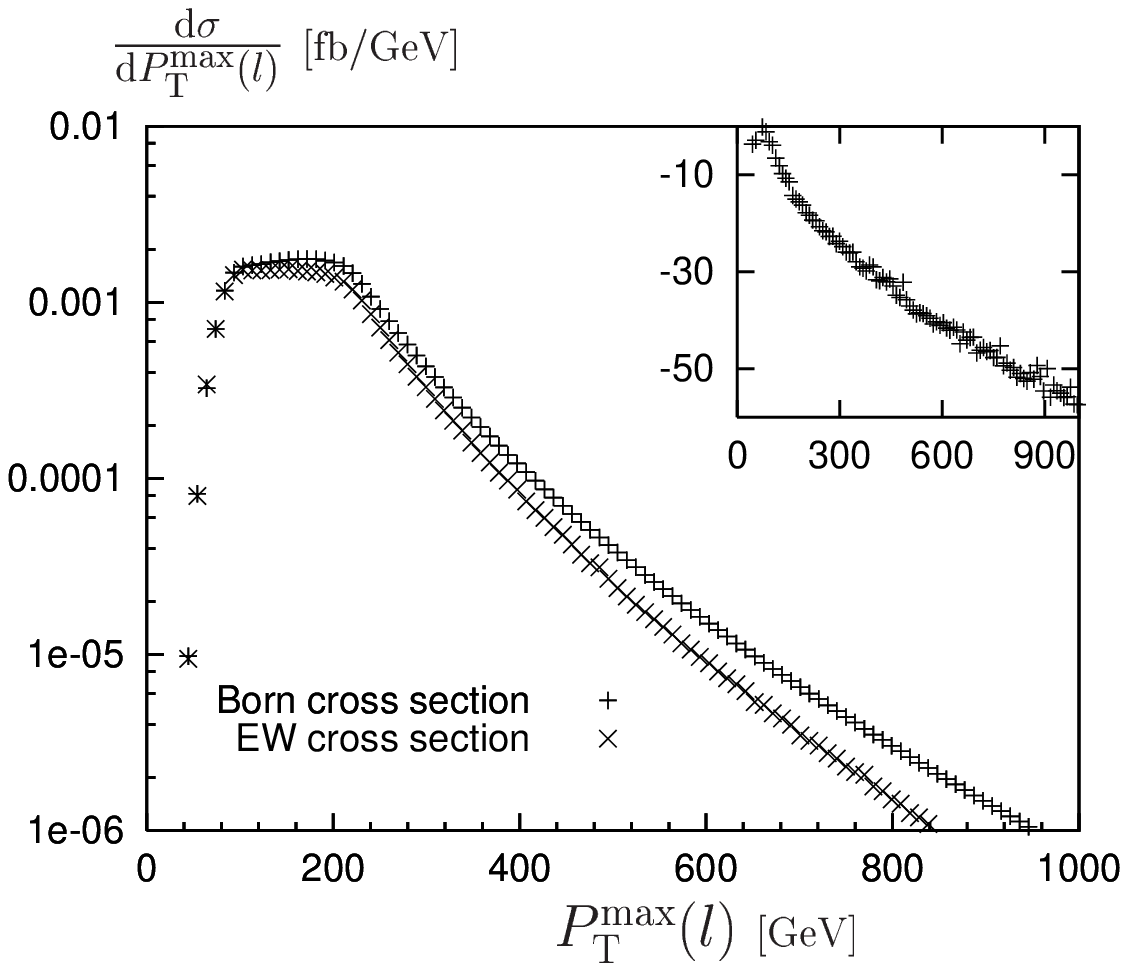,width=8cm}}
  \put(8,7){\epsfig{file=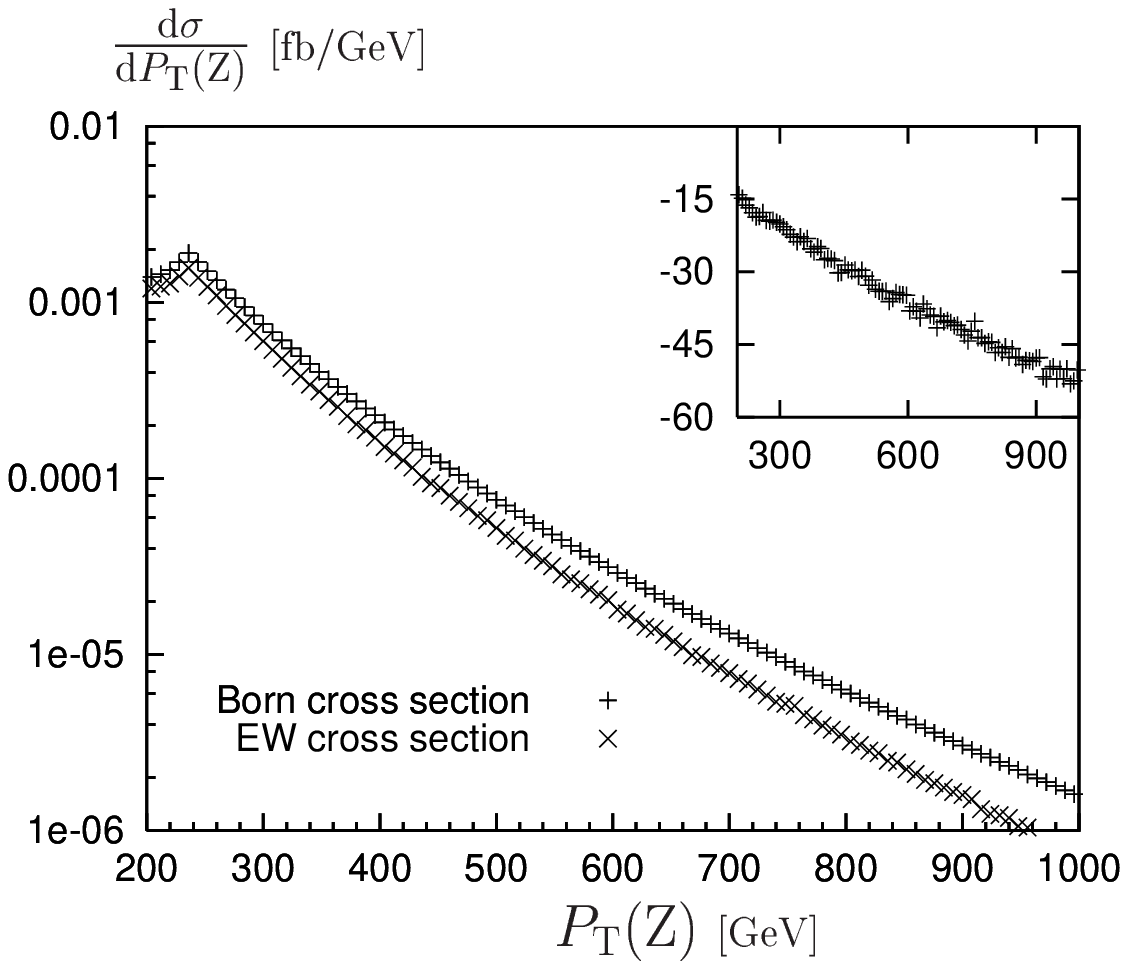,width=8cm}}
  \put(0,0){\epsfig{file=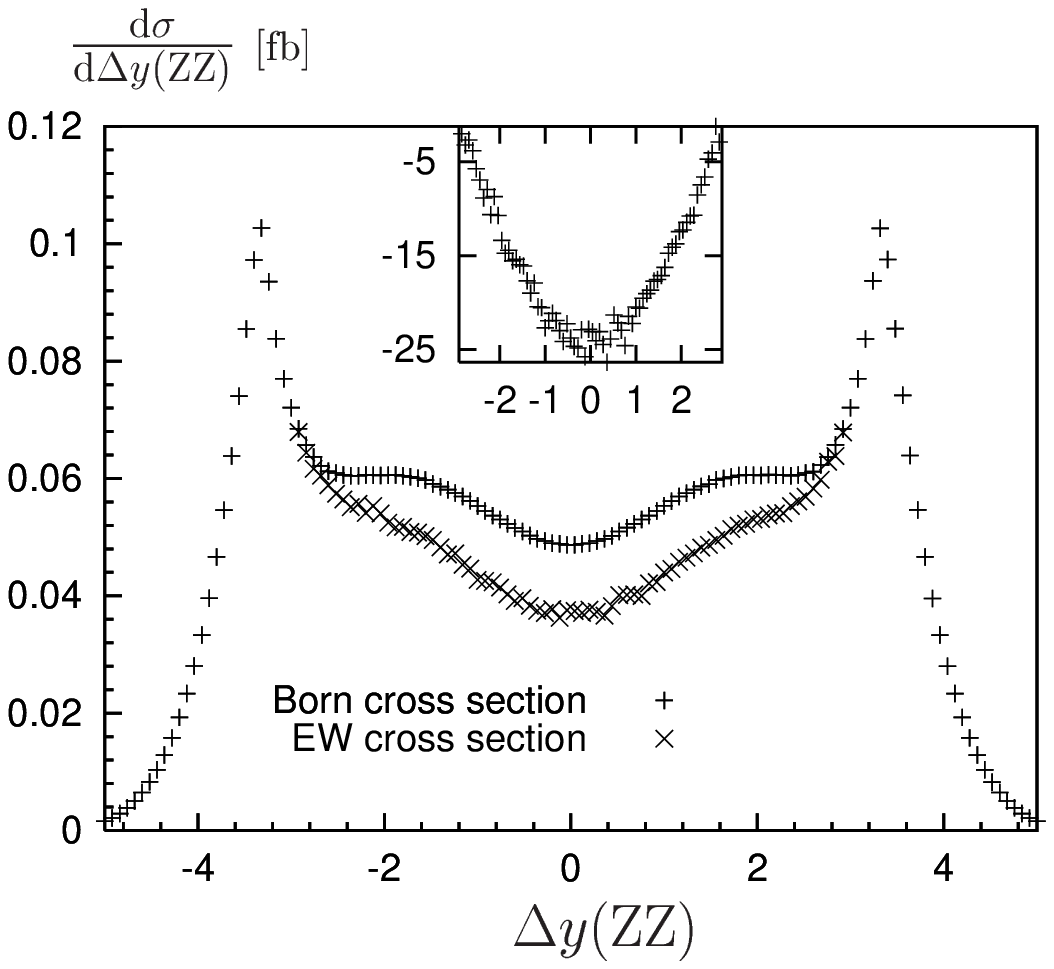,width=8cm}}
  \put(8,0){\epsfig{file=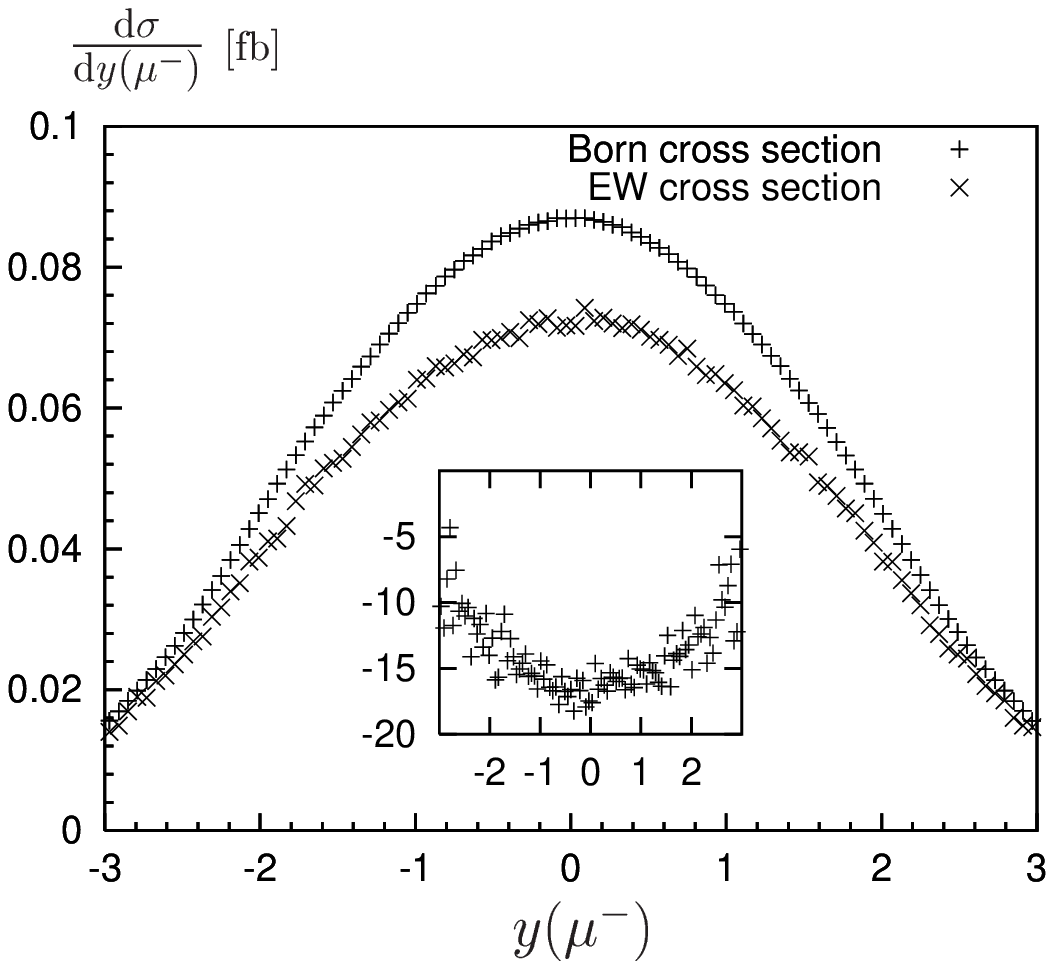,width=8cm}}
  \end{picture}
  \end{center}
\caption{Distributions for $\PZ\PZ$ production:
  (a) Maximal transverse momentum of the charged leptons. (b) Maximal
  transverse momentum of the reconstructed $\PZ$ bosons.  (c)
  Difference in rapidity between the two reconstructed $\PZ$ bosons.
  (d) Rapidity of the $\mu^-$.  The final states is
  $\Pe^-\Pep\mu^-\mu^+$, and standard cuts as well as $\Minv(l\bar
  ll^\prime\bar{l^\prime})> 500\GeV$ and $|\Delta y(\PZ\PZ)|< 3$ are
  applied.  The last cut is omitted for the $\De y(\PZ\PZ)$
  distribution in lowest order. The inset plots show the $\Oa$
  corrections relative to the Born results in per cent.}
\label{fi:ZZ_s2}
\end{figure}
As one can see in \reffi{fi:ZZ_s2}, EW corrections modify the Born
result in the same way as for $\PW\PZ$ production, but the effect is
typically a factor of 1.5 larger.  We note that $\PZ\PZ$ production at
tree level does not present any true or approximate radiation zero.
The dip in the distribution of the rapidity difference of the two
reconstructed $\PZ$ bosons results from the fact that the partonic
process $q\bar{q}\rightarrow \PZ\PZ$, dominated by the transversely
polarized $\PZ$ bosons, is peaked forward and backward. Moreover, the
dip is enhanced by the invariant mass cut in \refeq{eq:ZZscenarioII}.

In \refta{ta:ZZ_s2} we compare the relative correction $\Delta$ to the
Born cross section with the estimated experimental accuracy for some
values of the cut on the partonic CM energy
$\Minv(l\bar{l}l^\prime\bar{l^\prime})$.  To this purpose, we sum over
all three final states $\Pe^-\Pe^+\mu^-\mu^+$, $\Pe^-\Pe^+\Pe^-\Pe^+$,
and $\mu^-\mu^+\mu^-\mu^+$.  The entries in \refta{ta:ZZ_s2} are
defined as in the previous section.  One can see that, compared to
$\PW\PZ$ production, $\Oa$ corrections manifest the same behaviour on
the shown observables, but they are globally by a factor of
${\sim1.5}$ larger.  At modest $\PZ\PZ$ invariant masses, the effect
of the EW corrections can amount to two standard deviations, while it
becomes comparable to the experimental precision with increasing CM
energy. Of course, final states coming from $\PZ\PZ$ production
involving only charged leptons will not be copiously generated at the
LHC. A detailed study of their properties would be possible only
during a high-luminosity run. Although it yields higher statistics, we
have not investigated ZZ production leading to $l \bar{l}
\nu_{l^\prime} \bar{\nu}_{l^\prime}$ final states because there the
reconstruction of the Z~bosons is more problematic.
\begin{table}\centering
$$
\begin{array}{|c|c|c|c|c|c|c|}
\hline 
\multicolumn{7}{|c|}{\Pp\Pp\to
  l\bar{l}l^\prime\bar{l^\prime} \rule[-2ex]{0ex}{5ex}}\\
\hline
\Minvcut(l\bar{l}l^\prime\bar{l^\prime})~[\mathrm{GeV}] & 
\sigma_{\Born}~[\mathrm{fb}] & \sigma_{\AEWS}~[\mathrm{fb}] &
\sigma_{\virt}^{\finite}~[\mathrm{fb}] & 
~\sigma_{\EW} ~[\mathrm{fb}]~ &~\Delta~[\%]~~ & 1/\sqrt{2L\sigma_{\Born}}~[\%] \\
\hline
\hline
500 & 0.692 & 0.637 & 0.633 & 0.588 & -15.0 & 8.5 \\
\hline
600 & 0.356 & 0.314 & 0.312 & 0.291  & -18.3  & 11.9 \\
\hline
700 & 0.203 & 0.173 & 0.172 & 0.160 & -21.0 & 15.7 \\
\hline
800 & 0.123 & 0.102 & 0.101 & 0.094 & -23.8 & 20.1 \\
\hline
900 & 0.078 & 0.063 & 0.062 & 0.058 & -26.1 & 25.3 \\
\hline
1000 & 0.051 & 0.040 & 0.040 & 0.037 & -28.1 & 31.2 \\
\hline
\end{array}$$
\caption {Cross section for $\Pp\Pp\to \Pe^-\Pe^+\mu^-\mu^+$,
  $\Pe^-\Pe^+\Pe^-\Pe^+$, and $\mu^-\mu^+\mu^-\mu^+$
for $|\Delta y(\PZ\PZ)|< 3$ and various values of
$\Minvcut(l\bar{l}l^\prime\bar{l^\prime})$}  
\label{ta:ZZ_s2}
\end{table}

\subsection{$\PW\PW$ production}
\label{sec:WW}

Finally, we discuss the processes $\Pp\Pp\to
l\bar\nu_l\nu_{l^\prime}\bar{l^\prime}$ ($l,l'=\Pe$ or $\mu$).  This
channel contains information on the charged gauge-boson vertices
$\PW\PW\PZ$ and $\PW\PW\gamma$.  While LEP2 could establish the
non-abelian nature of the SM by measuring these couplings,
high-precision measurements are still missing.  At the LHC, the
precision will be sensitively improved, if the large background from
$\Pt\bar\Pt$ production can be properly controlled.  The $\PW\PW$
channel has in fact the largest cross section among all vector-boson
pair-production processes.  Despite the presence of two neutrinos,
which do not allow a clean and unambiguous reconstruction of the two
$\PW$ bosons, the sensitivity to anomalous couplings is not seriously
reduced. One can in fact consider the distribution in the missing
transverse momentum \cite{Baur:1995uv}.

For this channel, following the study of \citere{Dixon:1999di} on the
sensitivity to new-physics effects, we choose to discuss distributions
in the following variables:
\begin{description}
\item[\qquad$\PTmax(l)$:] maximal transverse momentum of the two
  charged leptons,

\item[\qquad$\PTmiss$:] missing transverse momentum,
  
\item[\qquad$\Delta y(l\bar{l^\prime})$:] rapidity difference between
  the charged leptons,

\item[\qquad$y(l^- )$:] rapidity of the negatively charged lepton.
\end{description}
Despite of the fact that we do not perform a reconstruction of the
\PW~bosons for these processes, the quality of the DPA is better than
10\%. Since we apply the DPA only to the corrections and these are
below 25\%, at least where the cross section is appreciable, this
introduces an error of only a few per cent. We consider the scenario
\begin{equation}\label{eq:WWscenarioII}
\Minv(l\bar{l^\prime})> 500\GeV,\qquad 
|\Delta y(l\bar{l^\prime})|< 3,
\end{equation}
which fulfils the conditions \refeq{HEA} for the validity of the
logarithmic high-energy approximation, as we have verified.
Possible $\PZ\PZ$ intermediate states are heavily suppressed by the
invariant-mass cut in \refeq{eq:WWscenarioII}. 
Therefore, we can safely neglect contributions of 
$l \bar{l} \nu_{l^\prime} \bar{\nu}_{l^\prime}$ final states
with $l \neq l^\prime$. 

\begin{figure}
  \unitlength 1cm
  \begin{center}
  \begin{picture}(16.,15.)
  \put(0,7){\epsfig{file=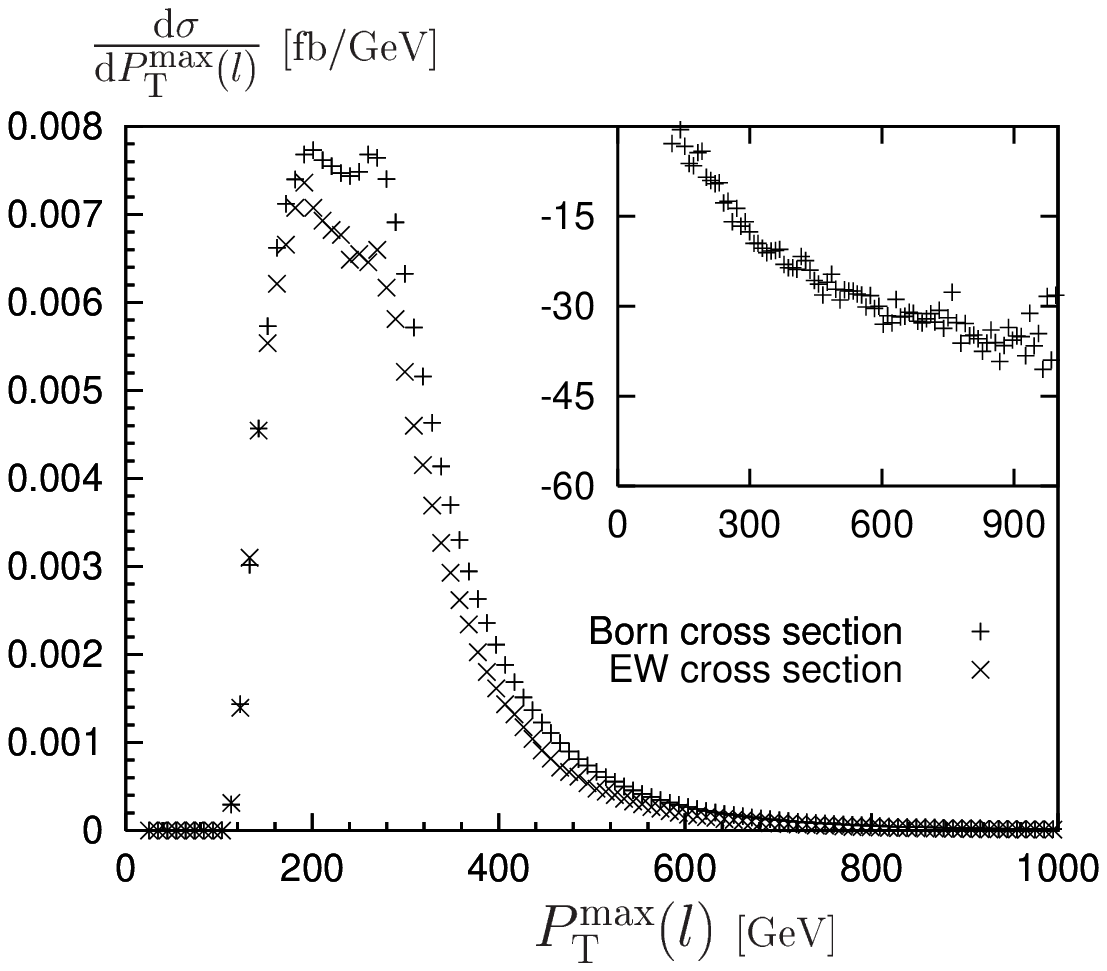,width=8cm}}
  \put(8,7){\epsfig{file=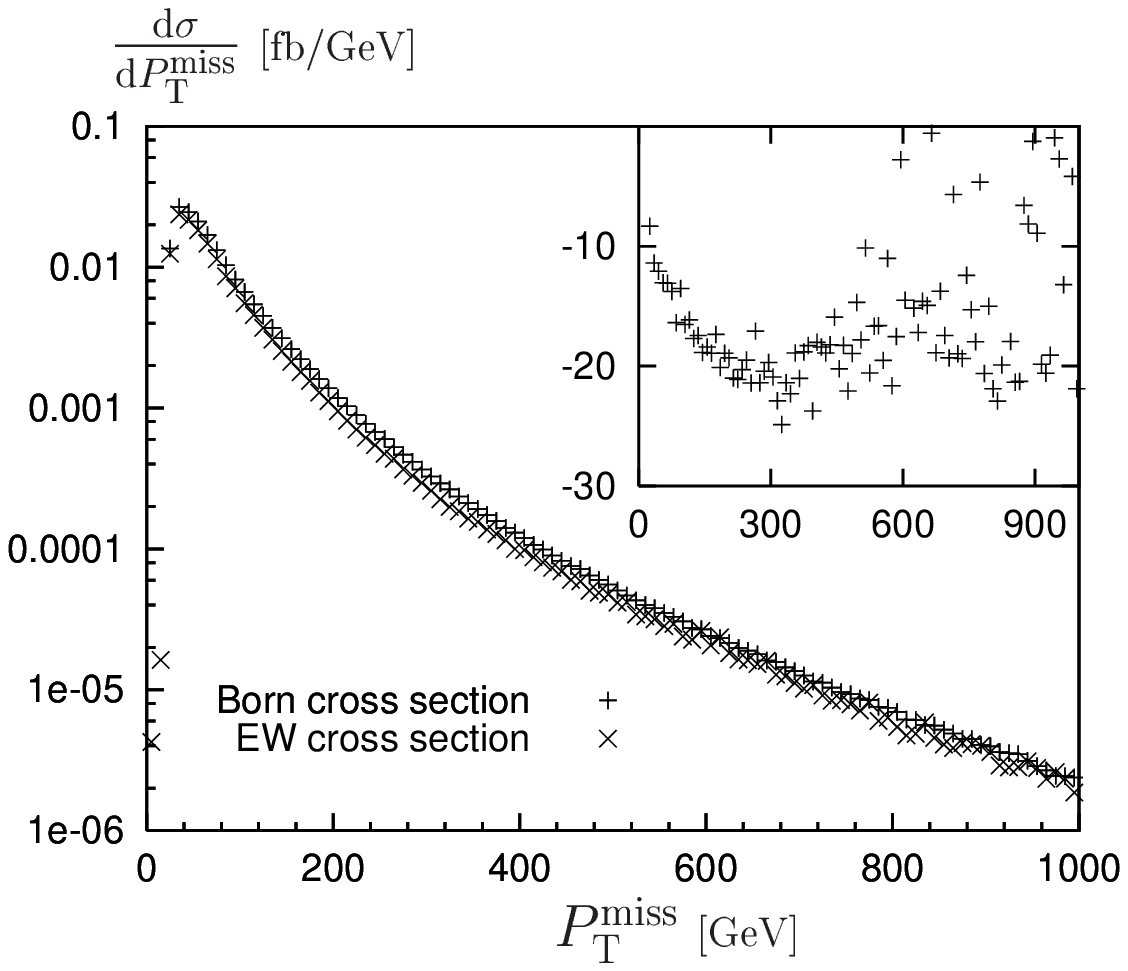,width=8cm}}
  \put(0,0){\epsfig{file=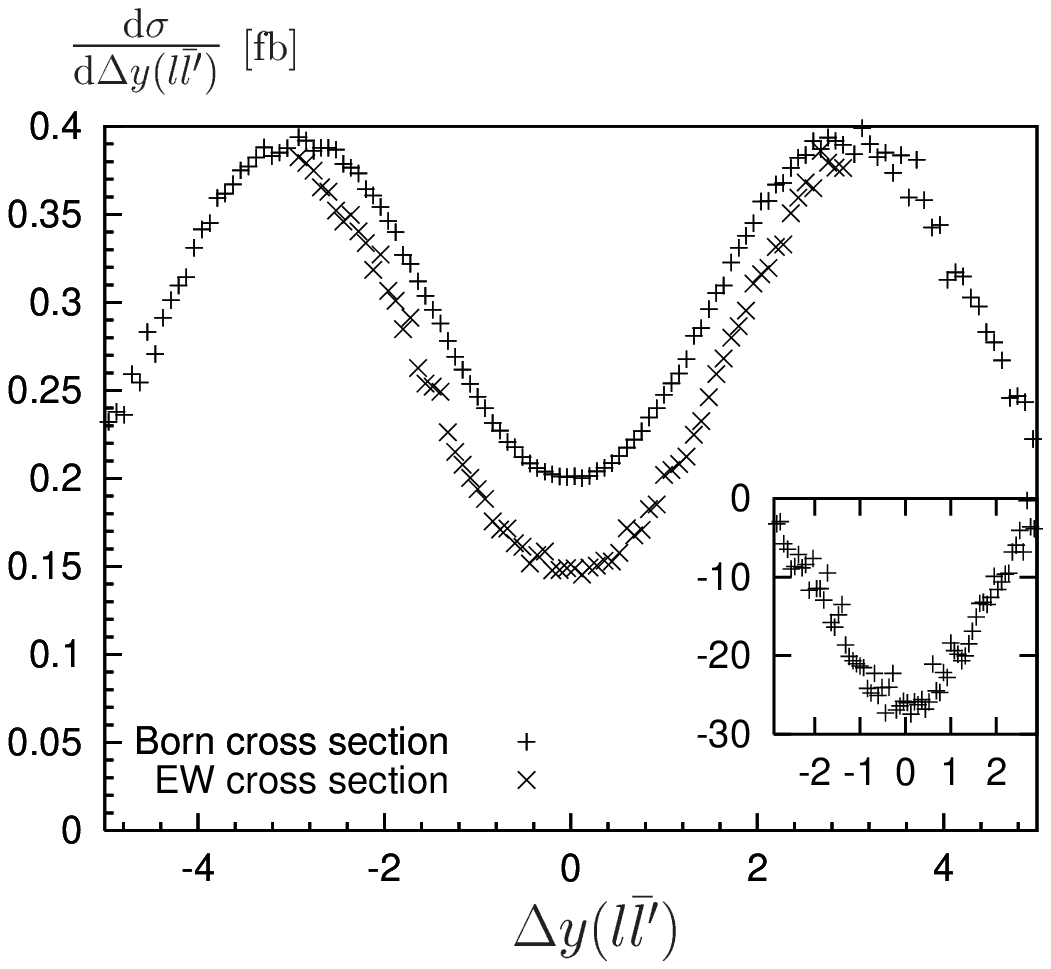,width=8cm}}
  \put(8,0){\epsfig{file=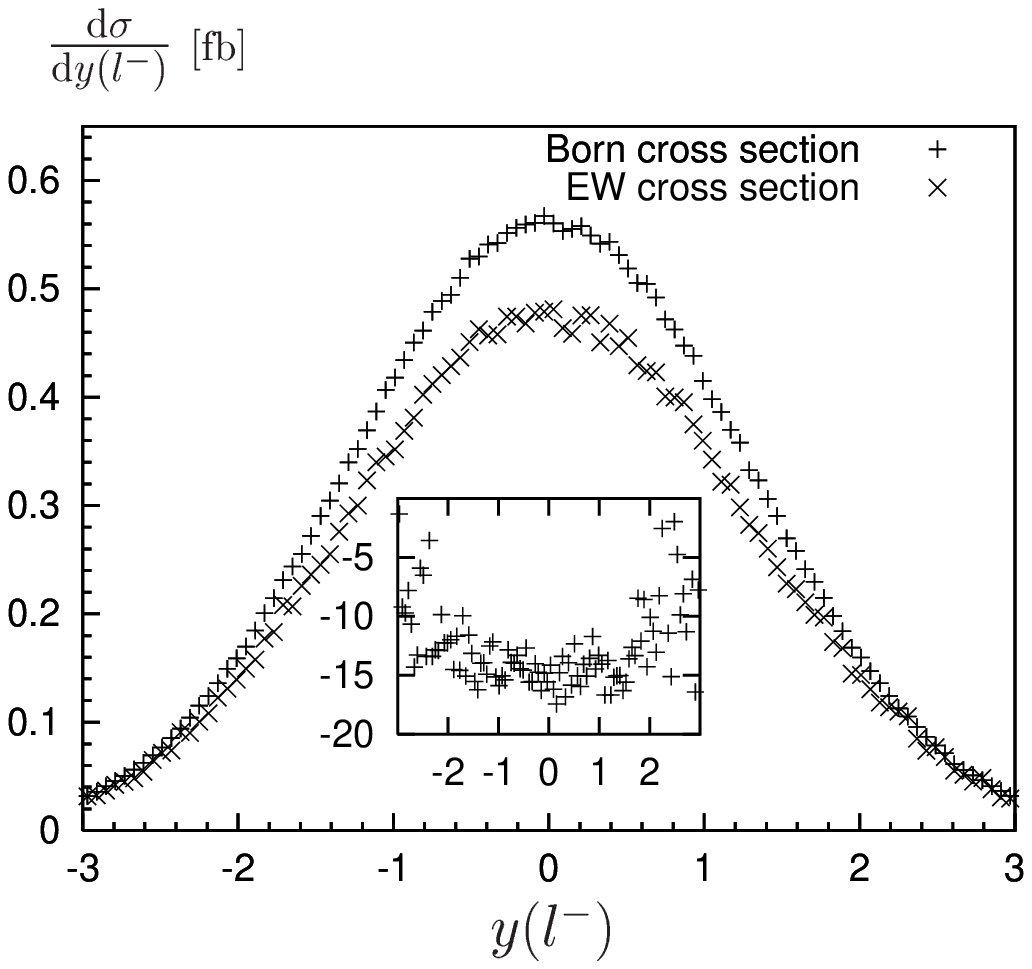,width=8cm}}
  \end{picture}
  \end{center}
\caption{Distributions for $\PW\PW$ production:
  (a) Maximal transverse momentum of the charged leptons. (b) Missing
  transverse momentum.  (c) Difference in rapidity between the two
  charged leptons.  (d) Rapidity of the negatively charged lepton.
  The final state is $\nu_e\Pep\mu^-\bar\nu_\mu$ with standard cuts as
  well as $\Minv(l\bar{l^\prime})> 500\GeV$ and $|\Delta
  y(l\bar{l^\prime})|< 3$ applied.  The last cut is omitted to the
  $\De y(l\bar{l^\prime})$ distribution in lowest order. The inset
  plots show the $\Oa$ corrections relative to the Born results in per
  cent.}
\label{fi:WW_s2}
\end{figure}
In \reffi{fi:WW_s2} we show the distributions for the final state
$\nu_e\Pep\mu^-\bar\nu_\mu$ with our standard cuts applied.  As in the
previous two cases, $\Oa$ corrections are enhanced at high energy and
large scattering angles.  This translates into larger radiative
corrections in the tails of transverse momentum distributions and in
the central region of rapidity distributions. Let us note that also in
this case the partonic process at Born level does not vanish for any
scattering angle, independently on the $\PW$-boson polarization. The
dip appearing in the distribution of the rapidity difference between
the two charged leptons is this time exclusively due to the chosen set
of cuts.  In absence of any kinematical cuts, the
$\Pp\Pp\rightarrow\PW^+\PW^-$ process is dominated by the $\Pu$-quark
contribution, and the rapidity-difference $\Delta y(l\bar{l^\prime})$
for the partonic process $\bar\Pu\Pu\rightarrow 4f$ is maximal and
symmetric around zero.  The requirement of having a large invariant
mass of the two charged leptons, forces the two leptons to be produced
at large separation angles.  This fact depletes the number of events
in the central region of $\Delta y(l\bar{l^\prime})$ and leaves events
with larger rapidity difference. This gives rise to the shape of
\reffi{fi:WW_s2}.

The general behaviour of EW corrections does not present novelties
compared to the previous cases. The interesting feature of $\PW\PW$
processes is the remarkable statistics of purely leptonic final
states.  As shown in \refta{ta:WW_s2}, where we sum over the four
final states $\Pe^-\bar\Pne\nu_\mu\mu^+$, $\mu^-\bar\nu_\mu\Pne\Pe^+$,
$\mu^-\bar\nu_\mu\nu_\mu\mu^+$, and $\Pe^-\bar\Pne\Pne\Pe^+$, the
estimated experimental precision is around a few per cent at CM
energies below 700$\GeV$. On the other hand, the deviation from the
Born result given by the $\Oa$ contributions ranges between
$-14$ and $-18\%$ in the same energy domain. At larger invariant
masses, the overall cross section decreases but radiative corrections
are still of order 2--3 standard deviations.  Thus, a reliable
analysis of these final states requires the inclusion of EW
corrections. Note also that, in contrast to previous processes,
$\Oa$ corrections can be relevant even in the low-luminosity
run ($L=30 \fba^{-1}$). They are about twice the standard deviation
for $\Minvcut(l\bar{l^\prime})\le 700\GeV$, and become comparable with
the experimental accuracy above that threshold.
\begin{table}\centering
$$
\begin{array}{|c|c|c|c|c|c|c|}
\hline 
\multicolumn{7}{|c|}{\Pp\Pp\to 
l\bar\nu_l\nu_{l^\prime}\bar{l^\prime}\rule[-2ex]{0ex}{5ex}}\\
\hline
\Minvcut(l\bar{l^\prime})~[\mathrm{GeV}] & 
\sigma_{\Born}~[\mathrm{fb}] & \sigma_{\AEWS}~[\mathrm{fb}] &
\sigma_{\virt}^{\finite}~[\mathrm{fb}] & 
~\sigma_{\EW} ~[\mathrm{fb}]~ &~\Delta~[\%]~~ & 1/\sqrt{2L\sigma_{\Born}}~[\%] \\
\hline
\hline
500  & 7.235 & 6.561 & 6.682 & 6.235 & -13.8 & 2.6  \\ \hline
600  & 3.723 & 3.280 & 3.350 & 3.131 & -15.9 & 3.7  \\ \hline
700  & 2.059 & 1.765 & 1.808 & 1.688 & -18.1 & 4.9  \\ \hline
800  & 1.201 & 1.003 & 1.031 & 0.959 & -20.2 & 6.5  \\ \hline
900  & 0.731 & 0.596 & 0.613 & 0.570 & -22.0 & 8.3  \\ \hline
1000 & 0.460 & 0.366 & 0.378 & 0.352 & -23.4 & 10.4 \\ \hline
\end{array}$$
\caption {Cross section for $\Pp\Pp\to
  \Pe^-\bar\Pne\nu_\mu\mu^+,\mu^-\bar\nu_\mu\Pne\Pe^+,
  \mu^-\bar\nu_\mu\nu_\mu\mu^+,\Pe^-\bar\Pne\Pne\Pe^+$ 
for $|\Delta y(l\bar{l^\prime})|< 3$ and various values of
$\Minvcut(l\bar{l^\prime})$}  
\label{ta:WW_s2}
\end{table}

\section{Conclusion}
\label{sec:concl}

At the LHC, gauge-boson production processes will be used for precise
measurements of the triple gauge-boson couplings. The relevant
processes to investigate are $\PW\PZ$, $\PZ\PZ$, and $\PW\PW$
production, and the physically interesting region is the one of high
di-boson invariant mass.

We have examined these processes by means of a complete four-fermion
calculation, \ie by taking into account the decays of the gauge
bosons, in the purely leptonic channels.  The primary aim of our
analysis was to investigate the influence of electroweak radiative
corrections on the di-boson production processes at the LHC. The
one-loop logarithmic corrections to the full four-fermion processes
have been calculated in double-pole approximation. This includes
corrections to the gauge-boson-pair-production processes, corrections
to the gauge-boson decays, as well as non-factorizable corrections.
In this study, we have included the full electromagnetic radiative
corrections in the logarithmic approximation, which involve also the
emission of real photons and therefore depend on the detector
resolution. We have verified that the double-pole approximation and
the high-energy approximation are applicable for the considered
phase-space regions of large transverse momentum or large invariant
mass of the gauge-boson pair. Thus, our approach is reliable in this
region. The corrections have been implemented in a Monte Carlo
program, so that arbitrary cuts and distributions can be studied.

In order to illustrate the behaviour and the size of $\Oa$
contributions, we have presented different cross sections and
distributions. For $\PW\PZ$-, $\PZ\PZ$-, and $\PW\PW$-production
processes, electroweak corrections turn out to be sizeable in the
high-energy region of the hard process, in particular for large
transverse momentum and small rapidity separation of the reconstructed
vector bosons, which is the kinematical range of maximal sensitivity
to new-physics phenomena.  Electroweak radiative corrections lower the
Born results for $\PW\PZ$, $\PZ\PZ$, and $\PW\PW$ production by
7--22\%, 15--28\%, and 14--24\%, in the region of experimental
sensitivity. Their size depends sensibly not only on the CM energy but
also on the applied cuts and varies according to the selected
observables and kinematical regions. Despite of the strong decrease of
the cross section with increasing di-boson invariant mass, radiative
effects are appreciable if compared with the expected experimental
precision. This depends of course on the available luminosity. For
$\PW\PZ$ and $\PZ\PZ$ production, these effects are only relevant for
a high-luminosity run of the LHC. Owing to their larger overall cross
section, $\PW\PW$-production processes can instead show a sensitivity
to radiative effects even at a low-luminosity run.

\section*{Acknowledgements}
We thank M.~Roth for his invaluable help concerning the Monte Carlo
generator and S.~Pozzorini for his contributions in the evaluation of
the logarithmic corrections.  This work was supported in part
by the Swiss Bundesamt f\"ur Bildung und Wissenschaft, by the
European Union under contract HPRN-CT-2000-00149, and by the Italian
Ministero dell'Istruzione, dell'Universit\`a e della Ricerca (MIUR)
under contract Decreto MIUR 26-01-2001 N.13.

\appendix

\section*{Appendix}

\section{Non-factorizable photonic corrections}
\label{sec:non-fact-phot}

In this appendix we generalize the results of
\citeres{Denner:2000bj,Denner:1997ia} for the virtual non-factorizable
corrections to a general class of processes.

\subsection{Conventions and notations}

We start by discussing non-factorizable corrections to the generic
process
\begin{equation}\label{process}
g_1(p_1) + g_2(p_2) \;\to\; 
\sum_{l=1}^\nr R_l(k_l) +  \sum _{j=1}^{n_0} f_{0j}(q_{0j}) 
\;\to\; \sum _{l=1}^\nr \sum_{i=1}^{n_l} f_{li}(q_{li}) 
+  \sum _{j=1}^{n_0} f_{0j}(q_{0j}).
\end{equation}
Two incoming particles $g_1$ and $g_2$ with momenta $p_1$ and $p_2$,
masses $m'_1$ and $m'_2$, and charges $Q'_1$ and $Q'_2$ scatter into
$\nr$ resonances $R_l$ with momenta $k_l$, masses $M_l$, decay widths
$\Ga_l$, and charges $Q_l$ and $n_0$
stable particles $f_{0j}$ with momenta $q_{0j}$, masses $m_{0j}$, and
charges $Q_{0j}$. Each resonance $R_l$ then decays into $n_l$ stable
massless particles with momenta $q_{li}$, masses $m_{li}$, and charges
$Q_{li}$. Whereas the charges $Q'_k$ are incoming, all charges $Q_l$
and $Q_{li}$ are assumed to be outgoing.  The masses of
the external particles, which are typically light fermions, are neglected,
except where this would lead to mass singularities.  The complex masses
squared of the resonances are denoted by
\beq
\Mbar_l^2 = M_l^2-\ri M_l \Gamma_l,
\eeq
and we introduce the off-shellness variables
\beq
K_l = k_l^2-M_l^2.
\eeq

We want to give the non-factorizable corrections to the process
\refeq{process} in leading-pole approximation (LPA). The LPA takes
into account only the leading terms in an expansion around the poles
originating from the propagators of the resonances.
For two resonances, the LPA is just the double-pole approximation used
in the main part of this paper.
In LPA, the lowest-order matrix element for process \refeq{process}
factorizes into the matrix element for the production of the $\nr$
on-shell resonances, $\M^{g_1g_2\to R_1\ldots R_N f_{01}\ldots
  f_{0n_0}}_\born(p_1,p_2,k_l,q_{0j})$, the propagators%
\footnote{For gauge bosons only the physical transverse parts enter.}
of these resonances, and the matrix elements for the decays of these
on-shell resonances, $\M^{R_l\to f_{l1}\ldots
  f_{ln_l}}_\born(k_l,q_{li})$:
\beq\label{mborn}
{\cal M}_{\born} = 
\sum_{\mathrm{pol}}
\M^{g_1g_2\to R_1\ldots R_{\nr} f_{01}\ldots f_{0n_0}}_\born 
\prod_{l=1}^{\nr}\frac{\M^{R_l\to f_{l1}\ldots f_{ln_l}}_\born}{K_l}.
\eeq
The sum runs over the physical polarizations of the resonances.

\subsection{Generic form of the correction factor}
\label{se:correctionfactors}

The non-factorizable EW corrections result exclusively from
the exchange of photons that connect the production and decay
subprocesses or two decay subprocesses
\cite{Denner:1997ia,Aeppli:1993rs,Beenakker:1997bp}. Only photons with
energies of the order of the decay widths or smaller are relevant so
that an extended soft-photon approximation, which takes into account
the dependence of the resonant propagators on the photon momenta, can
be used. Consequently, the non-factorizable corrections $\rd\si_\nf$
to the fully differential lowest-order cross section $\rd\si_\born$
resulting from the matrix element \refeq{mborn} take the form of a
correction factor to the lowest-order cross section:
\beq
\rd \si_{\nonfact,\LPA}^\virt = \de^{\virt}_{\nonfact,\LPA}\,
\rd\si_{\born,\LPA} .
\eeq
The non-factorizable corrections get contributions from virtual
photons exchanged 
between a resonance and its decay products ($\mf$), 
between the production process and a resonance ($\im$), 
between the production process and the decay products of a resonance
($\iif$), 
between two resonances ($\mmp$),
between a resonance and the decay products of another resonance ($\mfp$),  
between decay products of different resonances ($\ffp$),
and virtual photons attached to one resonance ($\mm$).
Examples can be found in Figures 1 and 2 of \citere{Denner:1997ia}
or in Figure 2 of \citere{Denner:2000bj}.

Upon splitting the contributions that result from photons coupled to
the charged resonances according to $Q_l=\sum_i Q_{li}$ into
contributions associated with definite final-state fermions and using
$Q'_1+Q'_2= \sum_{l=0}^{\nr} \sum_{i=1}^{n_l} Q_{li}$ to rewrite the
terms originating from the ($\mf$) and ($\mm$) contributions, the
complete correction factor to the lowest-order cross section can be
written as
\beqar\label{nfcorrfac}
\de^{\virt}_{\nonfact,\LPA}&=& -\sum_{l=1}^{N-1} \,\sum_{m=l+1}^{N}
\sum_{i=1}^{n_l}\sum_{j=1}^{n_m}  \, Q_{li} Q_{mj} \,
\frac{\alpha}{\pi} \, \Re\{\Delta_1(k_{l},q_{li};k_{m},q_{mj})\}
\nl&&{}-
\sum_{k=1}^2 \sum_{l=1}^{N} \,
\sum_{i=1}^{n_l} \, Q'_{k} Q_{li} \,
\frac{\alpha}{\pi} \, \Re\{\Delta_2(p_k;k_{l},q_{li})\}
\nl&&{}+
\sum_{j=1}^{n_0} \sum_{l=1}^{N} \,
\sum_{i=1}^{n_l} \, Q_{0j} Q_{li} \,
\frac{\alpha}{\pi} \, \Re\{\Delta_2(-q_{0j};k_{l},q_{li})\}
.
\eeqar

The quantity $\De_1$ gets contributions from $\Delta^\virt_{\mmp}$,
$\Delta^\virt_{\mfp}$, $\Delta^\virt_{\ffp}$, $\Delta^\virt_{\mf}$, and
$\Delta^\virt_{\mm}$,
\beqar\label{eq:Deltao}
\Delta_1(k_{l},q_{li};k_{m},q_{mj}) &=& 
(\Delta^\virt_{\mmp}+\Delta^\virt_{\mfp}+\Delta^\virt_{\ffp})(k_{l},q_{li};k_{m},q_{mj})
\nl&&{}-
(\Delta^\virt_{\mf}+\Delta^\virt_{\mm})(k_{l},q_{li})-
(\Delta^\virt_{\mf}+\Delta^\virt_{\mm})(k_{m},q_{mj}),
\eeqar
and $\De_2$ gets contributions from $\Delta^\virt_{\im}$,
$\Delta^\virt_{\iif}$, $\Delta^\virt_{\mf}$, and $\Delta^\virt_{\mm}$,
\beqar\label{eq:Deltat}
\Delta_2(p_k;k_{l},q_{li}) &=& 
(\Delta^\virt_{\im}+\Delta^\virt_{\iif})(p_k;k_{l},q_{li})+
(\Delta^\virt_{\mf}+\Delta^\virt_{\mm})(k_{l},q_{li}).
\eeqar

The contributions of the different types of diagrams are given by
\beqar\label{eq:Delta1}
\Delta^\virt_{\ffp} &\sim& {}
-2\scpr{q_{li}}{q_{mj}} K_lK_{m} E_0(-q_{mj},-k_{m},k_l,q_{li},\lambda,m_{mj},\Mbar_{m},\Mbar_l,m_{li}),
\nl[.5em]
\Delta^\virt_{\mfp} &\sim& {}
- 2\scpr{k_l}{q_{mj}}K_{m} D_0(-q_{mj},-k_{m},k_l,0,m_{mj},\Mbar_{m},\Mbar_l)
\nn\\ && {}
- 2\scpr{k_{m}}{q_{li}}K_l D_0(-q_{li},-k_l,k_{m},0,m_{li},\Mbar_l,\Mbar_{m}),
\nl[.5em]
\Delta^\virt_{\mathrm{if}} &\sim&{}
-2\scpr{p_k}{q_{li}}K_l D_0(p_k,k_l,q_{li},\lambda,m'_{k},\Mbar_l,m_{li}),
\nl[.5em]
\Delta^\virt_{\mmp} &\sim& {} -2\scpr{k_l}{k_{m}}\biggl\{ C_0(k_l,-k_{m},0,\Mbar_l,\Mbar_{m})
- \Bigl[C_0(k_l,-k_{m},\la,M_l,M_{m})\Bigr]_{k_{l,m}^2=M_{l,m}^2} \biggr\},
\hspace{2em}
\nl[.5em]
\Delta^\virt_{\mathrm{im}} &\sim& {}
-2\scpr{p_k}{k_l}\biggl\{C_0(p_k,k_l,0,m'_k,\Mbar_l)
-\Bigl[C_0(p_k,k_l,\lambda,m'_k,M_l)\Bigr]_{k_l^2=M_l^2} \biggr\},
\hspace{2em}\nl[.5em]
\Delta^\virt_{\mathrm{mf}} &\sim& {}
-2\scpr{k_l}{q_{li}}\biggl\{ C_0(k_l,q_{li},0,\Mbar_l,m_{li})
-\Bigl[C_0(k_l,q_{li},\lambda,M_l,m_{li})\Bigr]_{k_l^2=M_l^2}\biggr\} ,
\nl[.5em]
\Delta^\virt_{\mathrm{mm}} &\sim& {}
-2M_l^2\Biggl\{ \frac{B_0(k_l^2,0,\Mbar_l)-B_0(\Mbar_l^2,0,\Mbar_l)}{k_l^2-\Mbar_l^2}
-B'_0(M_l^2,\lambda,M_l) \Biggr\}.
\label{eq:Delta2}
\eeqar
Note that these definitions deviate partially in sign and form from
those used in \citeres{Denner:2000bj,Denner:1997ia}. The symbol
``$\sim$'' in \refeq{eq:Delta1} indicates that the limits $k_l^2\to
M_l^2$ and $\Ga_l\to 0$ are implicitly understood whenever possible.
The definition of the scalar integrals $B_0$, $C_0$, $D_0$, $E_0$ and
of their arguments can be found in \citeres{Denner:kt,Denner:1997ia}.
The explicit expressions of these functions have been given in
\citeres{Denner:2000bj,Denner:1997ia} for equal masses of the
resonances. The generalized expressions for arbitrary masses are
listed in \refapp{app:scalints}.

\subsection{Simplifications of the generic correction factor}

Using the explicit expression for the loop integrals given in
\refapp{app:scalintsgen}, the terms in the correction factor can be
simplified. The results given in the following are only valid for
$(k_l-q_i)^2=0$, which holds if the resonances decay into two massless
particles.

The sum $\Delta^\virt_{\mfp}+\Delta^\virt_{\ffp}$ can be
simplified by inserting the decompositions of the 5-point function
\refeq{vE5redpole}. In LPA this leads to
\beqar\label{denf1}
\lefteqn{(\Delta^\virt_{\mfp}+\Delta^\virt_{\ffp})(k_m,q_j;k_l,q_i) \sim
\frac{K_lK_ms_{ij}\det(Y_0)}{\det(Y)}D_0(0)}\quad
\nn\\ && {}
+\frac{K_l\det(Y_3)}{\det(Y)}
\left\{[K_l\stwo_{mi}+K_mM_l^2]D_0(1) + K_ms_{ij}D_0(3)\right\} 
\nn\\ && {}
+\frac{K_m\det(Y_2)}{\det(Y)}
\left\{[K_m\stwo_{lj}+K_lM_m^2]D_0(4) + K_ls_{ij}D_0(2)\right\}
,
\eeqar
where $s_{ij}$ and $\stwo_{lj}$ are defined in
\refeq{eq:shorthands} and $D_0(l)$ in \refeq{D0_ints}.
Note that $\Delta^\virt_{\mfp}$ is exactly cancelled by the contributions 
of the last two terms in \refeq{vE5redpole}. 
Inserting the expressions for the scalar integrals
into the different contributions, we find using the first relation in 
\refeq{reldetYipole}
\beqar\label{eq:simp}
\lefteqn{(\Delta^\virt_{\mfp}+\Delta^\virt_{\ffp})(k_m,q_j;k_l,q_i)
-\Delta^\virt_{\mf}(k_m,q_j)-\Delta^\virt_{\mf}(k_l,q_i)
 }\qquad
\nl&\sim&{}
\frac{K_lK_ms_{ij}\det(Y_0)}{\det(Y)}D_0(0) 
+ \frac{K_l \det(Y_3)}{\det( Y)} F(k_m,q_j;k_l,q_i)
\nl&&{}
+ \frac{K_m \det(Y_2)}{\det( Y)} F(k_l,q_i;k_m,q_j)
+\ln\biggl(\frac{\lambda^2}{M_lM_m}\biggr)
\ln\biggl(-\frac{s_{ij}}{M_lM_m}-\ri\epsilon\biggr)
\eeqar
with $D_0(0)$
given in \refeq{D00} and
\beqar\label{Fi}
\lefteqn{F(k_m,q_j;k_l,q_i)=
[K_l\stwo_{mi}+K_mM_l^2]D_0(1) + K_ms_{ij}D_0(3)
}\qquad\nl
&&{}+\ln\biggl(\frac{\lambda^2}{M_lM_m}\biggr)
\ln\biggl(-\frac{s_{ij}}{M_lM_m}-\ri\epsilon\biggr)\nl
&& {}-M_l^2\biggl\{ C_0(k_l,q_i,0,\Mbar_l,m_i)
- \Bigl[C_0(k_l,q_i,\la,M_l,m_i)\Bigr]_{k_l^2=M_l^2} \biggr\}\nl
&& {}-M_m^2\biggl\{ C_0(k_m,q_j,0,\Mbar_m,m_j)
- \Bigl[C_0(k_m,q_j,\la,M_m,m_j)\Bigr]_{k_m^2=M_m^2} \biggr\}
\nl&=&\sum\limits_{\tau=\pm 1}\biggl[
\cLi\biggl(\frac{K_lM_m}{K_mM_l},r_{lm}^\tau\biggr)-
\cLi\biggl(-\frac{M_lM_m}{\stwo_{mi}}+\ri\epsilon,r_{lm}^\tau\biggr)\biggr]
\nl && {}
-2\cLi\biggl(\frac{K_lM_m}{K_mM_l},-\frac{\stwo_{mi}}{M_lM_m}-\ri\epsilon\biggr)
-\Li\biggl(1-\frac{\stwo_{mi}}{s_{ij}}\biggr)
- \ln^2\biggl(-\frac{\stwo_{mi}}{M_lM_m}-\ri \epsilon\biggr)
\nl&&{}
+\ln\biggl(-\frac{s_{ij}}{M_lM_m}-\ri \epsilon\biggr)
 \biggl[ \ln\biggl(-\frac{K_m}{M_lM_m}\biggr)
+ \ln\biggl(-\frac{K_m}{M_m^2}\biggr)\biggr]
.
\end{eqnarray}
The dilogarithms $\Li$, $\cLi$ are defined in \refeqs{cLi} and
\refeqf{Li}.  The quantity $\De_1$ is then obtained from
\refeq{eq:simp} and
\beqar
\Delta^\virt_{\mm}(k_l,q_i) &=& -2\ln\biggl(\frac{\la M_l}{-K_l}\biggr)-2, \nl
\Delta^\virt_{\mmp}(k_m,q_j;k_l,q_i)&=& -\stwotwo_{lm}\biggl\{ C_0(k_l,-k_m,0,\Mbar_l,\Mbar_m)
- \Bigl[C_0(k_l,-k_m,\la,M_l,M_m)\Bigr]_{k_l^2=M_l^2} \biggr\}\nln
\eeqar
as 
\beqar\label{eq:Deltaores}
\Delta_1(k_m,q_{j};k_l,q_{i}) 
&\sim&
\frac{K_lK_ms_{ij}\det(Y_0)}{\det(Y)}D_0(0) 
+ \frac{K_l \det(Y_3)}{\det( Y)} F(k_m,q_j;k_l,q_i)
\nl&&{}
+ \frac{K_m \det(Y_2)}{\det( Y)} F(k_l,q_i;k_m,q_j)
+\ln\biggl(\frac{\lambda^2}{M_lM_m}\biggr)
\ln\biggl(-\frac{s_{ij}}{M_lM_m}-\ri\epsilon\biggr)\nl
&& {}-\stwotwo_{lm}\biggl\{ C_0(k_l,-k_m,0,\Mbar_l,\Mbar_m)
- \Bigl[C_0(k_l,-k_m,\la,M_l,M_m)\Bigr]_{k_l^2=M_l^2} \biggr\}\nl
&& {}+2\ln\biggl(\frac{\la M_l}{-K_l}\biggr) 
+2\ln\biggl(\frac{\la M_m}{-K_m}\biggr)+4 .
\eeqar

When inserting the explicit expressions for the integrals, the
quantity $\De_2$ simplifies to
\beqar\label{eq:Deltatres}
\Delta_2(p_k;k_l,q_i) &=& 
(\Delta^\virt_{\im}+\Delta^\virt_{\iif})(p_k;k_l,q_{i})+
(\Delta^\virt_{\mf}+\Delta^\virt_{\mm})(k_l,q_{i})\nl
&=&2\ln\biggl(\frac{\la
  M_l}{-K_l}\biggr)\biggl[\ln\biggl(\frac{\ttwo_{kl}}{t_{ki}}\biggr)-1\biggr] 
-2 -\Li\left(1-\frac{\ttwo_{kl}}{t_{ki}}\right)
\eeqar
with $t_{ki}$ and $\ttwo_{kl}$ defined in \refeq{eq:shorthands}.

\subsection{Generic correction factor in the high-energy limit}

Using the expressions for the scalar integrals in the
high-energy limit given in \refapp{se:scalint_he}, we find
\beqar\label{eq:Deltaoreshe}
\lefteqn{\Delta_1(k_m,q_j;k_l,q_i)\sim
\frac{1}{2}(s_{ij}\stwotwo_{lm}-\stwo_{lj}\stwo_{mi})
D_0(q_j-k_m,q_j+k_l,q_i+q_j,0,M_m,M_l,0)}\quad  \nl
&&{}+
\ln\biggl(\frac{K_mM_l}{K_lM_m}\biggr)
\ln\biggl(\frac{\stwo_{mi}}{\stwo_{lj}}\biggr)
+\biggl[2+\ln\biggl(\frac{s_{ij}}{\stwotwo_{lm}}\biggr)\biggl]
\biggl[\ln\biggl(\frac{\la M_m}{-K_m}\biggr)
+ \ln\biggl(\frac{\la M_l}{-K_l}\biggr)\biggr]
\eeqar
with $D_0$ defined in \refeq{D00he} and
\beqar\label{eq:Deltatreshe}
\Delta_2(p_k;k_l,q_i) &=& 
2\ln\biggl(\frac{\la M_l}{-K_l}\biggr)
\biggl[\ln\biggl(\frac{\ttwo_{kl}}{t_{ki}}\biggr)-1\biggr].
\eeqar
Note that we always assume that the resonances decay into a pair of
massless particles.

\section{Scalar integrals for non-factorizable corrections}
\label{app:scalints}

\subsection{Scalar integrals in pole approximation}
\label{app:scalintsgen}

In \refse{se:correctionfactors} we have given the virtual
non-factorizable corrections in terms of scalar one-loop integrals.
In this appendix we list the explicit expressions for these 
integrals.  We have the on-shell conditions for the external particles
\beq
p_k^2=(m'_k)^2,\qquad q_i^2=m_i^2,
\eeq
and all expression are given for $k_l^2\to M_l^2$ and $\Ga_l\to 0$,
\ie we neglect $k_l^2-M_l^2$ and $\Ga_l$ everywhere where this does not
give rise to singularities.
Moreover, we assume 
\beq
(k_l-q_i)^2=0,
\eeq
which holds if the resonances decay into a pair of massless particles.

We introduce the shorthand notations
\beqar\label{eq:shorthands}
K_l &=& k_l^2 - \Mbar_l^2\nl
 t_{ki}&=& -2\scpr{p_k}{q_i} = (p_k-q_i)^2, \qquad 
\ttwo_{kl} = -2\scpr{p_k}{k_l} \sim (p_k-k_l)^2 - M_l^2 , \qquad  \nl
s_{ij} &=&  2\scpr{q_i}{q_j} = (q_i+q_j)^2,\qquad
\stwo_{lj} = 2\scpr{k_l}{q_j} \sim (k_l+q_j)^2-M_l^2,\nl
\stwotwo_{lm} &=& 2\scpr{k_l}{k_m} \sim (k_l+k_m)^2-M_l^2-M_m^2, \nl
w_{lm} &=& \sqrt{\la[(k_l+k_m)^2,M_l^2,M_m^2]},\nl
r_{lm} &=& \frac{1}{2M_lM_m}(-\stwotwo_{lm}+w_{lm})
\left(1-\frac{\ri\eps}{w_{lm}}\right),\nl
\kappa_{lmij} &=&
\sqrt{\la[s_{ij}(\stwotwo_{lm}+s_{ij}-\stwo_{lj}-\stwo_{mi}),
(\stwo_{lj}-s_{ij})(\stwo_{mi}-s_{ji}),M_l^2M_m^2]}\nl
&=& \sqrt{\la[4\scpr{q_i}{q_j}\scpr{(k_l-q_i)}{(k_m-q_j)},
4\scpr{(k_l-q_i)}{q_j}\scpr{q_i}{(k_m-q_j)},M_l^2M_m^2]}\,,
\eeqar
where $\ri\eps$ is an infinitesimal imaginary part,
and use the definitions
\beq
\lambda(x,y,z) = x^2+y^2+z^2-2xy-2xz-2yz
\eeq
and
\beqar\label{cLi}
\cLi(x,y) &=& \Li(1-xy)+[\,\ln(xy)-\ln(x)-\ln(y)]\ln(1-xy),
\nl && 
|\arc(x)|,|\arc(y)|<\pi
\eeqar
with the usual dilogarithm
\beq\label{Li}
\Li(z) = -\int_0^z\,\frac{\rd t}{t}\,\ln(1-t),
\qquad |\arc(1-z)|<\pi. 
\eeq

The various combinations of scalar integrals read: \\
case $\mm$
\beqar
\frac{B_0(k_l^2,0,\Mbar_l)-B_0(\Mbar_l^2,0,\Mbar_l)}{k_l^2-\Mbar_l^2}
-B'_0(M_l^2,\lambda,M_l) \sim
\frac{1}{M_l^2}\biggl\{ \ln\biggl(\frac{\lambda M_l}{-K_l}\biggr)+1
\biggr\},
\qquad
\label{eq:B0}
\eeqar
case $\mf$
\beqar 
\lefteqn{C_0(k_l,q_i,0,\Mbar_l,m_i)
-\Bigl[C_0(k_l,q_i,\lambda,M_l,m_i)\Bigr]_{k_l^2=M_l^2}}\qquad&&\hspace{30em}
\nn\\
&\sim&
-\frac{1}{M_l^2} \biggl\{ 
\ln\biggl(\frac{m_i^2}{M_l^2}\biggr)
\ln\biggl(\frac{-K_l}{\lambda M_l}\biggr)
+\ln^2\biggl(\frac{m_i}{M_l}\biggr) + \frac{\pi^2}{6} \biggr\},
 \qquad
\eeqar
case $\im$
\beqar 
\lefteqn{C_0(p_k,k_l,0,m'_k,\Mbar_l)
-\Bigl[C_0(p_k,k_l,\lambda,m'_k,M_l)\Bigr]_{k_l^2=M_l^2}}\qquad &&\hspace*{30em}
\nn\\
&\sim&
\frac{1}{\ttwo_{kl}} \biggl\{ 
\ln\biggl(\frac{m'_kM_l}{-\ttwo_{kl}}+\ieps\biggr)
\biggl[ \ln\biggl(\frac{K_l}{\ttwo_{kl}}\biggr)
+\ln\biggl(\frac{-K_l}{\lambda^2}\biggr)
+\ln\biggl(\frac{m'_k}{M_l}\biggr) \biggr] + \frac{\pi^2}{6} \biggr\},
 \qquad
\eeqar
case \mmp\
\beqar \label{eq:Immp} 
\lefteqn{ C_0(k_l,-k_m,0,\Mbar_l,\Mbar_m) -
\Bigl[C_0(k_l,-k_m,\la,M_l,M_m)\Bigr]_{k_l^2=M_l^2,{k_m}^2=M_m^2} }
\qquad\nn &&\hspace*{30em}\\
&\sim& \frac{1}{w_{lm}}\biggl\{
\cLi\biggl(\frac{K_lM_m}{K_mM_l},\frac{1}{r_{lm}}\biggr)
-\cLi\biggl(\frac{K_lM_m}{K_mM_l},r_{lm}\biggr)
+\cLi(r_{lm},r_{lm})
\nl && {}
+\ln^2(r_{lm})
+2\ln(r_{lm})\ln\biggl(\frac{-K_m}{M_m\la}\biggr)
\biggr\},
\eeqar
case \iif
\beqar 
\lefteqn{
D_0(p_k,k_l,q_i,\lambda,m'_k,\Mbar_l,m_i) \sim
-\frac{1}{t_{ki} K_l} \biggl\{
2\ln\biggl(\frac{-t_{ki}}{m'_k m_i}-\ieps\biggr)
\ln\biggl(\frac{\lambda M_l}{-K_l}\biggr)
} \qquad &&\hspace*{30em}
\nn\\
&&{}
+\ln^2\biggl(\frac{-\ttwo_{kl}}{m'_k M_l}-\ieps\biggr)
+\ln^2\biggl(\frac{m_i}{M_l}\biggr)
+\frac{\pi^2}{3}+\Li\biggl(1-\frac{\ttwo_{kl}}{t_{ki}}\biggr)
\biggr\},
 \qquad
\eeqar
case \mfp
\beqar 
\lefteqn{D_0(1) = D_0(-q_i,-k_l,k_m,0,m_i,\Mbar_l,\Mbar_m) 
 = D_0(-k_m,k_l,q_i,0,\Mbar_m,\Mbar_l,m_i) }
\qquad &&\hspace*{30em}\nn\\
&\sim& \frac{1}{K_l \stwo_{mi}+K_m M_l^2} \biggl\{
\sum_{\tau=\pm 1} \biggl[
 \cLi\biggl(\frac{K_l M_m}{K_m M_l},r_{lm}^\tau\biggr)
-\cLi\biggl(-\frac{M_lM_m}{\stwo_{mi}}+\ri\epsilon,r_{lm}^\tau\biggr)
\biggr]
\nl && {}
-2\cLi\biggl(\frac{K_l M_m}{K_m M_l},-\frac{\stwo_{mi}}{M_lM_m}-\ri\epsilon\biggr)
\nl&& {}
-\ln\biggl(\frac{m_i^2}{M_l^2}\biggr) \biggl[
\ln\biggl(\frac{K_l M_m}{K_m M_l}\biggr)
+\ln\biggl(-\frac{\stwo_{mi}}{M_lM_m}-\ri\epsilon\biggr) \biggr]
\biggr\},
\eeqar
case \ffp
\beqar \label{D00}
\lefteqn{ D_0(0) = D_0(-k_m+q_j,k_l+q_j,q_i+q_j,0,M_m,M_l,0) }
\qquad &&\hspace*{30em}\nn\\
&\sim& \frac{1}{\kappa_{lmij}} \sum_{\si=1,2} (-1)^\si \biggl\{
 \cLi\biggl(-\frac{\stwo_{lj}}{M_lM_m}-\ri\epsilon,-x_\si\biggr)
+\cLi\biggl(-\frac{M_lM_m}{\stwo_{mi}}+\ri\epsilon,-x_\si\biggr)
\nn\\ && \qquad {}
-\cLi\biggl(r_{lm},-x_\si\biggr)
-\cLi\biggl(r_{lm}^{-1},-x_\si\biggr)
-\ln\biggl(\frac{\stwo_{mi}}{s_{ij}}\biggr)\ln(-x_\si)
\biggr\},
\\[.5em] &&
\mbox{with} \qquad
x_1 = \frac{(\stwo_{mi}-s_{ji})z}{M_lM_m}
-\frac{s_{ij}}{\kappa_{lmij}}\ri\eps , 
\quad x_2 =\frac{M_lM_m}{(\stwo_{lj}-s_{ij})z}
+\frac{s_{ij}}{\kappa_{lmij}}\ri\eps, \quad
\nl[.5em] &&\qquad\qquad
z =
\frac{M_l^2M_m^2+\stwo_{lj}\stwo_{mi}-\stwotwo_{lm}s_{ij}+\kappa_{lmij}}
{2(\stwo_{lj}-s_{ij})(\stwo_{mi}-s_{ji})},
\label{eq:xtilde}
\\[1em]
\lefteqn{ D_0(2) = D_0(-q_j,k_l,q_i,\la,m_j,\Mbar_l,m_i) }\qquad
\nn\\*
&\sim& -\frac{1}{K_ls_{ij}} \biggl\{
2\ln\biggl(-\frac{s_{ij}}{m_i m_j}-\ri\epsilon\biggr)
\ln\biggl(\frac{\la M_l}{-K_l}\biggr)
+\ln^2\biggl(-\frac{\stwo_{lj}}{m_jM_l}-\ri\epsilon\biggr)
\nn\\ && {}
+\ln^2\biggl(\frac{m_i}{M_l}\biggr)
+\frac{\pi^2}{3}
+\Li\biggl(1-\frac{\stwo_{lj}}{s_{ij}}\biggr)
\biggr\}.
\eeqar

The 5-point function 
\beq\label{E0}
E_0 = E_0(-q_j,-k_m,k_l,q_i,\la,m_j,\Mbar_m,\Mbar_l,m_i)
\eeq
can be reduced to the five 4-point functions
\beqar\label{D0_ints}
D_0(0) &=& D_0(-k_m+q_j,k_l+q_j,q_i+q_j,m_j,M_m,M_l,m_i),
\nl
D_0(1) &=& D_0(-k_m,k_l,q_i,0,\Mbar_m,\Mbar_l,m_i),
\nl
D_0(2) &=& D_0(-q_j,k_l,q_i,\la,m_j,\Mbar_l,m_i),
\nl
D_0(3) &=& D_0(-q_j,-k_m,q_i,\la,m_j,\Mbar_m,m_i)
\nlc
D_0(4) &=& D_0(-q_j,-k_m,k_l,0,m_j,\Mbar_m,\Mbar_l)
\eeqar
according to
\beq\label{E0redf}
E_0 = -\sum_{l=0}^4 \frac{\det(Y_l)}{\det Y} D_0(l).
\eeq
The symmetric matrix $Y$ reads (using $\stwo_{li}=M_l^2$)
\beqar\label{Yvpole}
Y &=& \left(\begin{array}{@{\ }c@{\ \ }c@{\ \ }c@{\ \ }c@{\ \ }c@{\ }}
{}0   & 0    & -K_m     & -K_l         & 0 \\
{}*   & 0    & M_m^2    & \;(-K_l-\stwo_{lj})\;        & -s_{ij} \\
{}*   & *    & 2M_m^2   & \; (-\stwotwo_{lm}-K_l-K_m)\;& (-K_m -\stwo_{mi})\\
{}*   & *    & *        & 2M_l^2       & M_l^2 \\
{}*   & *    & *        & *            & 0 
\end{array}\right),
\eeqar
and $Y_i$ is obtained from $Y$ by replacing all entries in the $i$th
column with~1.

Neglecting terms that do not contribute to the correction factor in
LPA, the corresponding determinants are given by
\beqar\label{detYvpole}
\det(Y) &\sim& 2s_{ij} \Bigl[
K_lK_m(s_{ij}\stwotwo_{lm}-\stwo_{lj}\stwo_{mi}-M_l^2M_m^2) 
\nl&&\qquad {}
+ K_l^2M_m^2(s_{ij}-\stwo_{mi}) + K_m^2M_l^2(s_{ij}-\stwo_{lj})\Bigr],
\nl
\det(Y_0) &\sim&
\la(s_{ij}\stwotwo_{lm},\stwo_{lj}\stwo_{mi},M_l^2M_m^2)
+4s_{ij}M_l^2M_m^2(\stwo_{lj}+\stwo_{mi}-s_{ij})
,
\nl{}
\det(Y_1)
&\sim& K_l\left[M_l^2M_m^2(\stwo_{mi}-2s_{ij})+
                \stwo_{mi}(s_{ij}\stwotwo_{lm}-\stwo_{lj}\stwo_{mi})
\right]
\nn\\ && {}
- K_m M_l^2(s_{ij}\stwotwo_{lm}+\stwo_{lj}\stwo_{mi}
-2s_{ij}\stwo_{mi}-M_l^2M_m^2),
\nl
\det(Y_2)
&\sim& s_{ij}\left[ K_l(\stwo_{lj}\stwo_{mi}-s_{ij}\stwotwo_{lm}
+M_l^2M_m^2)
                +2K_mM_l^2(\stwo_{lj}-s_{ij}) \right],
\nl
\det(Y_3) &=& \det(Y_2)\big|_{i\leftrightarrow j, l\leftrightarrow m},
\nl
\det(Y_4) &=& \det(Y_1)\big|_{i\leftrightarrow j, l\leftrightarrow m}.
\eeqar

The specific determinants \refeq{detYvpole} appearing in the reduction
for the IR-singular $E_0$ \refeq{E0} with massless external lines obey
the relations
\beqar\label{reldetYipole}
0 &=& \det(Y) + K_m\det(Y_2) +   K_l\det(Y_3), \nl
0 &=& -\stwo_{mi}\det(Y) + K_ms_{ij}\det(Y_1)
        - \left[ K_l\stwo_{mi}+K_mM_l^2 \right]\det(Y_3), \nl
0 &=& -\stwo_{lj}\det(Y) + K_ls_{ij}\det(Y_4)
        - \left[ K_m\stwo_{lj}+K_lM_m^2 \right]\det(Y_2).
\eeqar
These relations allow us to eliminate $\det(Y_1)$ and $\det(Y_4)$ from
\refeq{E0redf}, resulting in:
\begin{eqnarray}\label{vE5redpole}
\nonumber
\lefteqn{E_0(-q_j,-k_m,k_l,q_i,\la,m_j,M_m,M_l,m_i))=
-\frac{\det(Y_0)}
{\det(Y)} D_0(0)}\quad\\
\nonumber
&&{}- \frac{\det(Y_3)}{\det(Y)K_ms_{ij}}
\Bigl\{[K_l\stwo_{mi}+K_mM_l^2]D_0(1)+K_ms_{ij}D_0(3)\Bigr\}\\
\nonumber
&&{}- \frac{\det(Y_2)}{\det(Y)K_ls_{ij}}
\Bigl\{[K_m\stwo_{lj}+K_lM_m^2]D_0(4)+K_ls_{ij}D_0(2)\Bigr\}\\
&&
-\frac{\stwo_{lj}}{K_ls_{ij}} D_0(4)
-\frac{\stwo_{mi}}{K_ms_{ij}} D_0(1).
\end{eqnarray}

\subsection{Scalar integrals in high-energy approximation}
\label{se:scalint_he}

In this section we list the integrals with the additional
approximation that all invariants are large compared with the masses,
\ie
\beq
s_{ij},\stwo_{lj},\stwotwo_{lm},t_{ki},\ttwo_{kl} \gg M_l^2.
\eeq
We keep only the logarithmic terms and omit also the constant terms.

The various combinations of scalar integrals read: \\
case $\mm$
\beqar
\frac{B_0(k_l^2,0,\Mbar_l)-B_0(\Mbar_l^2,0,\Mbar_l)}{k_l^2-\Mbar_l^2}
-B'_0(M_l^2,\lambda,M_l) \sim
\frac{1}{M_l^2}\ln\biggl(\frac{\lambda M_l}{-K_l}\biggr),
\qquad
\label{eq:B0he}
\eeqar
case $\mf$
\beqar 
\lefteqn{C_0(k_l,q_i,0,\Mbar_l,m_i)
-\Bigl[C_0(k_l,q_i,\lambda,M_l,m_i)\Bigr]_{k_l^2=M_l^2}}\qquad&&\hspace{30em}
\nn\\
&\sim&
-\frac{1}{M_l^2} \ln\biggl(\frac{m_i}{M_l}\biggr)\biggl[ 
2\ln\biggl(\frac{-K_l}{\lambda M_l}\biggr) 
+\ln\biggl(\frac{m_i}{M_l}\biggr)
\biggr],
 \qquad
\eeqar
case $\im$
\beqar 
\lefteqn{C_0(p_k,k_l,0,m'_k,\Mbar_l)
-\Bigl[C_0(p_k,k_l,\lambda,m'_k,M_l)\Bigr]_{k_l^2=M_l^2}}\qquad &&\hspace*{30em}
\nn\\
&\sim&
\frac{1}{\ttwo_{kl}} 
\ln\biggl(\frac{m'_kM_l}{-\ttwo_{kl}}+\ri\eps\biggr)
\biggl[\ln\biggl(\frac{K_l}{\ttwo_{kl}}\biggr)
+\ln\biggl(\frac{-K_l}{\lambda^2}\biggr)
+\ln\biggl(\frac{m'_k}{M_l}\biggr) \biggr],
 \qquad
\eeqar
case \mmp\
\beqar 
\lefteqn{ C_0(k_l,-k_m,0,\Mbar_l,\Mbar_m) -
\Bigl[C_0(k_l,-k_m,\la,M_l,M_m)\Bigr]_{k_l^2=M_l^2,{k_m}^2=M_m^2} }
\qquad\nn &&\hspace*{30em}\\
&\sim& \frac{1}{\stwotwo_{lm}}\biggl\{
\ln\biggl(\frac{M_lM_m}{-\stwotwo_{lm}}+\ri\eps\biggr)
\ln\biggl(\frac{-\stwotwo_{lm}}{\la^2}-\ri\eps\biggr)
+\ln\biggl(\frac{\stwotwo_{lm}}{K_l}\biggr)
\ln\biggl(\frac{\stwotwo_{lm}}{K_m}\biggr)
-\frac{1}{2}\ln^2\biggl(\frac{M_l^2}{-K_l}\biggr)
\nl&&{}
-\frac{1}{2}\ln^2\biggl(\frac{M_m^2}{-K_m}\biggr)
+\frac{1}{4}\ln^2\biggl(\frac{-\stwotwo_{lm}}{M_l^2}-\ri\eps\biggr)
+\frac{1}{4}\ln^2\biggl(\frac{-\stwotwo_{lm}}{M_m^2}-\ri\eps\biggr)
\biggr\},
\eeqar
case \iif
\beqar 
\lefteqn{
D_0(p_k,k_l,q_i,\lambda,m'_k,\Mbar_l,m_i)} \qquad &&\hspace*{30em}\nl
&\sim&
-\frac{1}{t_{ki} K_l} \biggl\{
2\ln\biggl(\frac{-t_{ki}}{m'_k m_i}-\ri\eps\biggr)
\ln\biggl(\frac{\lambda M_l}{-K_l}\biggr)
+\ln^2\biggl(\frac{-t_{ki}}{m'_k m_i}-\ri\eps\biggr)
+\ln^2\biggl(\frac{m_i}{M_l}\biggr)
\biggr\},\nln
\eeqar
case \mfp
\beqar 
\lefteqn{D_0(1) = D_0(-q_i,-k_l,k_m,0,m_i,\Mbar_l,\Mbar_m) }
\qquad\nn\\
&\sim& \frac{1}{K_l \stwo_{mi}} \biggl\{
\ln^2\biggl(\frac{-\stwo_{mi}}{M_l^2}-\ri\eps\biggr)
-\frac{1}{2}\ln^2\biggl(\frac{-\stwotwo_{lm}}{M_l^2}-\ri\eps\biggr)
-\ln\biggl(\frac{-\stwo_{mi}}{M_l^2}-\ri\eps\biggr)
\ln\biggl(\frac{m_i^2}{M_l^2}\biggr)
\nl&&{}
+\frac{1}{2}\ln^2\biggl(\frac{K_l}{K_m}\biggr)
+\ln\biggl(\frac{K_l}{K_m}\biggr)
\biggl[\ln\biggl(\frac{-\stwo_{mi}}{m_i^2}-\ri\eps\biggr)
+\ln\biggl(\frac{\stwo_{mi}+\ri\eps}{\stwotwo_{lm}+\ri\eps}\biggr)
\biggr]
\biggr\},
\qquad
\eeqar
case \ffp
\beqar \label{D00he}
\lefteqn{ D_0(0) = D_0(-k_m+q_j,k_l+q_j,q_i+q_j,0,M_m,M_l,0) }
\qquad\nn\\
&\sim& -\frac{1}{\stwo_{lj}\stwo_{mi}-\stwotwo_{lm}s_{ij}} 
\biggl\{
\cLi\biggl(-\frac{M_lM_m}{\stwo_{mi}}+\ri\epsilon,-x_1\biggr)
+\cLi\biggl(-\frac{M_lM_m}{\stwo_{lj}}+\ri\epsilon,-\frac{1}{x_2}\biggr)
\nl && \qquad {}
-\cLi\biggl(-\frac{M_lM_m}{\stwotwo_{lm}}+\ri\eps,-x_1\biggr)
-\cLi\biggl(-\frac{M_lM_m}{\stwotwo_{lm}}+\ri\eps,-\frac{1}{x_2}\biggr)
\nl&& \qquad {}
-\biggl[\ln\biggl(\frac{\stwo_{mi}+\ri\eps}{s_{ij}+\ri\eps}\biggr)
+\ln\biggl(\frac{\stwo_{lj}+\ri\eps}{\stwotwo_{lm}+\ri\eps}\biggr)
\biggr]\Bigl(\ln(-x_1)-\ln(-x_2)\Bigr)
\biggr\}
\\[.5em] &&
\mbox{with} \qquad
x_1 = \frac{(\stwo_{mi}-s_{ij})z}{M_lM_m}(1-\ri\eps s_{ij}(\stwo_{lj}-s_{ij})),
\quad\nl 
&&\phantom{\mbox{with} \qquad}
x_2 = \frac{M_lM_m}{(\stwo_{lj}-s_{ij})z}(1+\ri\eps s_{ij}(\stwo_{mi}-s_{ij})), \quad
\nl[.5em] &&\qquad\qquad
z =
\frac{\stwo_{lj}\stwo_{mi}-\stwotwo_{lm}s_{ij}}
{(\stwo_{lj}-s_{ij})(\stwo_{mi}-s_{ij})},
\label{eq:xtildehe}
\\[1em]
\lefteqn{ D_0(2) = D_0(-q_j,k_l,q_i,\la,m_j,\Mbar_l,m_i) }\qquad
\nl*
&\sim& -\frac{1}{K_ls_{ij}} \biggl\{
2\ln\biggl(-\frac{s_{ij}}{m_i m_j}-\ri\epsilon\biggr)
\ln\biggl(\frac{\la M_l}{-K_l}\biggr)
+\ln^2\biggl(-\frac{\stwo_{lj}}{m_jM_l}-\ri\epsilon\biggr)
+\ln^2\biggl(\frac{m_i}{M_l}\biggr)
\biggr\}.\nln
\eeqar
We note that $D_0(0)$ does not involve large logarithms and could be
replaced by zero in the logarithmic approximation.

\section{Couplings}
\label{sec:couplings}

In this appendix we list the explicit values for the couplings
$I^{V_a}_{\phi_i\phi_{i'}}$ introduced in \citere{Denner:2001jv} 
required for our calculation. Note that a bar over a field indicates
the charge-conjugated field.

For quarks the couplings $I^{V}_{q_{i,\si} q_{j,\si}}$ depend on the
helicity $\si$ and involve the quark-mixing matrix $V_{ij}$:
\beqar \label{eq:F_couplings}
I^{A}_{q_- q_-} &=& I^{A}_{q_+ q_+} = -Q_q, \nl
I^{Z}_{q_- q_-} &=& \frac{I^3_q-\sw^2 Q_q}{\sw\cw},\qquad
I^{Z}_{q_+ q_+} = -\frac{\sw Q_q}{\cw},\nl
I^{W^+}_{u_{i,-} d_{j,-}} &=& \frac{1}{\sqrt{2}\sw}V_{ij} ,\qquad
I^{W^-}_{d_{j,-} u_{i,-}} = \frac{1}{\sqrt{2}\sw}V^*_{ij} . 
\eeqar
All other quark couplings vanish, and
the couplings for antiquarks are obtained by
\beq
I^{V}_{\bar q_{i,\si} \bar q_{j,\si}} = -I^{V}_{q_{j,-\si} q_{i,-\si}}.
\eeq
For gauge bosons the couplings $I^{V_1}_{\bar V_2V_3}$ are totally
antisymmetric in the field indices $V_1,V_2,V_3$ and read
\beqar \label{eq:V_couplings}
I^{A}_{W^-W^-} &=& I^{W^+}_{W^+A} = I^{W^-}_{AW^+} = 
-I^{A}_{W^+W^+} = -I^{W^-}_{W^-A} = -I^{W^+}_{AW^-} 
=1, \nl
I^{Z}_{W^-W^-} &=& I^{W^+}_{W^+Z} = I^{W^-}_{ZW^+} = 
-I^{Z}_{W^+W^+} = -I^{W^-}_{W^-Z} = -I^{W^+}_{ZW^-} 
= -\frac{\cw}{\sw}. 
\eeqar
For Higgs and would-be Goldstone bosons, the couplings $I^{V}_{\bar S_1S_2}$
  are antisymmetric in the scalar fields $S_1$ and $S_2$ and read
\beqar \label{eq:S_couplings}
I^{Z}_{\chi H} &=& -I^{Z}_{H\chi} = \frac{\ri}{2\cw\sw}, \nl
I^{A}_{\phi^-\phi^-} &=& -I^{A}_{\phi^+\phi^+} = 1, 
\qquad I^{Z}_{\phi^-\phi^-} = -I^{Z}_{\phi^+\phi^+} = 
-\frac{\cw^2-\sw^2}{2\cw\sw},\nl
I^{W^\pm}_{\phi^\pm H} &=& -I^{W^\pm}_{H\phi^\mp} =  \pm\frac{1}{2\sw}, \qquad 
I^{W^\pm}_{\phi^\pm \chi} = -I^{W^\pm}_{\chi\phi^\mp} = \frac{\ri}{2\sw}.
\eeqar

The couplings appearing in the lowest-order matrix elements for
four-fermion production can be read off from Appendix A of
\citere{Denner:kt} and are related to the quantities $I^{V}_{ff'}$
via
\beq\label{eq:rel_C_I}
C^{\si\si'}_{V\bar f f'} = \de_{\si,-\si'} I^{V}_{f_{\si'}f'_{\si'}}.
\eeq

\end{document}